\documentclass[11pt]{article}
\usepackage{fullpage}
\usepackage{times}
\usepackage{amsmath,amsfonts,amssymb,amsthm}
\usepackage{enumerate}
\usepackage{xcolor}
\usepackage{xspace}
\usepackage{tikz}
\usepackage{paralist}
\usepackage{framed}
\usepackage{xparse}
\usepackage{hyperref}
\usepackage{subcaption}
\usetikzlibrary{decorations.pathreplacing, decorations.pathmorphing}

\newtheorem{theorem}{Theorem}
\newtheorem{lemma}[theorem]{Lemma}
\newtheorem{claim}[theorem]{Claim}

\newtheorem{proposition}[theorem]{Proposition}

\newtheoremstyle{restate}{}{}{\itshape}{}{\bfseries}{~(restated).}{.5em}{\thmnote{#3}}
\theoremstyle{restate}
\newtheorem*{restate}{}

\theoremstyle{definition}
\newtheorem{definition}[theorem]{Definition}
\theoremstyle{remark}
\newtheorem*{remark}{Remark}

\allowdisplaybreaks

\newcommand{\cD}{\mathcal{D}}

\newcommand{\cR}{\mathcal{R}}
\newcommand{\cS}{\mathcal{S}}
\newcommand{\cT}{\mathcal{T}}

\newcommand{\cB}{\mathcal{B}}

\newcommand{\tT}{\tilde{T}}
\newcommand{\tS}{\tilde{S}}
\newcommand{\tR}{\tilde{R}}
\newcommand{\tm}{\tilde{m}}

\newcommand{\range}[1]{\mathrm{range}(#1)}

\newcommand{\ph}{\mathtt{perfHash}}

\newcommand{\flip}{\mathtt{flip}}
\newcommand{\skt}{\mathtt{sketch}}
\newcommand{\main}{\mathrm{main}}

\newcommand{\prim}{\mathrm{pr}}
\newcommand{\rand}{\mathrm{sc}}
\newcommand{\unk}{\mathrm{unk}}
\newcommand{\tbl}{\mathtt{table}}
\newcommand{\bad}{\mathrm{bad}}
\newcommand{\fra}{\mathrm{frac}}
\newcommand{\Mod}{\ \mathrm{mod}\ }
\newcommand{\Div}{\ \mathrm{div}\ }

\newcommand{\qalg}{\mathtt{qAlg}}
\newcommand{\rank}{\mathtt{rank}}

\newcommand{\rankPre}{\mathtt{prepRank}}
\newcommand{\rankPreL}{\mathtt{prepRankL}}
\newcommand{\rankPreS}{\mathtt{prepRankS}}
\newcommand{\tblr}{\mathtt{tableRank}}
\newcommand{\tblrL}{\mathtt{tableRankL}}
\newcommand{\tblrS}{\mathtt{tableRankS}}
\newcommand{\qalgr}{\mathtt{qAlgRank}}
\newcommand{\qalgrL}{\mathtt{qAlgRankL}}
\newcommand{\qalgrS}{\mathtt{qAlgRankS}}
\newcommand{\intPre}{\mathtt{prepIntoSet}}
\newcommand{\intPreT}{\mathtt{prepIntoTwo}}
\newcommand{\qalgi}{\mathtt{request}}
\newcommand{\qalgiT}{\mathtt{reqFromTwo}}
\newcommand{\tblint}{\mathtt{tableInt}}
\newcommand{\encset}{\mathtt{encSet}}
\newcommand{\decset}{\mathtt{decSet}}
\newcommand{\phB}{\mathtt{perfHashBlk}}
\newcommand{\tblB}{\mathtt{tableBlk}}
\newcommand{\qalgB}{\mathtt{qalgBlk}}
\newcommand{\vr}{\mathtt{valRet}}
\newcommand{\hash}{\mathtt{hash}}
\newcommand{\aux}{\mathrm{aux}}
\newcommand{\sm}{\mathrm{sc}}
\newcommand{\la}{\mathrm{last}}
\newcommand{\bl}{\mathrm{bl}}
\newcommand{\cl}{\mathrm{cl}}
\newcommand{\smain}{\mathtt{SIZE}_{\main}}

\newcommand{\phS}{\mathtt{perfHashS}}
\newcommand{\tblS}{\mathtt{tableS}}
\newcommand{\qalgS}{\mathtt{qalgS}}

\newcommand{\oh}{\overline{h}}
\newcommand{\new}{\textrm{new}}

\newcommand{\OPT}{\mathtt{OPT}}
\newcommand{\poly}{\mathrm{poly}}

\newsavebox{\mybox}
\NewDocumentEnvironment{code}{mO{Algorithm}O{()}O{#1}}
	{\begin{center}
	\begin{lrbox}{\mybox}\footnotesize%
	\begin{minipage}{5.5in}
	\ifcsname c@pctr#1\endcsname\else\newcounter{pctr#1}\textbf{#2 $\mathtt{#4}#3$}:\fi
	\setlength{\topsep}{0pt}
	\begin{compactenum}
	\setcounter{enumi}{\value{pctr#1}}
	}{\setcounter{pctr#1}{\value{enumi}}
	\end{compactenum}
	\end{minipage}
	\end{lrbox}\fbox{\usebox{\mybox}}
	\end{center}}

\newcommand{\ctn}{\hfill (to be cont'd)}

\begin{document}

\title{Nearly Optimal Static Las Vegas Succinct Dictionary}
\author{Huacheng Yu\thanks{Department of Computer Science, Princeton University. \texttt{yuhch123@gmail.com}}}
\date{}

\setcounter{page}{-1}
\maketitle
\thispagestyle{empty}
\begin{abstract}
Given a set $S$ of $n$ (distinct) keys from key space $[U]$, each associated with a value from $\Sigma$, the \emph{static dictionary} problem asks to preprocess these (key, value) pairs into a data structure, supporting value-retrieval queries: for any given $x\in [U]$, $\vr(x)$ must return the value associated with $x$ if $x\in S$, or return $\bot$ if $x\notin S$.
The special case where $|\Sigma|=1$ is called the \emph{membership} problem.
The ``textbook'' solution is to use a hash table, which occupies linear space and answers each query in constant time.
On the other hand, the minimum possible space to encode all (key, value) pairs is only $\OPT:= \lceil\lg_2\binom{U}{n}+n\lg_2|\Sigma|\rceil$ bits, which could be much less.

In this paper, we design a randomized dictionary data structure using
\[
	\OPT+\poly\lg n+O(\lg\lg\lg\lg\lg U)
\]
bits of space, and it has \emph{expected constant} query time, assuming the query algorithm can access an external lookup table of size $n^{0.001}$.
The lookup table depends only on $U$, $n$ and $|\Sigma|$, and not the input.
Previously, even for membership queries and $U\leq n^{O(1)}$, the best known data structure with constant query time requires $\OPT+n/\poly\lg n$ bits of space (Pagh~\cite{Pagh01} and P\v{a}tra\c{s}cu~\cite{Pat08});
the best known using $\OPT+n^{0.999}$ space has query time $O(\lg n)$;
the only known non-trivial data structure with $\OPT+n^{0.001}$ space has $O(\lg n)$ query time and requires a lookup table of size $\geq n^{2.99}$ (!).
Our new data structure answers open questions by P\v{a}tra\c{s}cu and Thorup~\cite{Pat08,Thorup13}.

We also present a scheme that compresses a sequence $X\in\Sigma^n$ to its zeroth order (empirical) entropy up to $|\Sigma|\cdot\poly\lg n$ extra bits, supporting decoding each $X_i$ in $O(\lg |\Sigma|)$ expected time.
\end{abstract}
\newpage
\thispagestyle{empty}

\tableofcontents

\newpage




\section{Introduction}
Given $n$ (key, value) pairs $\{(k_i,v_i)\}_{i=1,\ldots,n}$ with distinct keys $k_i\in\{0,\ldots,U-1\}$ and (possibly duplicated) values $v_i\in\{0,\ldots,\sigma-1\}$, the \emph{static dictionary} problem asks to preprocess them into a data structure, supporting value-retrieval queries
\begin{compactitem}
	\item $\vr(x)$: return $v_i$ if $x=k_i$, and return $\bot$ if $x\neq k_1,\ldots,k_n$.
\end{compactitem}
When $\sigma=1$, it is called the \emph{membership} problem, i.e., preprocessing a set $S$ of $n$ keys into a data structure, supporting queries of form ``is $x\in S$?''

Dictionaries are very fundamental data structures, which have been extensively studied in theory~\cite{CW79,TY79,Yao81a,FKS84,FNSS92,FN93,BM99,Pagh01,Pagh01b,FM95,Mil96,MNSW98,BMRV02}.
They are also one of the most basic data structures in practice, included in standard libraries for most of the popular programming languages (e.g., \verb!std::unordered_map! for \verb!C++!, \verb!HashMap! for \verb!Java!, and language-level built-in support for \verb!JavaScript!, \verb!Python!, \verb!Ruby!, etc).

The ``textbook'' implementation of a dictionary is to store a hash table: use a hash function to map all keys to $O(n)$ buckets, and store each (key, value) pair in the corresponding bucket.
Simple hash functions (e.g. $(kx\Mod p)\Mod n$) have low collision probabilities, and resolving collisions by chaining leads to a dictionary data structure with \emph{expected} constant query time.
Using perfect hashing (e.g.~\cite{FKS84}), one can further improve the query time to \emph{worst-case} constant.
However, such data structures use at least ${n\lg U+n\lg\sigma}$ bits of space, even just to write down all (key, value) pairs in the buckets, while the information theoretical space lower bound for this problem is only\footnote{Throughout the paper, $\lg$ is the binary logarithm.}
\[
	\OPT:=\lceil\lg\binom{U}{n}+n\lg\sigma\rceil
\]
bits, which is much less than $n\lg U+n\lg\sigma$ (note that $\lg\binom{U}{n}=n\lg (U/n)+O(n)$).

It turns out that it is possible to not \emph{explicitly} store all pairs, and beat $n\lg U+n\lg\sigma$ bits.
For \emph{membership} queries ($\sigma=1$), the previously best known data structure by Pagh~\cite{Pagh01} (and later improved by P\v{a}tra\c{s}cu~\cite{Pat08}) uses $\OPT+O(n/\poly\lg n+\lg\lg U)$ bits of space, and answers queries in constant time.
This data structure also gives a smooth tradeoff between time and space: for query time $O(t)$, it uses space
\[
	\OPT+n/r+O(\lg\lg U),
\]
where $r=(\frac{\lg n}{t})^{\Omega(t)}$.
To achieve this query time, it is assumed that the query algorithm has access to an external lookup table of size $\min\{n^3,r^6\}$, which depends only on $U$ and $n$, and not the input.
In particular, when $U=\poly\ n$, if the number of extra bits is $n^{0.99}$, the query time becomes $O(\lg n)$.
If we want the space to be very close to $\OPT$, the query time is $O(\lg n)$, but the lookup table size becomes about $n^3$ (it may even be larger than the data structure itself).
For $\sigma>1$, only $(\OPT+O(n+\lg\lg U))$-bit data structures were known~\cite{Pagh01}.
While these data structures have deterministic query algorithms (and worst-case query-time guarantee), no better zero-error randomized data structure was known.
To the best of our knowledge, data structures with Las Vegas query algorithms have never been the state-of-the-art for this problem since perfect hashing~\cite{FKS84}.\footnote{Monte Carlo algorithms, where the query is allowed to err with a small probability, would have a different space benchmark. Thus, they are not the focus of this paper.}
Therefore, it was unclear if randomization is even useful for static dictionaries.

\subsection{Our contributions}
In this paper, we show that if we allow randomization, near-optimal space and optimal time can be achieved simultaneously.
We design a dictionary data structure with $\poly\lg n+O(\lg\lg U)$ extra bits and expected constant query time, making a step towards the optimal static dictionary.
The query algorithm only needs to access a small lookup table.
\begin{theorem}\label{thm_main_informal}
	There is a randomized algorithm that preprocesses $n$ (key, value) pairs into a data structure with 
	\[
		\OPT+\poly\lg n+O(\lg\lg U)
	\]
	bits, such that for any given query $x$, the query algorithm answers $\vr(x)$ in expected constant time on a random access machine with word-size $w\geq {\Omega(\lg U+\lg\sigma)}$, assuming it can access an external lookup table of size $n^{\epsilon}$, for any constant $\epsilon>0$.
\end{theorem}
Same as the previous data structures, the lookup table depends only on $U$, $n$ and $\sigma$, and not the input.
In fact, the $\lg\lg U$ term can be improved to $\lg\lg\cdots\lg U$ for logarithm iterated any constant
number of times.
Hence, when $U$ is at most $2^{2^{\raisebox{1pt}{\scalebox{-0.7}[0.7]{$\ddots$}}^{\raisebox{-3pt}{\scalebox{0.6}{$2^n$}}}}}$ with $O(1)$ levels, this term can be removed.
In this case, among the $\OPT+\poly\lg n$ bits of the data structure, the first $\poly\lg n$ are the (plain) random bits used by the preprocessing algorithm, and the ``actual'' data structure only occupies the next (and last) $\OPT+1$ bits.
The expectation of the query time is taken over these random bits, which we assume the worst-case input data and query do not see.
Moreover, the query time is $O(1)$ with probability $1-o(1)$, and is $\poly\lg n$ in worst-case.

By storing the lookup table as part of the data structure, Theorem~\ref{thm_main_informal} implies a data structure with ${\OPT+n^{\epsilon}+O(\lg\lg U)}$ bits and expected constant query time, which is still an improvement over the previous best known.
In the \emph{cell-probe} model, where we only count how many times the query algorithm accesses the memory and the computation is free, the lookup table is not necessary, because it does not depend on the input and can be computed without accessing the data structure.

In the theorem, we assume that each (key, value) pair fits in $O(1)$ words, which is necessary to obtain constant query time on random access machines.
We will discuss larger keys or values in Section~\ref{sec_disc}.

\paragraph{Perfect hashing.}
In general, a perfect hashing maps $n$ input keys to distinct buckets, and it is called \emph{minimal} if it maps them to exactly $n$ distinct buckets, labeled from $0$ to $n-1$.
En route to the new dictionary data structure, the key component is a succinct membership data structure, equipped with a \emph{two-sided} minimal perfect hashing.
Specifically, given a set $S\subseteq [U]$ of size $n$, we would like to construct a data structure $\cD$, which not only supports membership queries, but also defines 
\begin{compactitem}
	\item a bijection $h$ between $S$ and $[n]$, and
	\item a bijection $\oh$ between $[U]\setminus S$ and $[U-n]$.
\end{compactitem}
That is, we want to perfectly hash all keys, as well as all non-keys.
The data structure must support $\hash(x)$ queries, which returns a pair $(b, v)$ such that
\begin{compactitem}
	\item if $x\in S$, $b=1$ and $v=h(x)$;
	\item if $x\notin S$, $b=0$ and $v=\oh(x)$.
\end{compactitem}

\begin{theorem}[informal]\label{thm_ph_informal}
	There is a randomized algorithm that preprocesses a set $S\in [U]$ of size $n$ into a data structure with
	\[
		\lg\binom{U}{n}+\poly\lg n+O(\lg\lg U)
	\]
	bits, such that it defines a bijection $h$ between $S$ and $[n]$ and a bijection between $[U]\setminus S$ and $[U-n]$.
	For any $x$, the query algorithm outputs $\hash(x)$ in expected constant time on a random access machine with word-size $w\geq \Omega(\lg U)$, assuming it has access to an external lookup table of size $n^{\epsilon}$, for any constant $\epsilon>0$.
\end{theorem}
See Section~\ref{sec_general} for the formal statement.

\paragraph{Locally decodable arithmetic codes.}
We also show that the above perfect hashing data structure can be applied to obtain a version of locally decodable arithmetic codes with a better space~\cite{Pat08}.
This problem asks to compress a sequence $X=(x_1,\ldots,x_n)\in\Sigma^n$ for some (small) alphabet set $\Sigma$, such that each symbol $x_i$ can be recovered efficiently from the compression.
We should think of a sequence $X$ that is sampled from some low entropy source, and the encoding should take much less than $n\lg |\Sigma|$ bits.
The arithmetic codes match the zeroth order entropy of $X$, i.e., if each symbol in the sequence has entropy $H$ (marginally), then the encoding has length $\sim n\cdot H$.
P\v{a}tra\c{s}cu~\cite{Pat08} gave a data structure whose size is 
\[
	\sum_{\sigma\in\Sigma}f_{\sigma}\lg\frac{n}{f_{\sigma}}+O(|\Sigma|\lg n)+n/\left(\frac{\lg n}{t}\right)^t+\tilde{O}(n^{3/4}),
\]
where $f_{\sigma}$ is the number of occurrences of $\sigma$.
It supports single-element access in $O(t)$ time on word RAM.
Note that when each symbol $x_i$ is sample independent from a source of entropy $H$, then the empirical entropy $\sum_{\sigma\in\Sigma}f_{\sigma}\lg\frac{n}{f_{\sigma}}\approx n\cdot H$ with high probability.

\begin{theorem}\label{thm_code_informal}
There is a randomized algorithm that preprocesses a sequence $(x_1,\ldots,x_n)\in\Sigma^n$ into a data structure with
\[
	\lg\left(\frac{n!}{f_{\sigma_1}! f_{\sigma_2}!\cdots}\right)+|\Sigma|\cdot\poly\lg n
\]
bits, where $f_{\sigma}$ is the number of occurrences of $\sigma$.
For any index $i$, the query algorithm recovers $x_i$ in $O(\lg|\Sigma|)$ time in expectation on a word RAM with word-size $w\geq \Omega(\lg n)$, assuming it has access to an external lookup table of size $n^{\epsilon}$, for any constant $\epsilon>0$.
\end{theorem}

Note that the first term in the space is the minimum possible space to store a sequence with frequencies $(f_\sigma)_{\sigma\in\Sigma}$, which is at most $\sum_{\sigma} f_{\sigma\in\Sigma}\lg \frac{n}{f_{\sigma}}$.

\subsection{Related work}
The perfect hashing scheme by Fredman, Koml\'{o}s and Szemer\'{e}di~\cite{FKS84} maps $[U]$ to $[n]$ such that for any given set $S$ of size $n$, there is a hash function $h$ that maps all elements in $S$ to different buckets (i.e., no hash collision) such that $h(x)$ can be evaluated in constant time.
This hash function takes $O(n\sqrt{\lg n}+\lg\lg U)$ bits to store, and it is later improved to $O(n+\lg\lg U)$ bits by Schmidt and Siegel~\cite{SS90}.
By storing the (key, value) pair in the corresponding bucket, the perfect hashing scheme solves the dictionary problem with $O(n)$ \emph{words} of space and constant query time.
Fiat, Naor, Schmidt and Siegel~\cite{FNSS92} showed that only $O(\lg n+\lg\lg U)$ extra bits are needed to store \emph{both} the hashing function and the table, obtaining space of $n\lceil\lg U\rceil+n\lceil\lg \sigma\rceil+O(\lg n+\lg\lg U)$.
Fiat and Naor~\cite{FN93} further removed the $O(\lg n)$ term, as well as the $O(\lg\lg U)$ term when $U$ is not too large.

The first dictionary data structure that achieves nearly optimal space is due to Brodnik and Munro~\cite{BM99}.
Their data structure uses $\OPT+O(\OPT/\lg\lg\lg U)$ bits, and it has constant query time.
Pagh~\cite{Pagh01} reduced the dictionary problem to the \emph{rank} problem (see below, also Section~\ref{sec_overview_pat} for definition of the rank problem), and designed a data structure for membership queries using $\OPT+O(n\lg^2\lg n/\lg n+\lg\lg U)$ bits for $n<U/\lg\lg U$, and $\OPT+O(U\lg\lg U/\lg U)$ for $n\geq U/\lg\lg U$.
Pagh's dictionary uses rank data structures as subroutines.
By improving the rank data structures, P\v{a}tra\c{s}cu~\cite{Pat08} improved the bound to $\OPT+n/\poly\lg n+O(\lg\lg U)$, as we mentioned earlier.
Such data structures using $\OPT+o(\OPT)$ bits are called \emph{succinct} data structures~\cite{Jacobson89}, where the number of extra bits $o(\OPT)$ is called the \emph{redundancy}.

It is worth mentioning that when $n=U$, i.e., when the input is a sequence of values $v_1,\ldots,v_U\in[\sigma]$, Dodis, P\v{a}tra\c{s}cu and Thorup~\cite{DPT10} designed a data structure using optimal space.
Their data structure uses a lookup table of $\poly\lg n$ size.
We get rid of this lookup table (see Lemma~\ref{lem_change_base} in Section~\ref{sec_block}) as an application of our new technique.

No non-trivial lower bounds are known without restrictions on the data structure or model.
Fich and Miltersen~\cite{FM95} and Miltersen~\cite{Mil96} proved $\Omega(\lg n)$ and $\Omega(\lg\lg n)$ lower bounds in the RAM model with restricted operations.
Buhrman, Miltersen, Radhakrishnan and Venkatesh~\cite{BMRV02} proved that in the \emph{bit-probe} model (where the word-size $w=1$), any data structure using $O(\OPT)$ space must have query time at least $O(\lg \frac U n)$.
Viola~\cite{Viola12a} proved a higher lower bound for the case where $U=3n$, that any bit-probe data structure with query time $q$ must use space $\OPT+n/2^{O(q)}-\log n$.

\bigskip

Raman, Raman and Rao~\cite{RRR07} considered the \emph{indexable dictionary} problem, which generalizes membership.
Given a set $S$ of $n$ keys, it supports rank and select queries: $\rank_S(x)$ returns $\bot$ if $x\notin S$, and returns $i$ if $x$ is the $i$-th smallest in $S$; $\mathtt{select}_S(i)$ returns the $i$-th smallest element in $S$.
They obtained a data structure using $\OPT+o(n)+O(\lg\lg U)$ bits and constant query time.
Grossi, Orlandi, Raman and Rao~\cite{GORR09} studied the \emph{fully indexable dictionary} problem.
It generalizes the indexable dictionary problem to let $\rank_S(x)$ return the number of elements in $S$ that are at most $x$ (also for $x\notin S$).
They obtained a data structure using space $\OPT+n^{1+\epsilon}+U^{\delta}\cdot n^{-1/\epsilon}$.
In fact, this problem is much harder.
It was observed in~\cite{PV10} that $\rank$ queries can be reduced from \emph{colored predecessor search}, which has a query time lower bound of $\Omega(\lg\lg n)$ even when the space is $O(n\lg U)$~\cite{PT06,PT07} (not to say the succinct regime).
When $U>n^2$, the problem requires $n^{1+\epsilon}$ space to get constant query time (when the word-size is $\lg n$), even only supporting $\rank$ queries.

\bigskip

The locally decodable source coding~\cite{MHMP15} studies (almost) the same problem as Theorem~\ref{thm_code_informal}, in a slightly different setting.
They consider $X$ that consists of i.i.d samples from a source of entropy $H$. 
However, the main focus is non-adaptive bit-probe query algorithms.
That is, the query algorithm has to decide which $t$ \emph{bits} of the encoding to access based only on the queried index $i$.
They studied the lossy case, where the encoding is equipped with the error correcting ability.



\subsection{Technical contributions}
We make two technical contributions to succinct data structures:
We summarize the ``spillover representation'', introduced by P\v{a}tra\c{s}cu~\cite{Pat08}, to define binary strings with \emph{fractional lengths} and build a ``toolkit'' of black-box operations;
we study the ``opposite'' of data structures, 
called the \emph{data interpretation}.
We believe they will have more applications to other problems in succinct data structures.

\subsubsection{Strings with fractional lengths}
A data structure is simply a bit string, and its length (or size) is the number of bits.
Under standard notions, an $s$-bit string is only well-defined for integer $s$.
Here, we show how to define such strings when $s$ is fractional. 
We will see why this notion is useful later (or see~\cite{Pat08}).

Let $(M,K)$ be a pair such that $M\in\{0,1\}^m$ is a bit string, and $K\in[R]$ is an integer.
Such a pair is viewed as a ``binary string'' of length $m+\lg_2 R$.
When $R$ is a power of two, this matches the standard notion of length, as we could simply write $K$ in its binary representation using $\lg R$ bits and append it to $M$.
As we increase $R$, such a pair could potentially represent more information.
Only when $R$ is increased by a factor of two, does the pair correspond to a string with one more bit.
That is, by restricting $R\in[2^\kappa,2^{\kappa+1})$ for some fixed parameter $\kappa$, we essentially ``insert'' $2^{\kappa}-1$ valid lengths between adjacent integers.
It makes the measure of space more fine-grained.
Also note that a uniformly random pair of this size has binary entropy exactly equal to $m+\lg_2 R$.
In this paper, $R$ is always set to $2^{O(w)}$ (i.e. $\kappa=O(w)$), where $w$ is the word-size.
Thus, $K$ is an $O(w)$-bit integer, and the algorithms are able to do arithmetic operations on $K$ in constant time.

We summarize a few black-box operations on fractional-length strings.
The two major ones are \emph{concatenation} and \emph{fusion}.

\paragraph{Concatenation.}
Given $B$ (fractional-length) strings $\cS_1,\ldots,\cS_B$ of lengths $s_1,\ldots,s_B$, we show that they can be ``concatenated'' into one string of length $s\approx s_1+\cdots+s_B$ (note that we do not get exactly $s_1+\cdots+s_B$, because the set of valid lengths is not closed under addition).
This is trivial for integral-length strings, as we could simply connect all strings.
Moreover for integral-lengths, suppose for a given $i$, $s_1+\cdots+s_{i-1}$ can be computed efficiently, then we will be able to find where $\cS_i$ starts, and access it within the long string.
Likewise, we prove the same is true for fractional-length strings.
That is, we show that if $s_1+\cdots+s_{i-1}$ can be \emph{well-approximated}, then $\cS_i$ can be \emph{decoded}, i.e., it may be accessed within the long string, as if it was stored independently.
We emphasize that decoding an input string $\cS_i=(M_i,K_i)$ \emph{does not} mean reconstructing the entire string.
Instead, the decoding algorithm only recovers $K_i$, and \emph{finds where} $M_i$ is located within the long string (where $M_i$ is guaranteed to be a consecutive substring).
Thus, the decoding algorithm can be \emph{very efficient}.
Nevertheless, after decoding, $\cS_i$ can still be accessed as if it was stored independently.
In particular, by storing the prefix sums $s_1+\cdots+s_i$ in a lookup table, we will be able to decode any $\cS_i$ in \emph{constant time}.

Concatenation is useful when the data structure needs multiple subordinates.
We could simply construct each subordinate separately and then concatenate them.
It also demonstrates, to some extent, why fractional-lengths are useful and necessary.
If we only use integral-length strings, then each $\cS_i$ will have length (at least) $\lceil s_i\rceil$.
The length of the concatenated string becomes $\lceil s_1\rceil+\cdots+\lceil s_B\rceil$, which could be $B-1$ bits longer than $\lceil s_1+\cdots+s_B\rceil$.


\paragraph{Fusion.}
The other major operation is to \emph{fuse} an integer into a string.
Roughly speaking, it is to jointly store a pair $(i, \cS_i)$, where $i\in [C]$ is an integer, and $\cS_i$ is a string of length $s_i$.
More specifically, let us first fix lengths $s_1,\ldots,s_C$.
We are then given a pair $(i, \cS_i)$ such that $\cS_i$ is guaranteed to have length $s_i$.
We show that such a pair can be encoded by a string $\cS$ of (fixed) length $s\approx \lg(2^{s_1}+\cdots+2^{s_C})$.
This length is the best possible, because there are $2^{s_i}$ different possible strings of length $\cS_i$.
Therefore, there are $2^{s_1}+\cdots+2^{s_C}$ different pairs $(i, \cS_i)$ in total.
To encode such a pair, $\lg(2^{s_1}+\cdots+2^{s_C})$ bits are necessary.
Furthermore, suppose for \emph{every} $i$, $\lg(2^{s_1}+\cdots+2^{s_i})$ can be \emph{well-approximated}, then we will be able to recover the value of $i$ and decode $\cS_i$.

The fusion operation is useful when we study different cases of the input, and construct a data structure for each case separately.
For example, suppose we wish to construct a data structure on a subset $S\subseteq [n]$ (of arbitrary size), using close to $n$ bits (and supporting some queries).
Now suppose when $|S|=i$, we already know how to construct a data structure using $\approx\lg\binom{n}{i}$ bits, such that the queries can be answered efficiently when the query algorithm is \emph{given the value of $i$} for free.
If this is the case for every $i$, then by applying the fusion operation, we automatically obtain a data structure for subsets of \emph{arbitrary} size using $\approx n$ bits, \emph{without} giving $|S|$ to the query algorithm for free.
To see this, let case $i$ be all $S$ with exactly $i$ elements.
Then we are able to construct a data structure $\cS_i$ just for all inputs in case $i$.
The final data structure is the pair $(i,\cS_i)$ fused into one single data structure.
The space bound guarantees that if each $\cS_i$ has nearly optimal size $\approx \lg\binom{n}{i}$, then the final data structure also has nearly optimal size of $\approx n$ bits, since $\sum_i\binom{n}{i}=2^n$.
Given a query, we first retrieve the value of $i$ and decode $\cS_i$, then run the query algorithm for inputs in case $i$, given the value of $i$.
We may also view $\cS_i$ as the data structure ``conditioned on'' $i$.
Suppose all ``conditional'' data structures almost match their ``conditional'' optimal sizes, then they can be combined into one single data structure matching the overall optimal space.
See Section~\ref{sec_overview_pat} for a more concrete example.

\bigskip

By including a few other operations, we build a ``toolkit'' for operating on fractional-length strings.
The view of fractional-length strings makes the ``spillover representation'' of P\v{a}tra\c{s}cu~\cite{Pat08} more explicit.
The original paper needs huge lookup tables to store truth tables for $O(w)$-bit bizarre word operations.
The new view assigns semantic meanings to those operations, so that a major part can be efficiently computed \emph{without} lookup tables.
This is the main reason why we can reduce the lookup table size.

\subsubsection{Data interpretation}
For a data structure problem, we preprocess a combinatorial object into a binary string.
Then this string is stored in memory, which is divided into $w$-bit words.
In each time step, a query algorithm may access a memory word (i.e. a $w$-bit substring), or do local computation.
Finally, it computes some function of the input object.
The concept of \emph{data interpretation} is to perform the above procedure in the opposite direction.
Given a binary string, we preprocess it into a combinatorial object.
In each time step, a query algorithm may query an oracle for some function of the object, or do local computation.
Finally, it reconstructs a $w$-bit substring of the input string.

Since this paper concerns data structures with space almost matching the information theoretical lower bound, we will also make data interpretations space-efficient.
We design a data interpretation algorithm which preprocesses an input string of (fractional-)length $\approx \lg\binom{V}{m}$ into a set $S\subseteq [V]$ of size $m$, such that assuming there is a $\rank$ oracle for $S$ (recall that $\rank_S(x)$ returns the number of elements in $S$ that are at most $x$), any designated $w$ consecutive bits of the input string can be reconstructed in $\poly\lg V$ time (see Section~\ref{sec_data_interpret}).
It might not be obvious at this moment why it is beneficial to convert a string (data structure) back to a set, but it turns out to be a key subroutine in our data structure.
See Section~\ref{sec_overview} for more details.




\subsection{Overview of P\v{a}tra\c{s}cu's data structure}\label{sec_overview_pat}
In this subsection, we summarize how P\v{a}tra\c{s}cu's rank data structure~\cite{Pat08} works, which has important ideas to be used in our data structure.
We will ``rephrase'' this data structure using fractional-length strings, which is a non-trivial simplification.

Given a set $S\subseteq [U]$ of size $n$, the rank data structure preprocesses it into $\approx \lg\binom{U}{n}$ bits, such that for any given query $x$, the number of elements in $S$ that are at most $x$ can be computed in $O(\lg U)$ time.
The idea is to recursively construct data structures for smaller universes, and then merge the subordinate using concatenation and fusion.
Suppose $S$ has $i$ elements in $\{0,\ldots,U/2-1\}$, the first half of the universe, and it has $n-i$ elements in the second half.
We first recursively construct (fractional-length) data structures for both halves, using space $\approx\lg\binom{U/2}{i}$ and $\approx\lg\binom{U/2}{n-i}$ respectively.
Next, we concatenate two data structures, and obtain one single data structure $\cS_i$, which has length $\approx \lg\left(\binom{U/2}{i}\binom{U/2}{n-i}\right)$.
Note that the data structure $\cS_i$ encodes an input set $S$ with exactly $i$ elements in the first half (and $n-i$ in the second half), and it \emph{does not} encode the value of $i$ (likewise in the desired final data structure, the value of $n$ is assumed to be known, and is not encoded).
Finally, we encode the value of $i$ by fusing it into $\cS_i$, i.e., we jointly store the pair $(i, \cS_i)$.
The fusion operation guarantees that the pair can be stored using approximately
\[
	\lg\left(\sum_{i=0}^{n} 2^{|\cS_i|} \right)\approx \lg \left(\sum_{i=0}^n \binom{U/2}{i}\binom{U/2}{n-i}\right)=\lg \binom{U}{n}
\]
bits.

This recursion terminates at sets of size $n=0$ or $n=U$, in which case there is nothing to store (again we assume $n$ does not need encoding, so it is clear which case we are in).
We guarantee that both concatenation and fusion are implemented such that each operation only causes an overhead of no more than $O(1/U^2)$ bits.
Therefore the overall space is no more than $\lg\binom{U}{n}+O(1/U)$.
For the final (fractional-length) data structure $(M, K)$, we simply write $K$ in its binary representation and append it to $M$.
This gives us an integral-length data structure using at most $\lceil\lg\binom{U}{n}\rceil+1$ bits.

It is then straightforward to answer a $\rank$ query on this data structure.
Given a query $x$, we first recover the value of $i$, and decode $\cS_i$ (again, decoding $\cS_i$ does not mean reconstructing it).
Then we further decode $\cS_i$ into the two data structures for the two halves.
It can be done in constant time using a lookup table.
Next, if $x<U/2$, we recurse into the first half.
If $x\geq U/2$, we recurse into the second half (and add $i$ to the final answer).
Since each time $U$ decreases by a factor of two, the query time is $O(\lg U)$.

In~\cite{Pat08}, it is also shown that when $U$ is small, we can do a $B$-way divide-and-conquer, as long as $B\lg U\leq O(w)$ (recall that $w$ is the word-size).
Therefore when $U\leq w^{O(1)}$, we can afford to set $B=w^{1/2}$ and have only constant depth of recursion (rather than $O(\lg U)$).
This gives us a rank data structure with constant query time for small $U$.
In this paper, we show that it is possible to further improve it, and we design a constant-query-time data structure when only $n$ is bounded by $w^{O(1)}$ (and $U$ could be still as large as $2^{\Theta(w)}$).
This will be the starting point of our new data structure.




\section{Overview}\label{sec_overview}
In this section, we overview our new static dictionary.
For simplicity, we will first focus on the membership queries (i.e., $\sigma=1$), and assume $U=\poly\, n$.
In this case, all previous solutions use hash functions in their main construction, to map the keys into buckets.
Our data structure is conceptually different: Instead of random hash functions, we consider \emph{random inputs}.
While our data structure works for worst-case inputs, let us first think of the input set being $n$ uniformly random (distinct) keys.
Then with high probability, the input already has the properties we wanted from a random hash function, e.g., by dividing the key space into buckets in some fixed way, we have the sizes of buckets roughly balanced, etc.
We first construct a data structure just for those ``random-looking'' inputs.
On the other hand, with low probability, the input may look ``non-typical,'' e.g., some bucket may have size much larger than average.
However, ``with low probability'' means that only a small fraction of all possible inputs have these non-typical features.
Suppose the total number of such inputs is, say $\frac{1}{n^2}\cdot 2^{\OPT}$, then only $\OPT-2\lg n$ bits are needed for the optimal encoding.
This suggests that we can afford to spend more extra bits on them.
Suppose we use $\OPT-\lg n$ bits ($\lg n$ extra bits) to encoding these non-typical inputs, it is still negligible overall --- among all $O(2^{\OPT})$ possible data structures (memory states), such an encoding only wastes $2^{\OPT}\cdot (\frac{1}{n}-\frac{1}{n^2})$.
Another useful way to view it is that if we use $x$ extra bits for such rare cases, then those $x$ bits ``start'' at the $(\OPT-2\lg n)$-th bit, rather than the $\OPT$-th bit.
The more non-typical the input is, the more extra bits we can afford to spend.
Finally, we will use the fusion operation to fuse all cases together.
Similar strategies for constructing succinct data structures, where we consider random inputs and/or non-typical inputs, have been used in~\cite{BL13,Yu19,VWY20}.

In the following, we show how to handle the ``random-looking'' case and the ``non-typical'' cases for membership.

\subsection{Random inputs}
We partition the universe into $n/\lg^4 n$ buckets of size $V$.
Then for a ``random-looking'' input set $S$, there are $\lg^4 n\pm \lg^3 n$ keys in \emph{every} block.
As we mentioned in Section~\ref{sec_overview_pat}, for $\poly\lg n=\poly\ w$ keys, we can construct a rank data structure with only $O(1/U)$ extra bits, such that given the number of keys, a query algorithm answers $\rank$ queries in constant time.
In particular, it supports membership queries (e.g., by asking $\rank_S(x)$ and $\rank_S(x-1)$).
The high-level idea is to construct a rank data structure for each block, then \emph{concatenate} them.
In order to answer a query in block $i$, we need to
\begin{compactitem}
	\item recover the number of keys in block $i$ (as the rank data structure assumes this number is known), and
	\item approximate the total length of data structures for first $i-1$ blocks (to decode the $i$-th data structure).
\end{compactitem}
That is, besides the $n/\lg^4 n$ rank data structures, we need to store their lengths such that any prefix sum can be approximated.
Unfortunately, any data structure supporting prefix sum queries cannot simultaneously have ``low'' query time and ``small'' space, due to a lower bound of P\v{a}tra\c{s}cu and Viola~\cite{PV10}.
The underlying issue in this approach is that the data structure for each block has a variable length (the length depends on the number of keys in the block, which varies based on the input).
In order to locate the $i$-th data structure from the concatenated string, computing a prefix sum on a sequence of variables seems inevitable.
The P\v{a}tra\c{s}cu-Viola lower bound even prevents us from supporting prefix sums \emph{implicitly}.
That is, not only separately storing a prefix sum data structure for the lengths requires ``high'' query time or ``large'' space, there is also no ``clever'' way to jointly store the lengths together with the data structures for the blocks.
Hence, this ``variable-length encoding'' issue is the primary problem we need to tackle for ``random-looking'' inputs.


To this end, observe that although the number of keys in each block is not fixed, its deviation is actually small compared to the number, i.e., the number of keys cannot be too different for different inputs.
Then the main idea is to construct \emph{two} data structures for each block, consisting of
\begin{compactitem}
	\item \emph{a main data structure}, which stores ``most of the information'' about the block, and importantly, has a fixed length (independent of the number of keys), and
	\item \emph{an auxiliary data structure}, which stores all ``remaining information'' about the block (and unavoidably has variable length).
\end{compactitem}
Furthermore, we wish that with high probability, a given query can be answered by only accessing the main data structure (in constant time) \emph{without knowing the number of keys}.
If it is possible, then to construct the final data structure, we
\begin{compactitem}
	\item concatenate all main data structures,
	\item concatenate the auxiliary data structures, and store them together with a prefix sum structure,
	\item finally concatenate the two.
\end{compactitem}
Now, since all main data structures have fixed lengths, each one can be decoded in constant time without a prefix sum structure (the total length of the first $i-1$ data structures is simply $i-1$ times the length of a single one).
Then to answer a query in block $i$, we first decode the $i$-th main data structure, and query it in constant time.
With high probability, the answer to the query is already found, and we are done.
Otherwise, we decode the $i$-th auxiliary data structure by querying the prefix sum structure, and query the data structures to find the answer.
It may take a longer time, but if the probability that we have to decode the auxiliary data structure is sufficiently low, then the expected query time is still constant.

Next, we describe an approach to construct such two data structures for a block, which uses more space than what we aim for, but exhibits the main idea.
For each block of size $V$, we pick $\lg^4 n-\lg^3 n$ \emph{random} keys in the block to store in the main data structure.
We also pick $V-(\lg^4 n+\lg^3 n)$ \emph{random non-keys} (i.e. the elements in the key space but not in the input set), and store them in the main data structure.
This is always possible because there are at least $\lg^4n-\lg^3 n$ and at most $\lg^4n+\lg^3 n$ keys in each block for ``random-looking'' inputs.
Hence, only $2\lg^3 n$ elements are ``unknown'' from the main data structure.
Then we show that such a separation of the block into $\lg^4n-\lg^3 n$ keys, $V-(\lg^4 n+\lg^3n)$ non-keys and $2\lg^3 n$ unknowns can be jointly stored using the near-optimal $\approx\lg\binom{V}{\lg^4 n-\lg^3 n, 2\lg^3 n}$ bits (this is an easy application of the rank data structures).\footnote{$\binom{n}{k_1,k_2}=n!/(k_1!k_2!(n-k_1-k_2)!)$.}
Its size is independent of the actual input.
Then in the auxiliary data structure, we store the remaining information about the block, i.e., among the unknowns, which ones are the keys.
For a block with $m$ keys, it takes $\approx\lg\binom{2\lg^3 n}{m-(\lg^{4} n-\lg^3 n)}$ bits.
Then for each query, the answer can be found in the main data structure with probability at least $1-O(1/\lg n)$.
Only when the main data structure returns ``unknown'', does the query algorithm need to access the auxiliary data structure.

The above construction has all the desired properties, except that it uses too much space.
The inherent reason is that it implicitly stores the \emph{randomness} used in deciding which keys and non-keys to store in the main data structure.
If we sum up the sizes of the main and auxiliary data structures,
\begin{align*}
	&\ \lg\binom{V}{\lg^4 n-\lg^3 n,2\lg^3 n}+\lg\binom{2\lg^3 n}{m-(\lg^4 n-\lg^3 n)} \\
	=&\ \lg\binom{V}{m}+\lg\binom{m}{\lg^4n-\lg^3n}+\lg\binom{V-m}{V-(\lg^4n+\lg^3 n)}.
\end{align*}
Unsurprisingly, the number of extra bits $\lg\binom{m}{\lg^4n-\lg^3n}+\lg\binom{V-m}{V-(\lg^4n+\lg^3 n)}$ is exactly how much is needed to decide which keys and non-keys to store in the main data structure.
These ``random bits'' are not part of the input, and implicitly storing them causes a large amount of redundancy.

However, when the inputs are uniform, we do not really need any external randomness to decide the two subsets, since the entire data structure is close to a random string.
This suggests that for each block, we should treat the data structure from other part of the inputs as the ``randomness''.
That is, we use the opportunity of implicitly storing the random bits, to store other information that needs to be stored.
This is where we use \emph{data interpretation}.
We convert existing data structures back to subsets of certain sizes, which correspond to the keys and non-keys in the main data structure.
The details are presented in the next subsection.

\subsection{Using data interpretation}
To implement this idea, we will have to slightly modify the construction.
Now, the universe is partitioned into \emph{pairs of blocks}.
Each pair consists of a \emph{primary} block and a \emph{secondary} block, such that for a ``random-looking'' input, the {primary} block contains $\lg^{2c}n\pm\lg^{c+1} n$ keys, and the {secondary} block contains $\Theta(\lg^{c+1} n)$ keys (which plays the role of the ``randomness''), for some constant $c$.
Fix a block pair, let $V$ be the size of the primary block, $m$ be the number of keys in the primary block, $V_{\sm}$ be the size of the secondary block, and $m_{\sm}$ be the number of keys in the secondary block.
The goal is to construct two data structures using $\approx\lg\binom{V}{m}+\lg\binom{V_{\sm}}{m_{\sm}}$ bits in total.

We first construct a rank data structure for the secondary block using $\approx\lg\binom{V_{\sm}}{m_{\sm}}$ bits.
We then \emph{divide} this data structure into three substrings of lengths approximately $\lg\binom{m}{\lg^{2c}n-\lg^{c+1}n}$, $\lg\binom{V-m}{V-(\lg^{2c}n+\lg^{c+1} n)}$ and $\lg\binom{V_{\sm}}{m_{\sm}}-\lg\binom{m}{\lg^{2c}n-\lg^{c+1}n}-\lg\binom{V-m}{V-(\lg^{2c}n+\lg^{c+1} n)}$ (we show divisions can also be done for fractional-length strings).
Then $m_{\sm}= \Theta(\lg^{c+1} n)$ guarantees that there are enough bits and such division is possible.
Next, we apply a data interpretation algorithm to interpret the first string of length $\lg\binom{m}{\lg^{2c}n-\lg^{c+1}n}$ as a set of size $\lg^{2c}n-\lg^{c+1}n$ over a universe of size $m$, indicating which of the $m$ keys in the primary block should be stored in the main data structure.
We also interpret the second string as a subset indicating which of the $V-(\lg^{2c}n+\lg^{c+1} n)$ non-keys should be stored in the main data structure.
Moreover, we show that the data interpretation algorithm guarantees that any consecutive $w$ bits of the original string can be recovered in $\lg^{O(1)} n$ time, assuming there is a $\rank$ oracle of the set generated from the interpretation.
Therefore, there is no need to store the first two strings, as they can be implicitly accessed efficiently.

The main data structure is the same as what we stated in the previous subsection: storing $\lg^{2c}n-\lg^{c+1}n$ keys, $V-(\lg^{2c}n+\lg^{c+1} n)$ non-keys and $2\lg^{c+1}n$ ``unknowns'', supporting $\rank$ queries in constant time.
The auxiliary data structure now consists of two parts: 
\begin{compactitem}
	\item among the $2\lg^{c+1} n$ unknowns, which $m-(\lg^{2c}n-\lg^{c+1} n)$ are keys, and
	\item the third substring from above.
\end{compactitem}
One may verify that the sizes of the two data structures is what we claimed.
This leads to our main technical lemma.
\begin{lemma}[main technical lemma, informal]\label{lem_block_informal}
	For $V\leq \poly\, n$, given $S\subseteq [V]$ of size $m$ and $S_{\sm}\subseteq[V_\sm]$ of size $m_{\sm}$, we can construct a main data structure $\cD_{\main}$ of size
	\[
		\approx \lg\binom{V}{\lg^{2c}n-\lg^{c+1} n,2\lg^{c+1} n}
	\]
	and an auxiliary data structure $\cD_\aux$ of size
	\[
		\approx \lg\binom{V}{m}+\lg\binom{V_\sm}{m_\sm}-\lg\binom{V}{\lg^{2c}n-\lg^{c+1} n,2\lg^{c+1} n},
	\]
	such that 
	\begin{compactitem}
		\item any given query ``$x\stackrel{?}{\in} S$'' can be answered in constant time by accessing only $\cD_\main$ with probability $1-O(\lg^{-c+1} n)$;
		\item any given query ``$x\stackrel{?}{\in} S$'' or ``$x\stackrel{?}{\in} S_\sm$'' can be answered in $\poly\lg n$ time by accessing both $\cD_\main$ and $\cD_\aux$ in worst-case.
	\end{compactitem}
\end{lemma}

See Section~\ref{sec_block} for the formal statement.
Then, the final data structure will be the concatenation of all main and auxiliary data structures, in a similar way to what we stated in the previous subsection.
The auxiliary data structure needs to be decoded only when the main data structure returns ``unknown'' or the query lands in a secondary block.
By randomly shifting the universe, we bound the probability of needing the auxiliary data structure by $O(\lg^{-c+1}n)$.
By setting $c$ to be a sufficiently large constant, the expected query time is constant.

\subsection{Constructing data interpretation}
Next, we briefly describe how to design such a data interpretation algorithm, i.e., to convert a (fractional-length) string to a set.
The high-level idea is similar to the rank data structure described in Section~\ref{sec_overview_pat}, with all steps done in the opposite direction.
The data structure uses concatenation and fusion of fractional-length strings.
We first show how to do the opposite of the two operations.
More concretely, we show that
\begin{compactitem}
	\item given a string $\cD$, it can be \emph{divided} into two substrings $\cD_1,\cD_2$ of given lengths;
	\item given a string $\cD$, it can be viewed as a pair $(i, \cD_i)$, where $\cD_i$ has a given length $s_i$, i.e., we \emph{extract} an integer $i$ from $\cD$ and let $\cD_i$ be the rest.
\end{compactitem}
Both division and extraction are done with negligible space overhead.

Then, given a string $\cD$ of length $\approx\lg\binom{V}{m}$, to interpret it as a set of size $m$, we first \emph{extract} an integer $i$ from $\cD$ such that $i\in\{0,\ldots,m\}$ and $\cD_i$ has length $s_i\approx \lg\binom{V/2}{i}\binom{V/2}{m-i}$.
Then we divide $\cD_i$ of length $s_i$ into two substrings $\cD_a$ and $\cD_b$ of lengths $\approx \lg\binom{V/2}{i}$ and $\approx\lg\binom{V/2}{m-i}$ respectively.
The integer $i$ will represent the number of keys in the first half of the universe, and $m-i$ is the number of keys in the second half.
We recursively construct sets $S_a,S_b\subseteq [V/2]$ from $\cD_a$ and $\cD_b$ of sizes $i$ and $m-i$ respectively.
Then the final set is $S_a\cup (S_b+V/2)$.

To access $w$ consecutive bits of $\cD$ given a $\rank_S(\cdot)$ oracle, we first ask the oracle $\rank_S(V/2)$, i.e., the number of keys in the first half.
This determines the value of $i$, and hence the lengths of $\cD_a$ and $\cD_b$, which in turn determines whether the $w$ consecutive bits are entirely in $\cD_a$, or entirely in $\cD_b$, or split across the two substrings.
If it is entirely contained in one substring, we simply recurse into the corresponding half of the universe.
On the other hand, it is possible to show that splitting across the two substrings does not happen too many times, and when it happens, we recurse into both halves.
The recursion has depth $O(\lg V)$.
More details can be found in Section~\ref{sec_data_interpret}.

\subsection{Worst-case input}
Applying the above data structure to worst-case input has the following two issues:
\begin{compactenum}
	\item for each primary block, the sets stored in the main data structure are no longer random, hence, the expected query time may not be constant;
	\item some primary block may not have $\lg^{2c} n\pm\lg^{c+1} n$ keys, and some secondary block may not have $\Theta(\lg^{c+1} n)$ keys.
\end{compactenum}
The first issue is easy to resolve.
We simply sample a uniformly random string $\cR$, and ``XOR'' it to the substrings before we do data interpretation.
Thus, every string to be interpreted as a set will be a uniformly random string.
In particular, for each primary block, the subsets being stored in the main data structure are uniformly random subsets marginally.
This is sufficient to guarantee the constant query time.
We use the same $\cR$ for all blocks.
Storing $\cR$ in the data structure introduces $\poly\lg n$ extra bits of space.

For the second issue, we use the earlier argument.
If the expected number of keys in a primary block is $\lg^{2c} n$, then the probability that a random set has some primary block with more than $\lg^{2c} n+\lg^{c+1} n$ is at most $\exp(-\Theta(\lg^2 n))$.
Then we can afford to use more extra bits to encode such inputs.
Suppose $S$ has a block with, say $\approx 2\lg^{2c} n$ keys.
Then we could simply spend $O(\lg n)$ extra bits to encode which block it is, the number of keys in it, as well as a pointer to a separate rank data structure for this block.
The space usage for such inputs is still $\OPT-\Omega(\lg^2 n)$.

Similarly, we can show that the probability that $S$ has $N$ blocks with too many or too few keys is at most $\exp(-\Theta(N\lg^2 n))$, suggesting that we can afford to use $O(N\lg^2 n)$ extra bits (which we will use to store a perfect hash table for these ``bad'' blocks); the probability that $S$ has a block with $m$ keys for $m>\lg^{3c} n$ is $\exp(-\Theta(m\lg m))$, suggesting that we may at least use $O(m)$ extra bits for such a block (which is sufficient to store the previously known membership data structure).
By computing the probability that every particular case happens, we estimate how many extra bits we can afford.
The more ``non-typical'' the input is, the more extra bits we may use.
We then construct a data structure within the allowed extra bits.
It turns out that overall, the total space usage for inputs with \emph{at least} one ``bad'' block is at most $\OPT-\Theta(\lg^2 n)$.
Finally, we apply the fusion operation to combine the ``random-looking'' inputs and these ``non-typical'' inputs: if every block pair has number of keys close to the expectation, we set $b:=1$ and construct the data structure as in the previous subsection using $\approx\OPT$ bits; if at least one block pair has too many or too few keys, we set $b:=2$ and construct a data structure using $\OPT-\Theta(\lg^2 n)$ bits; then we fuse $b$ into the data structure.
By the guarantee of the fusion operation, the final space is bounded by
\[
	\approx \lg(2^{\OPT}+2^{\OPT-\Theta(\lg^2 n)})=\OPT+\lg(1+2^{-\Theta(\lg^2 n)})=\OPT+o(1).
\]

\bigskip

Generalizing the data structure to $\sigma>1$ (associating each key with a value) can be done by generalizing Lemma~\ref{lem_block_informal}.
Since the underlying data structure supports $\rank$, it naturally maps $m$ keys to $\{1,\ldots,m\}$.
Then we use~\cite{DPT10}, or simply concatenation, to store the list of $m$ values.
To retrieve the value of a key $x$, we first find its rank $x_r$.
Then retrieve the $x_r$-th value in the list.
This generalizes the data structure to the dictionary problem at essentially no extra cost.

\subsection{Organization}
In Section~\ref{sec_prelim}, we define notations and the model of computation.
In Section~\ref{sec_frac_length}, we formally define fractional-length strings, and state the black-box operations (the proofs are deferred to the appendix).
In Section~\ref{sec_red_ph}, we show how to construct the succinct dictionary and locally decodable arithmetic codes using perfech hashing.
In Section~\ref{sec_medium}, we design the data structure for the case where $U=\poly\ n$ using the main technical lemma.
Then we prove the main technical lemma in Section~\ref{sec_block}, and generalize to all $n$ and $U$ in Section~\ref{sec_general}.


\section{Preliminaries and Notations}\label{sec_prelim}
\subsection{Random access machine}
A random access machine (RAM) has a memory divided into $w$-bit \emph{words}, where $w$ is called the \emph{word-size}.
Typically, we assume the number of words in the memory is at most $2^w$, and they are indexed by ${\{0,\ldots,2^w-1\}}$.
In each time step, an algorithm may load one memory word to one of its $O(1)$ CPU registers, write the content of a CPU register to one memory word, or compute (limited) word operations on the CPU registers.

The standard word operations are the four basic arithmetic operations (addition, subtraction, multiplication and division) on $w$-bit integers, bit-wise operators (AND, OR, XOR), and comparison.
In this paper, we also assume that the machine supports \emph{floating-point} numbers.
A floating-point number is represented by two registers in the form of $a\cdot 2^b$, and the arithmetic operations extend to these numbers as well (possibly with rounding errors).
This is without loss of generality, as they can be simulated using the standard operations.
Finally, we assume it is possible to compute $2^x$ up to $1\pm2^{-w}$ multiplicative error, and $\lg_2 x$ up to additive $\pm2^{-w}$ error.
We further assume that the error can be arbitrary but has to be deterministic, i.e., for any given $x$, $2^x$ and $\lg_2 x$ always compute to the same result within the desired range.
By expanding into the Taylor series, the two can be computed in $O(w)$ time using only arithmetic operations, which is already sufficient for our application.
On the other hand, using a lookup table of size $2^{\epsilon w}$ (we already have lookup tables of this size), the computational time can be reduced to constant.

\subsection{Notations}
In this paper, let $X \Div Y$ denote $\lfloor X/Y \rfloor$, $X\Mod Y$ denote $X-Y\cdot (X\Div Y)$.
Let $[R]$ denote the set $\{0,1,\ldots,R-1\}$.
Let $\fra(x)$ denote $x-\lfloor x\rfloor$.
Throughout the paper, $\lg x$ is the binary logarithm $\lg_2 x$, $\tilde{O}(f)=f\cdot \poly\lg f$.


\section{Fractional-length Strings}\label{sec_frac_length}
In this subsection, we formally define binary strings with fractional lengths using the spillover representation of~\cite{Pat08}, and state block-box operations.
Throughout the paper, let $\kappa$ be the \emph{fineness parameter}, which characterizes the gaps between adjacent valid lengths, and determines the space loss when doing the operations.
It is an integer parameter that is specified by the algorithm designer and will be hardwired to the preprocessing and query algorithms.
In the following, we will see that each operation loses $O(2^{-\kappa})$ bits; and on the other hand, the algorithms will have to perform arithmetic operations on $\kappa$-bit integers.
In our data structure construction, $\kappa$ will be set to $\Theta(\lg U)=O(w)$, so that each operation loses negligible space, and $\kappa$-bit arithmetic operations can still be performed in constant time.

\begin{definition}[fractional-length strings]\label{def_frac_space}
	Let $\cS=(M, K)$ be a pair such that $M\in\{0,1\}^m$ and $K\in[R]$, where $m$ is a nonnegative integer and $R\in[2^\kappa,2^{\kappa+1})$ is an integer, or $m=0$ and integer $R\in[1, 2^{\kappa})$.
	Then $\cS$ is \emph{a binary string of length $m+\lg R$}, and 
	\[
		\cS[i]:=\begin{cases} M[i]& i\in[m], \\ K & i=m. \end{cases}
	\]
	Let $|\cS|$ denote the length of $\cS$.
	Let $\cS[i, j]$ denote the sequence (substring) $(\cS[i],\ldots,\cS[j])$.
	Let $\range{K}:=R$ be the size of range of $K$.
\end{definition}

\begin{remark}
	Note the following facts about fractional-length strings:
	\begin{compactitem}
		\item When $s$ is an integer, by writing $K$ in its binary representation, a binary string of length $s$ from Definition~\ref{def_frac_space} is a standard binary string of $s$ bits;
		\item A uniformly random string of length $s$ has entropy exactly $s$;
		\item $|\cS|$ uniquely determines $|M|$ and $\range{K}$, i.e., if $|\cS|<\kappa$, $|M|=0$ and $\range{K}=2^s$, otherwise, $|M|=\lfloor |\cS|\rfloor -\kappa$ and $\range{K}=2^{\kappa+\fra(|\cS|)}$;
		\item The length of a string may be an irrational number, but it can always be succinctly encoded, e.g., by encoding $|M|$ and $\range{K}$;
		\item One should \emph{not} view $\range{K}$ as a function of $K$, it is indeed a parameter of the \emph{variable} $K$.
	\end{compactitem}
\end{remark}

Since the word-size is $\Omega(\kappa)$, any $O(\kappa)$ consecutive bits of a string can be retrieved using $O(1)$ memory accesses, which suggests how a fractional-length string is accessed.
Formally, we define an access as follows.
\begin{definition}[access]\label{def_access}
Let $\cS$ be a (fractional-length) string, an \emph{access} to $\cS$ is to retrieve $\cS[i, j]$ for $j-i\leq O(\kappa)$.
\end{definition}
When $j<|M|$, an access is to retrieve $j-i+1$ bits of $M$.
When $j=|M|$, it is to retrieve $j-i$ bits and the integer $K$.

\bigskip

In the following, we show how to operate on fractional-length strings.
All the proofs are deferred to Appendix~\ref{sec_frac_length_app}.
Firstly, the strings can be \emph{concatenated}.

\newcommand{\contpropconcatapprox}{
	Let $s_1,\ldots,s_B\geq \kappa$.
	Suppose for any given $i$, $s_1+\cdots+s_i$ can be approximated (deterministically) in $O(t)$ time with an additive error of at most $2^{-\kappa}$.
	Then given $B$ strings $\cS_1,\ldots,\cS_B$, where $\cS_i=(M_i,K_i)$ has length $s_i$, they can be concatenated into one string $\cS=(M,K)$ of length at most
	\[
		s_1+\cdots+s_B+(B-1)\cdot 2^{-\kappa+4},
	\]
	so that each $M_i$ is a (consecutive) substring of $M$.
	Moreover, for any given $i$, $\cS_i$ can be \emph{decoded} using $O(t)$ time and two accesses to $\cS$, i.e., a decoding algorithm recovers $K_i$, and finds the starting location of $M_i$ using $O(t)$ time and two accesses to $\cS$.	
}
\begin{proposition}[concatenation]\label{prop_concat_approx}
	\contpropconcatapprox
\end{proposition}
In particular, by storing approximations of all $B$ prefix sums in a lookup table of size $O(B)$, each $\cS_i$ can be decoded in $O(1)$ time.
Note that this lookup table \emph{does not} depend on the $B$ strings.
\newcommand{\contpropconcat}{
	Let $s_1,\ldots,s_B\geq 0$.
	There is a lookup table of size $O(B)$.
	Given $B$ strings $\cS_1,\ldots,\cS_B$, where $\cS_i=(M_i,K_i)$ has length $s_i$, they can be concatenated into one string $\cS=(M,K)$ of length at most 
	\[
		s_1+\cdots+s_B+(B-1)2^{-\kappa+4},
	\]
	so that each $M_i$ is a (consecutive) substring of $M$.
	Moreover, assuming we can make random accesses to the lookup table, $\cS_i$ can be \emph{decoded} using constant time and two accesses to $\cS$, i.e., a decoding algorithm recovers $K_i$, and finds the starting location of $M_i$ using constant time and two accesses to $\cS$.	
}
\begin{proposition}[concatenation]\label{prop_concat}
	\contpropconcat
\end{proposition}

Next, an integer $i\in[C]$ can be \emph{fused} into a string.
\newcommand{\contpropfuseapprox}{
	Let $s_1,\ldots,s_C\geq 0$.
	Suppose for any given $i$, $2^{s_1}+\cdots+2^{s_i}$ can be approximated (deterministically) in $O(t)$ time with an additive error of at most $(2^{s_1}+\cdots+2^{s_C})\cdot 2^{-\kappa-3}$.
	Then given $i\in \{1,\ldots,C\}$ and string $\cS_i=(M_i, K_i)$ of length $s_i$, the pair $(i, \cS_i)$ can be stored in $\cS=(M,K)$ of length at most
	\[
		\lg(2^{s_1}+\cdots+2^{s_C})+C\cdot 2^{-\kappa+4},
	\]
	so that $M_i$ is a (consecutive) substring of $M$.
	Moreover, we can recover the value of $i$ and \emph{decode} $\cS_i$ using $O(t\lg C)$ time and two accesses to $\cS$, i.e., a decoding algorithm recovers $i$, $K_i$, and finds the starting location of $M_i$ using $O(t\lg C)$ time and two accesses to $\cS$.
}
\begin{proposition}[fusion]\label{prop_fuse_approx}
	\contpropfuseapprox
\end{proposition}
Note that the error term is always proportional to $2^{s_1}+\cdots+2^{s_C}$, for every $i$.
In particular, it is possible that for some (small) $i$, the error term dominates the value, making the assumption easy to satisfy (for that $i$).
The decoding algorithm can take constant time if we use a lookup table of size $O(C)$.
Again, the lookup table does not depend on the string.
\newcommand{\contpropfuse}{
	Let $s_1,\ldots,s_C\geq 0$.
	There is a lookup table of size $O(C)$.
	Given $i\in \{1,\ldots,C\}$ and string $\cS_i=(M_i, K_i)$ of length $s_i$, the pair $(i, \cS_i)$ can be stored in $\cS=(M,K)$ of length
	\[
		\lg(2^{s_1}+\cdots+2^{s_C})+C\cdot 2^{-\kappa+2},
	\]
	so that $M_i$ is a (consecutive) substring of $M$.
	Moreover, assuming we can make random accesses to the lookup table, the value of $i$ can be recovered and $\cS_i$ can be \emph{decoded} using constant time and two accesses to $\cS$, i.e., a decoding algorithm recovers $i$, $K_i$, and finds the starting location of $M_i$ using constant time and two accesses to $\cS$.
}
\begin{proposition}[fusion]\label{prop_fuse}
	\contpropfuse
\end{proposition}
\bigskip

Fractional-length strings have ``one of its ends'' encoded using an integer.
For technical reasons, we also need the following notion of \emph{double-ended} binary strings.

\begin{definition}[double-ended strings]
	Let $\cS=(K_h,M,K_t)$ be a triple such that $M\in\{0,1\}^m$, $K_h\in[R_h]$ and $K_t\in[R_t]$, where $m$ is a nonnegative integer and $R_h,R_t\in[2^\kappa,2^{\kappa+1})$ are integers.
	Then $\cS$ is a \emph{double-ended binary string of length $m+\lg R_h+\lg R_t$}, and
	\[
		\cS[i]:=\begin{cases} K_h & i=-1, \\ M[i] & i\in[m], \\ K_t & i=m. \end{cases}
	\]
	Let $|\cS|$ denote the length of $\cS$.
	Let $\cS[i, j]$ denote the substring $(\cS[i],\ldots,\cS[j])$.
	Let $\range{K_h}:=R_h,\range{K_t}:=R_t$ be the sizes of ranges of $K_h$ and $K_t$ respectively.
\end{definition}
\begin{remark} Note the following facts:
	\begin{compactitem}
		\item Unlike the (single-ended) fraction-length strings, the length of a double-ended string does not necessarily determine $\range{K_h}$, $\range{K_t}$, or even $|M|$;
		\item For $s\geq 2\kappa$, any $s$-bit string $(M,K)$ can be viewed as a double-ended string by taking the first $\kappa$ bits of $M$ as $K_h$ and letting $K_t$ be $K$;
		\item For simplicity, in this paper, we do not define double-ended strings with length shorter than $2\kappa$;
	\end{compactitem}
\end{remark}
Double-ended strings are accessed in the same way.
\begin{definition}[access]
Let $\cS$ be a double-ended string, an \emph{access} to $\cS$ is to retrieve $\cS[i, j]$ for $j-i\leq O(\kappa)$.
\end{definition}

Prefixes and suffixes of a double-ended string are defined in the natural way, as follows.
\begin{definition}[prefix/suffix]\label{def_pre_suf}
Let $\cS=(K_h,M,K_t)$ be a double-ended string.
Then $\cS[-1, j]$ is a \emph{prefix} of $\cS$ for any $j\leq|M|$, 
$\cS[i, |M|]$ is a \emph{suffix} of $\cS$ for any $i\geq -1$.
\end{definition}


Using double-ended strings, it is possible to divide a binary string into two substrings. 
\newcommand{\contpropdivide}{
	Let $s_1,s_2,s\geq 3\kappa$ and $s\leq s_1+s_2-2^{-\kappa+2}$.
	Then given a double-ended string $\cS=(K_h,M,K_t)$ of length $s$, a division algorithm outputs two double-ended strings $\cS_1=(K_{1,h},M_1,K_{1,t})$ and $\cS_2=(K_{2,h},M_2,K_{2,t})$ of lengths at most $s_1$ and $s_2$ respectively.
	Moreover, $(K_{1,h},M_1)$ is a \emph{prefix} of $\cS$, $(M_2,K_{2,t})$ is a \emph{suffix} of $\cS$, and $K_{1,t}$ and $K_{2,h}$ together determine $M\left[|M_1|,|M|-|M_2|-1\right]$, i.e., the remaining bits of $M$.
	$\range{K_{i,h}}$, $\range{K_{i,t}}$ and $|M_i|$ can be computed in $O(1)$ time given $\range{K_h}$, $\range{K_t}$, $|M|$ and $s_1,s_2$, for $i=1,2$.
}
\begin{proposition}[divide]\label{prop_divide}
	\contpropdivide
\end{proposition}
\begin{remark}
	Proposition~\ref{prop_divide} guarantees that each access to $\cS$ can be implemented using at most two accesses to $\cS_1$ and $\cS_2$.
	Moreover, accessing a (short) prefix of $\cS$ requires only accessing the prefix of $\cS_1$ of the same length.
	Likewise, accessing a suffix of $\cS$ requires only accessing the suffix of $\cS_2$ of the same length.
\end{remark}

Finally, the ``inverse'' of fusion can also be done efficiently.
\newcommand{\contpropsep}{
	Let $s_1,\ldots,s_C\geq 0$, $R_h,R_t\in[2^\kappa,2^{\kappa+1})$ and $m\geq \kappa$, let $s=m+\lg R_h+\lg R_t$, and $s\leq \lg(2^{s_1}+\cdots+2^{s_C})-C\cdot 2^{-\kappa+2}$, there is a lookup table of size $O(C)$.
	Given a double-ended string $\cS=(K_h,M,K_t)$ such that $\range{K_h}=R_h$, $\range{K_t}=R_t$ and $|M|=m$, there is an extraction algorithm that generates a pair $(i, \cS_i)$ such that $i\in\{1,\ldots,C\}$, and $\cS_i=(K_{i,h},M_i,K_{i,t})$ has length at most $s_i$.
	Moreover, $(M_i,K_{i,t})$ is a suffix of $\cS$, and given $i$ and $K_{i,h}$, the rest of $\cS$ (i.e., $\cS[-1,|M|-|M_i|-1]$) can be recovered in constant time, assuming random access to the lookup table.	
	$\range{K_{i,h}}$, $\range{K_{i,t}}$ and $|M_i|$ does not depend on $\cS$, and can be stored in the lookup table.
}
\begin{proposition}[extraction]\label{prop_sep}
	\contpropsep
\end{proposition}
\begin{remark}
	We can safely omit any $i$ with $|M|-|M_i|>\kappa+1$, since removing such $s_i$ from the list $(s_1,\ldots,s_C)$ (and decrease $C$ by one) could only increase the upper bound on $s$, $\lg(2^{s_1}+\cdots+2^{s_C})-C\cdot 2^{-\kappa+2}$.
	That is, the extraction algorithm may never generate a pair with this $i$.
	Therefore, we may assume that $\cS[-1,|M|-|M_i|-1]$ has length at most $O(\kappa)$, taking constant time to output.
\end{remark}





\section{Reductions to Perfect Hashing}\label{sec_red_ph}
Now, we show how to design succinct dictionary and compress low entropy sequence with local decodability using the membership and perfect hashing data structure in Theorem~\ref{thm_ph_informal}.
\paragraph{Succinct dictionary.}
For dictionary, we shall use the following lemma by Dodis, P\v{a}tra\c{s}cu and Thorup~\cite{DPT10}.
\begin{lemma}[\cite{DPT10}]\label{lem_change_base}
	There is an algorithm that preprocesses a given sequence $(x_1,\ldots,x_n)\in [\sigma]^n$ for $\sigma\leq 2^\kappa$ into a data structure of length at most $n\lg\sigma+(n-1)2^{-\kappa+5}$, such that given any $i$, $x_i$ can be retrieved in constant time.
\end{lemma}
The data structure in~\cite{DPT10} requires a lookup table of $O(\lg n)$ words.
We show that by using the view of fractional-length strings, we can completely remove the lookup table.
\begin{proof}
	Let $b=\lceil 2\kappa/\lg\sigma\rceil$.
	We partition the sequence into $n/b$ chunks of $b$ symbols each, then combine each chunk into one single character in $[\sigma^b]$ (if $n$ is not a multiple of $b$, then the last group will have more than $b$ symbols). 
	Since $\sigma^b=2^{O(\kappa)}=2^{O(w)}$, each $x_i$ can be decode in constant time given the character.
	Then compute $m=\lfloor \lg\sigma^b\rfloor -\kappa$ and $R=\lceil\sigma^b\cdot 2^{-m}\rceil$, and view each character in $[\sigma^b]$ as a data structure of size $m+\lg R$.
	Note that $m+\lg R-b\lg \sigma\leq \lg (\sigma^b+2^m)-b\lg\sigma\leq\lg(1+2^{-\kappa})\leq 2^{-\kappa+1}$.
	Then we apply Proposition~\ref{prop_concat_approx} to concatenate all $n/b$ data structure.
	Since $m$ and $R$ can both be computed in constant time, $m+\lg R$ can be approximated in constant time, hence Proposition~\ref{prop_concat_approx} guarantees that there is a data structure of size
	\[
		(m+\lg R)\cdot (n/b)+(n/b-1)\cdot 2^{-\kappa+4}\leq n\lg\sigma+(n-1)\cdot 2^{-\kappa+5},
	\]
	supporting symbol retrieval in constant time.
	This proves the lemma.
\end{proof}

To store a set of $n$ key-value pairs for keys in $[U]$ and values in $[\sigma]$, we first apply Theorem~\ref{thm_ph_informal} on the set of keys.
It produces a data structure of size $\lg\binom{U}{n}+\poly\lg n+O(\lg\lg U)$ bits, which defines a bijection $h$ between the keys and $[n]$.
Next, we apply Lemma~\ref{lem_change_base} to store the values.
Specifically, we construct the sequence $(v_1,\ldots,v_n)$ such that if $(x, u)$ is an input key-value pair, then $v_{h(x)+1}=u$.
This sequence can be stored in space $n\lg \sigma+O(1)$ by Lemma~\ref{lem_change_base}.
Hence, the total space of the data structure is 
\[
	\lg\binom{U}{n}+n\lg\sigma+\poly\lg n+O(\lg\lg U)=\OPT+\poly\lg n+O(\lg\lg U),
\]
as claimed in Theorem~\ref{thm_main_informal}.

To answer a query $\vr(x)$, we first query the membership data structure.
If $x$ is not a key, we return $\bot$.
Otherwise, we retrieve and return the $(h(x)+1)$-th value in the sequence.
The total query time is constant in expectation and with high probability.
This proves Theorem~\ref{thm_main_informal}.

\paragraph{Compression to zeroth order entropy with local decodability.}
Given a sequence $(x_1,\ldots,x_n)\in\Sigma^n$ such that each $\sigma\in\Sigma$ appears $f_\sigma$ times, we construct a data structure \emph{recursively on} $\Sigma$.
We first arbitrarily partition $\Sigma$ into $\Sigma_1\cup\Sigma_2$ such that $|\Sigma_1|=\lfloor|\Sigma|/2\rfloor$ and $|\Sigma_2|=\lceil|\Sigma|/2\rceil$.
For any set $\Gamma\subseteq\Sigma$, define $S_{\Gamma}:=\{i\in[n]: x_i\in \Gamma\}$.
Then we apply Theorem~\ref{thm_ph_informal} to construct a perfect hashing for $S_{\Sigma_1}$, which uses space
\[
	\lg\binom{n}{|S_{\Sigma_1}|}+\poly\lg n=\lg\left(\frac{n!}{|S_{\Sigma_1}|!\cdot|S_{\Sigma_2}|!}\right)+\poly\lg n,
\]
and defines a bijection $h$ that maps all coordinates in $S_{\Sigma_1}$ to $[|S_{\Sigma_1}|]$, and a bijection $\oh$ that maps all $S_{\Sigma_2}$ to $[|S_{\Sigma_2}|]$.
We recursively construct a data structure for $\Sigma_1$ over $h(S_{\Sigma_1})$, and a data structure for $\Sigma_2$ over $\oh(|S_{\Sigma_2}|)$.

In general, each node in the recursion tree corresponds to a subset $\Gamma$ of the alphabet such that $\Gamma_1$ and $\Gamma_2$ are the subsets corresponding to the left and the right child respectively.
In this node, we store
\begin{compactitem}
	\item the size of subset of its left child $|\Gamma_1|$,
	\item the perfect hashing data structure for $S_{\Gamma_1}$,
	\item two pointers to the data structures in its left and its right children.
\end{compactitem}
For $\Gamma$ of size one, we store nothing.
Thus, we obtain a final data structure of size
\begin{align*}
	\lg\left(\frac{n!}{f_{\sigma_1}!f_{\sigma_2}!\cdots}\right)+|\Sigma|\cdot\poly\lg n.
\end{align*}

To answer a query $i$, we first retrieve the size of $\Sigma_1$ and query if $i\in S_{\Sigma_1}$.
If $i\in S_{\Sigma_1}$, we go to the left child and recursively query $h(i)$.
If $i\notin S_{\Sigma_1}$, we go to the right child and recursively query $\oh(i)$.
Finally, when the current subset $|\Gamma|=1$, we return the only element in $\Gamma$.
Since in each level of the recursion, the perfect hashing data structure takes constant query time in expectation, and the size of $\Gamma$ reduces by a factor two, the total query time is $O(\lg|\Sigma|)$ in expectation.
This proves Theorem~\ref{thm_code_informal}.
\section{Perfect Hashing for Medium-Sized Sets}\label{sec_medium}

In this section, we present the minimal perfect hashing and membership data structure when the number of keys $n$ is neither too large nor too small, focusing  on the case where $n\geq U^{1/{12}}$ and $n\leq U-U^{1/{12}}$.
Generalizing to all $n$ involves less new ideas, and we defer the proof of the main theorem to Section~\ref{sec_general}.

Recall that we wish to preprocess a set of $n$ keys $S\subseteq [U]$, such that the data structure defines a bijection $h$ between $S$ and $[n]$ and a bijection $\oh$ between $[U]\setminus S$ and $[U-n]$.
A query $\hash(x)$ returns a pair $(b, v)$ such that
\begin{compactitem}
	\item if $x\in S$, then $b=1$ and $v=h(x)$;
	\item if $x\notin S$, then $b=0$ and $v=\oh(x)$.
\end{compactitem}

We partition the universe $[U]$ into pairs of blocks.
For each pair, we construct a \emph{main} data structure and an \emph{auxiliary} data structure, such that the main data structure contains ``most'' of the information in the block and has fixed length, and the auxiliary data structure stores the remaining information (which unavoidably has variable length).
Finally, we concatenate all data structures for all blocks.

Our main technical lemma is to construct such two (fractional-length) data structures for a pair of blocks of sizes $V$ and $V_{\sm}$.

\newcommand{\contlemblock}{
	Let $\kappa$ be the fineness parameter for fractional-length strings, and $c$ be a constant positive integer.
	Let $V\in [2\kappa^{2c-3},2^{\kappa/2}]$ and $V_{\sm}\geq 4\kappa^{c+1}$.
	For any constant $\epsilon>0$, there is a preprocessing algorithm $\phB$, query algorithms $\qalgB_{\main}$, $\qalgB$ and lookup tables $\tblB_{V,V_{\sm}}$ of size $\tilde{O}(2^{\epsilon \kappa})$.
	Given 
	\begin{compactitem}
	 	\item a set $S\subseteq [V]$ such that $m:=|S|\in [\kappa^{2c-3}+\kappa^c/3,\kappa^{2c-3}+2\kappa^c/3]$,
	 	\item a set $S_{\sm}\subseteq \{V,\ldots,V+V_{\sm}-1\}$ and $m_{\sm}:=|S_{\sm}|\in [\kappa^{c+1}, 3\kappa^{c+1}]$,
	 	\item a random string $\cR$ of $\kappa^{c+1}$ bits,
	\end{compactitem}
	$\phB$ preprocesses $S$ and $S_{\sm}$ into a pair of two (fractional-length) data structures $\cD_{\main}$ and $\cD_{\aux}$, such that
	\begin{compactenum}[(i)]
		\item $\cD_{\main}$ has length at most $$\lg\binom{V}{\kappa^{2c-3},\kappa^c}
		+\kappa^{2c-3}\cdot 2^{-\kappa/2+1};$$
		\item $\cD_{\aux}$ has length at most 
		\[
			\lg\binom{V}{m}+\lg\binom{V_{\sm}}{m_{\sm}}-\lg\binom{V}{\kappa^{2c-3},\kappa^c}+\kappa^{c+1}2^{-\kappa/2+2};
		\]
		\item $\cD_{\main}$ and $\cD_{\aux}$ together define a \emph{bijection} $h$ between 
		\[
			S\cup S_{\sm}\quad \textrm{and}\quad [m+m_{\sm}],
		\]
		and a \emph{bijection} $\oh$ between 
		\[
			[V+V_{\sm}]\setminus (S\cup S_{\sm})\quad \textrm{and} \quad [(V+V_{\sm})-(m+m_{\sm})],
		\]
		such that $h(S)\supset [\kappa^{2c-3}]$ and $\oh([V]\setminus S)\supset [V-\kappa^{2c-3}-\kappa^c]$;
		\item given any $x\in [V]$, $\qalgB_{\main}(V,x)$ outputs $\hash(x)$ when $x\in S$ and $h(x)\in [\kappa^{2c-3}]$, or when $x\notin S$ and $\oh(x)\in [V-\kappa^{2c-3}-\kappa^c]$, otherwise it outputs ``unknown'';
		moreover, it only accesses $\cD_{\main}$, $\cR$ and the lookup table $\tblB_{V,V_{\sm}}$, and it runs in constant time in the worst case;
		\item for any $x\in [V]$, the probability that $\qalgB_{\main}(V, x)$ outputs ``unknown'' is at most $O(\kappa^{-c+3})$ over the randomness of $\cR$;
		\item given any $x\in[V+V_{\sm}]$, $\qalgB(V,m,V_{\sm},m_{\sm},x)$ computes $\hash(x)$; it accesses $\cD_{\main}$, $\cD_{\aux}$, $\cR$ and the lookup table $\tblB_{V,V_{\sm}}$, and it runs in $O(\kappa^4)$ time.
	\end{compactenum}
}

\begin{lemma}[main technical lemma]\label{lem_block}
	\contlemblock
\end{lemma}
\begin{remark}
	Note that the size of $\cD_\main$ does not depend on $m$ or $m_{\sm}$, and $\qalgB_\main$ also does not need to know $m$ or $m_{\sm}$ to answer a query.
	The total size of $\cD_\main$ and $\cD_\aux$ is approximately $\lg\binom{V}{m}+\lg\binom{V_{\sm}}{m_\sm}$, close to the optimal space \emph{given} $m$ and $m_\sm$.
	The size of $\cD_{\main}$ is also close to the optimum, as $\qalgB_\main$ has to identify a set of $\kappa^{2c-3}$ keys and $V-\kappa^{2c-3}-\kappa^c$ non-keys by only accessing $\cD_\main$, which takes exactly $\lg \binom{V}{\kappa^{2c-3},V-\kappa^{2c-3}-\kappa^c}=\lg\binom{V}{\kappa^{2c-3},\kappa^c}$ bits.
\end{remark}

The proof of the lemma is deferred to Section~\ref{sec_block}.
In this following, we present our data structure assuming this lemma.
%
%
For simplicity of the notations, let 
\[
	\OPT_{V,m}:=\lg \binom{V}{m}
\]
be the information theoretical \emph{optimal} space when storing a set of $m$ keys over key space of size $V$.

\begin{theorem}\label{thm_medium_set}
	For any constant $\epsilon>0$ and constant integer $c>0$, there is a preprocessing algorithm $\ph$, a query algorithm $\qalg$ and lookup tables $\tbl_{U, n}$ of size $n^{\epsilon}$, such that given 
	\begin{compactitem}
		\item a set $S$ of $n$ keys over the key space $[U]$, where $n\geq U^{1/{12}}$ and $n\leq U-U^{1/{12}}$,
		\item a uniformly random string $\cR$ of length $O(\lg^{c+1} n)$,
	\end{compactitem}
	$\ph$ preprocesses $S$ into a data structure $\cD$ of (worst-case) length
	\[
		\OPT_{U,n}+U^{-1},
	\]
	such that $\cD$ defines a \emph{bijection} $h$ between $S$ and $[n]$ and a \emph{bijection} $\oh$ between $[U]\setminus S$ and $[U-n]$.
	Given access to $\cD$, $\cR$ and $\tbl_{U,n}$, for any key $x\in [U]$, $\qalg(U,n,x)$ outputs $\hash(x)$ on a RAM with word-size $w\geq \Omega(\lg U)$, in time
	\begin{compactitem}
		\item $O(1)$ with probability $1-O(\lg^{-c+4}U)$ and
		\item $O(\lg^7 U)$ in worst-case,
	\end{compactitem}
	where the probability is taken over the random $\cR$.
	In particular, the query time is constant in expectation and with high probability by setting $c=11$.
\end{theorem}
\begin{remark}
	When $n<U^{1/{12}}$, we could use a hash function to map the keys to $n^2$ buckets with no collisions.
	We could apply this theorem with the new key space being all buckets, and the keys being the non-empty buckets. 
	By further storing for each non-empty bucket, the key within it, it extends the membership query to $n<U^{1/{12}}$, using $O(\lg n+\lg\lg U)$ extra bits.
	We will see a more generic approach in Section~\ref{sec_general} (which works for perfect hashing and improves the $\lg\lg U$ term).
\end{remark}

Without loss of generality, we may assume $n\leq U/2$, since otherwise, we could simply take the complement of $S$.
Let $\kappa:=\lceil4\lg U\rceil$ be the fineness parameter, and $c$ be a (large) constant positive integer to be specified later.
We partition the universe $[U]$ into pairs of blocks.
Each block pair consists of a larger \emph{primary} block containing roughly $\kappa^{2c-3}+\kappa^c/2$ keys, and a smaller \emph{secondary} block containing roughly $2\kappa^{c+1}$ keys.
Formally, let 
\[
	V_{\prim}:=\lfloor\frac{(\kappa^{2c-3}+\kappa^c/2)U}{n}\rfloor,
\]
\[
	V_\rand:=\lfloor\frac{2\kappa^{c+1}U}{n}\rfloor
\]
and $V_\bl=V_\prim+V_\rand$.
Each primary block has size $V_\prim$ and each secondary block has size $V_\rand$.
Every block pair has size $V_\bl$.
For simplicity, let us first consider the case where $U$ is a multiple of $V_\bl$, and $U= V_\bl \cdot N_\bl$.
We will show how to handle general $U$ later.

Thus, we partition $U$ into $N_\bl$ block pairs in the natural way, where the $i$-th primary block
\[
	\cB_\prim^i:= \left\{x\in[U]:(i-1) V_\bl \leq x<V_\prim+(i-1)V_\bl\right\}
\]
and the $i$-th secondary block
\[
	\cB_\rand^i:= \left\{x\in[U]:V_\prim+(i-1)V_\bl\leq x<i V_\bl \right\}.
\]


We call the $i$-th block pair \emph{good}, if the numbers of keys in the primary and secondary blocks are close to the average:
\[
	|S\cap \cB_{\prim}^i|\in[\kappa^{2c-3}+\kappa^c/3,\kappa^{2c-3}+2\kappa^c/3],
\]
and
\[
	|S\cap \cB_{\rand}^i|\in[\kappa^{c+1},3\kappa^{c+1}].
\]
The pair is \emph{bad} if at least one of the two blocks has the number of keys outside the range.

In the following, we show that for inputs $S$ with no bad blocks, we can construct a good data structure.
The goal is to design a data structure using space close to $\OPT_{N_\bl V_\bl ,n}$.

\subsection{No bad block pair}\label{sec_good_block}
\begin{lemma}\label{lem_no_bad_block}
	If $U$ is a multiple of $ V_\bl $, then there is a data structure with the guarantees in Theorem~\ref{thm_medium_set}, for all sets $S$ with no bad block pair.
	Moreover, the size of the data structure is
	\[
		\OPT_{U,n}+n\cdot 2^{-\kappa/2+2}.
	\]
\end{lemma}

Before starting the preprocessing, we view the last $O(\lg U)$ bits of the random bits $\cR$ as a random number $\Delta\in[U]$.
We shift the entire universe according to $\Delta$, i.e., $x\mapsto (x+\Delta)\Mod U$.
It is applied to input $S$, and will be applied to the queries too (which guarantees that the query is in a primary block with good probability).

The preprocessing algorithm is based on recursion.
The following algorithm $\mathtt{dict\_rec}$ preprocesses $S$ restricted to the $i$-th to $j$-th blocks $\cB_{\prim}^{i},\cB_{\rand}^{i},\ldots,\cB_{\prim}^{j},\cB_{\rand}^{j}$, and outputs $j-i+2$ data structures $\cD_{\main}^{i},\ldots,\cD_{\main}^{j}$ and $\cD_{\aux}$.
We will inductively prove upper bounds on the sizes of the data structures: the length of each $\cD_{\main}^i$ is at most
\begin{equation}\label{eqn_size_main}
	\smain:=\lg\binom{V_\prim}{\kappa^{2c-3},\kappa^c}+\kappa^{2c-3}\cdot 2^{-\kappa/2+1},
\end{equation}
and the length of $\cD_\aux$ generated from $i,\ldots,j$-th block pair is at most
\begin{equation}\label{eqn_size_aux}
	\OPT_{(j-i+1) V_\bl ,m}-(j-i+1)\smain+(m-1)2^{-\kappa/2+2},
\end{equation}
where $m$ is the number of keys in blocks $i$ to $j$.
In the base case with only one block pair, we simply apply Lemma~\ref{lem_block}.
\begin{code}{dictrec}[preprocessing algorithm][(i,j,m,S,\cR)][dict\_rec]
	\item if $i=j$
	\item\quad let $S_\prim\subseteq S$ be the set of keys in the $i$-th primary block
	\item\quad let $S_\rand\subseteq S$ be the set of keys in the $i$-th secondary block
	\item\quad $m_\prim:=|S_\prim|$ and $m_{\rand}:=|S_\rand|$
	\item\quad $(\cD_{\main}^i,\cD_{\aux}'):=\phB(V_\prim,m_\prim,V_\rand,m_\rand,S_\prim,S_\rand,\cR)$ (from Lemma~\ref{lem_block})
	\item\label{step_dictrec_fuse_mpr}\quad apply Proposition~\ref{prop_fuse} to fuse $m_\prim$ into $\cD_{\aux}'$, and obtain $\cD_{\aux}$
	\item\quad return $(\cD_{\main}^i,\cD_{\aux})$ \ctn
\end{code}
\begin{claim}
	If $i=j$, $|\cD_\main^i|\leq \smain$ and $|\cD_\aux|\leq \OPT_{ V_\bl ,m}-\smain+(m-1)\cdot 2^{-\kappa/2+2}$.
\end{claim}
To prove the claim, note that the premises of Lemma~\ref{lem_block} are satisfied: since $2n\leq U$, $V_\prim\geq 2\kappa^{2c-3}$ and $V_\prim\leq U\leq 2^{\kappa/2}$; $V_\rand\geq 4\kappa^{c+1}$; by assumption, every primary block has between $\kappa^{2c-3}+\kappa^c/3$ and $\kappa^{2c-3}+2\kappa^c/3$ keys, and every secondary block has between $\kappa^{c+1}$ and $3\kappa^{c+1}$ keys.
Therefore, by Lemma~\ref{lem_block}, the size of $\cD_\main^i$ is at most $\smain$, and the size of $\cD_\aux'$ is at most
\[
	\lg\binom{V_\prim}{m_\prim}+\lg\binom{V_\rand}{m-m_\prim}-\smain+\kappa^{2c-3}\cdot 2^{-\kappa/2+2}.
\]
By fusing the value of $m_\prim$ into the data structure, the size of $\cD_{\aux}$ is at most
\begin{align*}
	\OPT_{ V_\bl ,m}-\smain+(m-1)\cdot 2^{-\kappa/2+2},
\end{align*}
due to the fact that $m\geq \kappa^{2c-3}+\kappa^{c+1}$ and $\sum_{m_\prim}\binom{V_\prim}{m_\prim}\binom{V_\rand}{m-m_\prim}=\lg\binom{ V_\bl }{m}=\OPT_{V_\bl,m}$.
$\cD_\main^i$ and $\cD_\aux$ both have sizes as claimed in \eqref{eqn_size_main} and \eqref{eqn_size_aux}.
Also, note that we give the same random string $\cR$ to \emph{all} block pairs.
Thus, the total number of random bits needed is $\kappa^{c+1}$ by Lemma~\ref{lem_block}.
\bigskip

Next, when $i<j$, we recurse on the two halves of the block pairs, and merge them.
\begin{code}{dictrec}
	\item $k:=\lfloor (i+j)/2\rfloor$
	\item let $m_1$ be the number of keys in the $i$-th$,\ldots,k$-th block pair
	\item let $m_2$ be the number of keys in the $(k+1)$-th$,\ldots,j$-th block pair
	\item recurse on the two halves: \\
	\phantom\quad $(\cD_{\main}^i,\ldots,\cD_{\main}^k, \cD_{\aux,1}):=\mathtt{dict\_rec}(i,k,m_1,S,\cR)$\\
	\phantom\quad $(\cD_{\main}^{k+1},\ldots,\cD_{\main}^j,\cD_{\aux,2}):=\mathtt{dict\_rec}(k+1,j,m_2,S,\cR)$
	\item\label{step_dictrec_concat} apply Proposition~\ref{prop_concat_approx} to concatenate $\cD_{\aux,1}$ and $\cD_{\aux,2}$, and obtain $\cD_{\aux}'$
	\item\label{step_dictrec_fuse} apply Proposition~\ref{prop_fuse_approx} to fuse the value of $m_1$ into $\cD_{\aux}'$ for $m_1\in \{0,\ldots,m\}$, and obtain $\cD_{\aux}$
	\item return $(\cD_{\main}^i,\ldots,\cD_{\main}^j, \cD_{\aux})$
\end{code}
\begin{claim}\label{cl_induct_space}
	$|\cD_\main^i|\leq \smain$ for all $i$, and $|\cD_\aux|\leq \OPT_{(j-i+1)V_\bl,m}-\smain+(j-i+1)(m-1)\cdot 2^{-\kappa/2+2}$.
\end{claim}
We have already showed that each $|\cD_\main^i|\leq \smain$ above.
To prove the bound on $|\cD_\aux|$, by inductive hypothesis, we know that $\cD_{\aux,1}$ has size at most
\[
	\OPT_{(k-i+1) V_\bl ,m_1}-(k-i+1)\smain+(m_1-1)2^{-\kappa/2+2}
\]
and $\cD_{\aux,2}$ has size at most
\[
	\OPT_{(j-k) V_\bl ,m_2}-(j-k)\smain+(m_2-1)2^{-\kappa/2+2}.
\]
To apply Proposition~\ref{prop_concat_approx} in line~\ref{step_dictrec_concat}, it requires us to approximates the data structure sizes.
The following claim implies that the premises can be satisfied.
\newcommand{\contcldictrecconcat}{
	Both $\OPT_{(k-i+1) V_\bl ,m_1}-(k-i+1)\smain+(m_1-1)2^{-\kappa/2+2}$ and $\OPT_{(j-k) V_\bl ,m_2}-(j-k)\smain+(m_2-1)2^{-\kappa/2+2}$ can be approximated with an additive error of at most $2^{-\kappa}$ in $O(1)$ time.
}
\begin{claim}\label{cl_dictrec_concat}
	\contcldictrecconcat
\end{claim}
Assuming Claim~\ref{cl_dictrec_concat}, Proposition~\ref{prop_concat_approx} concatenates $\cD_{\aux,1}$ and $\cD_{\aux,2}$ into a data structure $\cD_\aux'$ of length at most
\[
	\OPT_{(k-i+1) V_\bl ,m_1}+\OPT_{(j-k) V_\bl ,m-m_1}-(j-i+1)\smain+(m-2)2^{-\kappa/2+2}+2^{-\kappa+4}.
\]
The following claim implies that the premises of Proposition~\ref{prop_fuse_approx} from line~\ref{step_dictrec_fuse} can be satisfied, because $-(j-i+1)\smain+(m-2)2^{-\kappa/2+2}+2^{-\kappa+4}$ does not depend on $m_1$, and can be computed efficiently.
\newcommand{\contcldictrecfuse}{
	For any $V_1,V_2,m\geq 0$, and $0\leq l\leq m$, $\sum_{i=0}^l 2^{\OPT_{V_1,i}+\OPT_{V_2,m-i}}$ can be approximated up to an additive error of at most $2^{-\kappa-3}\cdot\sum_{i=0}^m 2^{\OPT_{V_1,i}+\OPT_{V_2,m-i}}$ in $O(\kappa^5)$ time.
}
\begin{claim}\label{cl_dictrec_fuse}
	\contcldictrecfuse
\end{claim}
The proofs of both claims are deferred to Appendix~\ref{sec_binom_approx}.
Assuming Claim~\ref{cl_dictrec_fuse}, Proposition~\ref{prop_fuse_approx} fuses $m_1$ into $\cD_\aux'$, and obtains $\cD_\aux$ of length at most
\begin{align*}
	&\ \lg\left(\sum_{l=0}^m 2^{\OPT_{(k-i+1) V_\bl ,l}+\OPT_{(j-k) V_\bl ,m-l}-(j-i+1)\smain+(m-2)2^{-\kappa/2+2}+2^{-\kappa+4}}\right)+ (m+1)\cdot 2^{-\kappa+4} \\
	\leq&\ \lg\left(\sum_{l=0}^m 2^{\OPT_{(k-i+1) V_\bl ,l}+\OPT_{(j-k) V_\bl ,m-l}}\right)-(j-i+1)\smain+(m-1)2^{-\kappa/2+2} \\
	=&\ \OPT_{(j-i+1)V_\bl,m}-(j-i+1)\smain+(m-1)2^{-\kappa/2+2}
\end{align*}
This proves Claim~\ref{cl_induct_space}.

Thus, by induction, $\mathtt{dict\_rec}$ outputs $\cD_\main^1,\ldots,\cD_\main^{N_\bl}$ of length $\smain$ and $\cD_\aux$ of length
\[
	\OPT_{N_\bl V_\bl ,n}-N_\bl\cdot \smain+(n-1)2^{-\kappa/2+2}.
\]

Finally, we apply Proposition~\ref{prop_concat_approx} again to concatenate all $N_\bl+1$ data structures.
By storing approximations of sizes of $\cD_\main^i$ and $\cD_\aux$ in the lookup table, we obtain a data structure of length at most
\begin{align*}
	&\, N_\bl\cdot\smain+(\OPT_{N_\bl V_\bl ,n}-N_\bl\cdot \smain+(n-1)2^{-\kappa/2+2})+N_\bl\cdot 2^{-\kappa+3} \\
	\leq&\, \OPT_{N_\bl V_\bl ,n}+n\cdot 2^{-\kappa/2+2}.
\end{align*}
This proves the space bound in Lemma~\ref{lem_no_bad_block}.

\paragraph{Hash functions.} Let $h_i$ and $\oh_i$ be the bijections obtained by Lemma~\ref{lem_block} for blocks $\cB_{\prim}^i$ and $\cB_{\rand}^i$.
We define the bijections $h$ and $\oh$ as follows:
\begin{compactitem}
	\item for key $x\in S\cap (\cB_{\prim}^i\cup \cB_{\rand}^i)$, if $h_i(x)<\kappa^{2c-3}$, let $h(x):=(i-1)\kappa^{2c-3}+h_i(x)$, \\
	otherwise, let $h(x):=(N_\bl-i)\cdot\kappa^{2c-3}+\sum_{j<i}|(\cB_{\prim}^j\cup \cB_{\rand}^j)\cap S|+h_i(x)$;
	\item for non-key $x\notin S$, if $\oh_i(x)<V_{\prim}-\kappa^{2c-3}-\kappa^c$, let $\oh(x):=(i-1)(V_{\prim}-\kappa^{2c-3}-\kappa^c)+\oh_i(x)$, \\
	otherwise, let $\oh(x):=(N_\bl-i)\cdot(V_{\prim}-\kappa^{2c-3}-\kappa^c)+\sum_{j<i}|(\cB_{\prim}^j\cup \cB_{\rand}^j)\setminus S|+\oh_i(x)$.
\end{compactitem}
Essentially, the smallest hash values will be those with $h_i(x)<\kappa^{2c-3}$ or $\oh_i(x)<V_{\prim}-\kappa^{2c-3}-\kappa^c$, ordered according to $i$ and $h_i(x)$ or $\oh_i(x)$.
Then the rest take larger values ordered according to $i$ and $h_i(x)$ or $\oh_i(x)$.
By definition, they are both bijections.

\paragraph{Lookup tables.} We store the following information in the lookup table.
\begin{code}{tbl}[lookup table][]
	\item $\tblB_{V_{\prim},V_{\rand}}$ from Lemma~\ref{lem_block}
	\item the lookup table for line~\ref{step_dictrec_fuse_mpr} from Proposition~\ref{prop_fuse} for all valid values $m_\prim$ and $m_{\rand}$
	\item approximated value of $\smain$ and the (final) size of $\cD_\aux$, up to $O(\kappa)$ bits of precision
\end{code}
By Lemma~\ref{lem_block}, the lookup table size is $2^{\epsilon\kappa}$.
Since $\kappa=O(\lg U)$ and $n\geq U^{1/{12}}$, by readjusting the constant $\epsilon$, the lookup table size is at most $n^\epsilon$.

\paragraph{Query algorithm.}
Now, we show how to answer $\hash$ queries.
Given a query $x\in[U]$, we first shift it according to $\Delta$, as we did at preprocessing, $x\mapsto (x+\Delta)\Mod U$.
If $x$ is in a primary block, we query the corresponding main data structure.
If the main data structure does not return the answer, or $x$ is not in a primary block, we recursively decode the corresponding auxiliary data structure, and run $\qalgB$.
\begin{code}{qalgG}[query algorithm][(U,n,x)]
	\item if $x$ is in the $i$-th primary block
	\item\quad apply Proposition~\ref{prop_concat_approx} to decode $\cD_\main^i$
	\item\quad if $(b,v):=\cD_\main^i.\qalgB_\main(V_\prim,x)\neq \textrm{``unknown''}$ (from Lemma~\ref{lem_block})
	\item\quad\quad if $b=1$, return $(1, (i-1)\kappa^{2c-3}+v)$
	\item\quad\quad if $b=0$, return $(0, (i-1)(V_\prim-\kappa^{2c-3}-\kappa^c)+v)$
	\item decode $\cD_\aux$ and return $\cD_\aux.\mathtt{qalg\_rec}(1,N_\bl,0,n,x)$
\end{code}
Since $V_\prim/V_\rand=O(\kappa^{c-4})$ and we randomly shifted the universe, $x$ is in a primary block with probability $1-O(\kappa^{-c+4})$.
Also, by Lemma~\ref{lem_block}, $\qalgB_\main$ runs in constant time.
It returns ``unknown'' with probability at most $O(\kappa^{-c+3})$ for a uniformly random $\cR$, and returns $\hash(x)$ otherwise.
Therefore, the probability that $\mathtt{qalgG}$ terminates before reaching the last line is $1-O(\kappa^{-c+4})$.
Since $\kappa=\Theta(\lg U)$, it computes $\hash(x)$ in constant time with probability $1-O(\lg^{-c+4} U)$.

Next, we show how to implement $\mathtt{qalg\_rec}(i, j, s, m, x)$, which takes as parameters 
\begin{compactitem}
	\item $(i, j)$: a range of blocks,
	\item $s$: the total number of keys before block $i$,
	\item $m$, the total number of keys in blocks $i$ to $j$, and 
	\item $x$, the element being queried.
\end{compactitem}
We will prove that its worst-case running time is $O(\lg^7 U)$.
\begin{code}{qalgrec}[query algorithm][(i,j,s,m,x)][qalg\_rec]
	\item if $i=j$
	\item \quad apply Proposition~\ref{prop_fuse} to decode $m_\prim$ and $\cD_\aux'$
	\item \quad $(b,v):=(\cD_\main^i,\cD_\aux').\qalgB(V_\prim,m_\prim,V_\rand,m-m_\prim,x-(i-1)(V_{\prim}+V_\rand))$ \\
	\phantom{}\hfill (from Lemma~\ref{lem_block})
	\item \quad if $b=1$, return $(1, (N_{\bl}-i)\cdot\kappa^{2c-3}+s+v)$
	\item \quad if $b=0$, return $(0, (N_{\bl}-i)\cdot(V_\prim-\kappa^{2c-3}-\kappa^c)+((i-1)\cdot V_\bl-s)+v)$ \ctn
\end{code}
In the base case with only one block, we simply decode the value of $m_{\prim}$ as well as the corresponding $\cD_\aux'$ from the $i$-th block pair.
By running $\qalgB$ from Lemma~\ref{lem_block} to query within the block pair, we compute $\hash(x)$ according to its definition in $O(\kappa^4)$ time.

\begin{code}{qalgrec}
	\item $k:=\left\lfloor (i+j)/2\right\rfloor$
	\item apply Proposition~\ref{prop_fuse_approx} and Claim~\ref{cl_dictrec_fuse} to decode $m_1$ and $\cD_\aux'$
	\item apply Proposition~\ref{prop_concat_approx} and Claim~\ref{cl_dictrec_concat} to decode $\cD_{\aux,1}$ and $\cD_{\aux,2}$
	\item if $x$ is in $i$-th$,\ldots,k$-th block pair
	\item\quad return $\cD_{\aux,1}.\mathtt{qalgrec}(i,k,s,m_1,x)$
	\item else
	\item\quad return $\cD_{\aux,2}.\mathtt{qalgrec}(k+1,j,s+m_1,m-m_1,x)$
\end{code}
In general, we decode $m_1$, the number of elements in the first half of the blocks.
Then we decode the data structures for the two halves.
Depending on where the query is, we recurse into one of the two data structures.
Proposition~\ref{prop_concat_approx}, Proposition~\ref{prop_fuse_approx}, Claim~\ref{cl_dictrec_concat} and Claim~\ref{cl_dictrec_fuse} guarantee that the decoding takes $O(\kappa^6)$ time. 
The recursion has at most $O(\lg n)\leq \kappa$ levels.
Thus, the total running time of $\mathtt{qalgrec}$ is at most $O(\kappa^7)$.
This proves the claim on the query time, and hence, it proves Lemma~\ref{lem_no_bad_block}.

\subsection{At least one bad block pair}\label{sec_bad_block}
Now, let us show how to handle sets with at least one bad block.
We will show that the space usage for such sets is $\OPT_{U,n}-\Omega(\kappa^3)$.
\begin{lemma}\label{lem_with_bad_block}
	If $U$ is a multiple of $ V_\bl $, then there is a data structure with guarantees in Theorem~\ref{thm_medium_set}, for all sets with at least one bad block pair.
	Moreover, the size of the data structure is at most
	\[
		\OPT_{U,n}-\Omega(\kappa^3).
	\]
\end{lemma}
Note that this is possible, because by Chernoff bound, there are only at most $2^{-\Omega(\kappa^3)}$ fraction such sets, and we can even afford to spend at least $O(\kappa^3)=O(\lg^3 U)$ extra bits.
The first $\lceil\lg N_\bl\rceil$ bits are used to encode the number of bad block pairs $N_\bad$.
It turns out that the fraction of input sets with $N_\bad$ bad pairs is $2^{-\Omega(\kappa^3 N_\bad)}$, as we mentioned in Section~\ref{sec_overview}.
By the argument there, we can afford to use $O(\kappa^3 N_\bad)$ extra bits.

The idea is to construct a mapping which maps all good block pairs to the first $N_\bl-N_\bad$ pairs, construct a data structure using the above algorithm for good blocks, and finally handle the bad pairs separately.

To construct such a mapping, observe that the following two numbers are equal:
\begin{compactenum}[(a)]
	\item the number of \emph{good} pairs among the last $N_\bad$ pairs, and
	\item the number of \emph{bad} pairs among the first $N_\bl-N_\bad$ pairs.
\end{compactenum}
Hence, in the mapping, we map all the good pairs among the last $N_\bad$ to all bad pairs among the first $N_\bl-N_\bad$.
The good pairs in the first $N_\bl-N_\bad$ pairs will be mapped to themselves.
To store such a mapping, we spend $O(N_\bad\cdot\lg N_\bl)$ bits to store a hash table of all the bad pairs using the FKS hashing.
Then we spend $O(N_\bad\cdot\lg N_\bl)$ bits to store for each pair in the last $N_\bad$ pairs, whether it is a good pair and if it is, which bad pair it will be mapped to.
The mapping takes $O(N_\bad\cdot\lg N_\bl)$ bits to store in total (which is much smaller than $\kappa^3 N_\bad$).
It takes constant time to evaluate.

This mapping maps all good pairs to the first $N_\bl-N_\bad$ pairs.
Then we apply $\mathtt{dict\_rec}$ for good pairs from Lemma~\ref{lem_no_bad_block} to construct a data structure using
\[
	\OPT_{(N_\bl-N_\bad) V_\bl ,n-n_{\bad}}+(n-n_{\bad})\cdot 2^{-\kappa/2+2}\leq \left\lceil \OPT_{(N_\bl-N_\bad) V_\bl ,n-n_{\bad}}\right\rceil+1
\]
bits, where $n_{\bad}$ is the number of keys in the bad block pairs.

Next, we construct data structures for the bad pairs.
Consider a bad pair with $m_\prim$ keys in the primary block and $m_\rand$ keys in the secondary block.
Thus, either $m_\prim\notin[\kappa^{2c-3}+\kappa^c/3,\kappa^{2c-3}+2\kappa^c/3]$, or $m_\rand\notin[\kappa^{c+1},3\kappa^{c+1}]$.
We construct two separate data structures, one for the primary block and one for secondary block (note that it might be the case that the number of keys in the primary block is within the above range, but the block pair is bad due to the secondary block, or vice versa, we still construct two separate data structures for \emph{both} of them using the following argument).
It turns out that if the number of keys in the block is at most $\kappa^{O(1)}$, then there is a data structure using only $O(1)$ extra bit, answering queries in constant time.
\newcommand{\contlemsmallbadblock}{
	Let $c$ be any constant positive integer and $\epsilon$ be any positive constant.
	There is a preprocessing algorithm $\phS$, query algorithm $\qalgS$ and lookup tables $\tblS_{V,m}$ of sizes $\tilde{O}(2^{\epsilon\kappa})$, such that for any $V\leq 2^{\kappa/2}$ and $m\leq \kappa^c$, given a set $S\subset [V]$ of $m$ keys, $\phS$ preprocesses $S$ into a data structure of size at most
	\[
		\OPT_{V,m}+(m-1)\cdot 2^{-\kappa/2+1},
	\]
	such that it defines a bijection $h$ between $S$ and $[m]$ and a bijection $\oh$ between $[V]\setminus S$ and $[V-m]$.
	Given any $x\in[V]$, $\qalgS$ answers $\hash(x)$ in constant time, by accessing the data structure and $\tblS_{V,m}$.	
}
\begin{lemma}\label{lem_small_bad_block}
	\contlemsmallbadblock
\end{lemma}
In particular, the size is at most $\OPT_{V,m}+O(1)$.
The lemma is an immediate corollary of Lemma~\ref{lem_block_rank} in Section~\ref{sec_small_set}.
Its proof is deferred to Section~\ref{sec_small_set}.

On the other hand, sets that have a block with more than $\kappa^{3c}$ keys are even more rare.
By Chernoff bound, we can estimate that the fraction of sets with at least one block with $m>\kappa^{3c}$ keys is at most $2^{-\Omega(m\lg m)}$.
This suggests that for every (bad) block with $m>\kappa^{3c}$ keys, we can afford to spend $O(m\lg m)$ extra bits.
A simple modification to \cite{Pagh01} gives such a data structure.
\newcommand{\contlemlargebadblock}{
	Given a set $S\subset [V]$ of $m$ keys, there is a data structure of size
	\[
		\OPT_{V,m}+O(m+\lg\lg V),
	\]
	such that it defines a bijection $h$ between $S$ and $[m]$ and a bijection $\oh$ between $[V]\setminus S$ and $[V-m]$.
	It supports $\hash$ queries in constant time.	
}
\begin{lemma}\label{lem_large_bad_block}
	\contlemlargebadblock
\end{lemma}
Note that $\lg\lg V\leq \lg\kappa$, and $m\geq \kappa^{3c}$.
The number of extra bits is simply $O(m)$.
We prove this lemma in Appendix~\ref{sec_proof_large_bad_block}.

For each bad block pair, we write down the two numbers $m_\prim$ and $m_\rand$ using $O(\lg n)$ bits.
Then if $m_\prim\leq \kappa^{3c}$, we apply Lemma~\ref{lem_small_bad_block}, and obtain a data structure with
\[
	\OPT_{V_\prim,m_\prim}+O(1)
\]
bits.
If $m_\prim>\kappa^{3c}$, we apply Lemma~\ref{lem_large_bad_block}, and obtain a data structure with
\[
	\OPT_{V_\prim,m_\prim}+O(m_\prim)
\]
bits.
Likewise for the secondary block, we obtain a data structure with
\[
	\OPT_{V_\rand,m_\rand}+O(1)
\]
bits if $m_\rand\leq \kappa^{3c}$, and
\[
	\OPT_{V_\rand,m_\rand}+O(m_\rand)
\]
bits if $m_\rand>\kappa^{3c}$.

Finally, we concatenate all data structures (which now all have integer lengths), and for each bad pair, we further store a pointer pointing to its corresponding data structure, as well as the total number of keys in all \emph{bad} blocks prior to it (which helps us compute the hash values).

\paragraph{Space usage.}
Let us first bound the space usage for the data structures for a bad block pair.
Consider the $i$-th bad block pair, suppose it has $m_{\prim,i}$ keys in the primary block and $m_{\rand,i}$ keys in the secondary block.
Then note that we have
\begin{align}
	\lg\binom{V_\prim}{m_{\prim,i}}&=\lg\frac{V_\prim!}{m_{\prim,i}!(V_\prim-m_{\prim,i})!} \nonumber\\
	\intertext{which by Sterling's formula, is at most}
	&\leq\lg\left(\frac{V_\prim}{e}\right)^{V_\prim}-\lg\left(\frac{m_{\prim,i}}{e}\right)^{m_{\prim,i}}-\lg\left(\frac{V_\prim-m_{\prim,i}}{e}\right)^{V_\prim-m_{\prim,i}}+O(\lg V_\prim) \nonumber\\
	&={V_\prim}\lg V_\prim-{m_{\prim,i}}\lg m_{\prim,i}-(V_\prim-m_{\prim,i})\lg\left(V_\prim-m_{\prim,i}\right)+O(\lg V_\prim).\label{eqn_bad_prim}
\end{align}
In the following, we are going to compare \eqref{eqn_bad_prim} with 
\begin{equation}\label{eqn_bad_prim_2}
	{V_\prim}\lg V_\prim-{m_{\prim,i}}\lg \overline{m_{\prim}}-(V_\prim-m_{\prim,i})\lg\left(V_\prim-\overline{m_{\prim}}\right)+O(\lg V_\prim),
\end{equation}
where $\overline{m_\prim}=\kappa^{2c-3}+\kappa^c/2$ is the average number of keys in a primary block.
First observe that $f(x)=m\lg x+(V-m)\lg (V-x)$ achieves its maximum at $x=m$, thus, $\eqref{eqn_bad_prim} \leq \eqref{eqn_bad_prim_2}$.
On the other hand, if $m_{\prim,i}\notin[\kappa^{2c-3}+\kappa^c/3,\kappa^{2c-3}+2\kappa^c/3]$, i.e., $m_{\prim,i}$ is far from $\overline{m_{\prim}}$, \eqref{eqn_bad_prim} is even smaller.
\begin{claim}\label{cl_eqn2_eqn1}
	We have $\eqref{eqn_bad_prim_2}\geq\eqref{eqn_bad_prim}$.
	Moreover, if $m_{\prim,i}\notin[\kappa^{2c-3}+\kappa^c/3,\kappa^{2c-3}+2\kappa^c/3]$, 
	\[
		\eqref{eqn_bad_prim_2}-\eqref{eqn_bad_prim}\geq \begin{cases} \Omega(\kappa^3) & m_{\prim,i}\leq \kappa^{3c}, \\ m_{\prim,i}\lg \kappa & m_{\prim,i}>\kappa^{3c}. \end{cases}
	\]
\end{claim}
\begin{proof}
\begin{align*}
	\eqref{eqn_bad_prim_2}-\eqref{eqn_bad_prim}&=-m_{\prim,i}\lg\frac{\overline{m_\prim}}{m_{\prim,i}}-(V_\prim-m_{\prim,i})\lg\frac{V_\prim-\overline{m_{\prim}}}{V_\prim-m_{\prim,i}} \\
	&=-m_{\prim,i}\lg\left(1+\frac{\overline{m_\prim}-m_{\prim,i}}{m_{\prim,i}}\right)-(V_\prim-m_{\prim,i})\lg\left(1+\frac{m_{\prim,i}-\overline{m_{\prim}}}{V_\prim-m_{\prim,i}}\right).
\end{align*}
By the facts that $\ln(1+x)\leq x$ for $x>-1$, $\ln(1+x)\leq x-\frac{1}{4}x^2$ for $|x|\leq 1$, and $\ln(1+x)\leq 3x/4$ for $x>1$, when $m_{\prim,i}\leq \overline{m_{\prim}}/2$, we have
\begin{align*}
	\eqref{eqn_bad_prim_2}-\eqref{eqn_bad_prim}&=-m_{\prim,i}\lg\left(1+\frac{\overline{m_\prim}-m_{\prim,i}}{m_{\prim,i}}\right)-(V_\prim-m_{\prim,i})\lg\left(1+\frac{m_{\prim,i}-\overline{m_{\prim}}}{V_\prim-m_{\prim,i}}\right) \\
	&\geq -m_{\prim,i}\cdot\frac{3}{4}\cdot\frac{\overline{m_\prim}-m_{\prim,i}}{m_{\prim,i}}\lg e-(V_\prim-m_{\prim,i})\cdot \frac{m_{\prim,i}-\overline{m_{\prim}}}{V_\prim-m_{\prim,i}}\lg e \\
	&=\frac{\lg e}{4}(\overline{m_\prim}-m_{\prim,i})\\
	&\geq \Omega(\kappa^{2c-3}).
\end{align*}

When $m_{\prim,i}\geq \overline{m_{\prim}}/2$ and $m_{\prim,i}\leq \kappa^{3c}$, we have
\begin{align*}
	\eqref{eqn_bad_prim_2}-\eqref{eqn_bad_prim}&=-m_{\prim,i}\lg\left(1+\frac{\overline{m_\prim}-m_{\prim,i}}{m_{\prim,i}}\right)-(V_\prim-m_{\prim,i})\lg\left(1+\frac{m_{\prim,i}-\overline{m_{\prim}}}{V_\prim-m_{\prim,i}}\right) \\
	&\geq -m_{\prim,i}\left(\frac{\overline{m_\prim}-m_{\prim,i}}{m_{\prim,i}}-\frac{1}{4}\left(\frac{\overline{m_\prim}-m_{\prim,i}}{m_{\prim,i}}\right)^2\right)\lg e \\
	&\quad -(V_\prim-m_{\prim,i})\cdot\frac{m_{\prim,i}-\overline{m_{\prim}}}{V_\prim-m_{\prim,i}}\lg e \\
	&=\frac{\lg e}{4}\cdot \frac{(m_{\prim,i}-\overline{m_\prim})^2}{m_{\prim,i}} \\
	&\geq \Omega(\kappa^3),
\end{align*}
where the last inequality uses the fact that $m_{\prim,i}\notin[\kappa^{2c-3}+\kappa^c/3,\kappa^{2c-3}+2\kappa^c/3]$.

When $m_{\prim,i}\geq \kappa^{3c}$, we have
\begin{align*}
	\eqref{eqn_bad_prim_2}-\eqref{eqn_bad_prim}&=-m_{\prim,i}\lg\left(1+\frac{\overline{m_\prim}-m_{\prim,i}}{m_{\prim,i}}\right)-(V_\prim-m_{\prim,i})\lg\left(1+\frac{m_{\prim,i}-\overline{m_{\prim}}}{V_\prim-m_{\prim,i}}\right) \\
	&\geq -m_{\prim,i}\lg\frac{\overline{m_\prim}}{m_{\prim,i}}-(m_{\prim,i}-\overline{m_{\prim}})\lg e \\
	&\geq m_{\prim,i}\lg\frac{m_{\prim,i}}{e\cdot \overline{m_\prim}} \\
	&\geq m_{\prim,i}\lg \kappa.
\end{align*}
Combining the three cases, we conclude that
\[
	\eqref{eqn_bad_prim_2}-\eqref{eqn_bad_prim}\geq \begin{cases} \Omega(\kappa^3) & m_{\prim,i}\leq \kappa^{3c}, \\ m_{\prim,i}\lg \kappa & m_{\prim,i}>\kappa^{3c}, \end{cases}
\]
proving the claim.
\end{proof}

On the other hand, since $V_\prim=U\cdot \frac{\overline{m_\prim}}{n}\pm O(1)$, we have
\[
	\eqref{eqn_bad_prim_2}=V_\prim\lg U-m_{\prim,i}\lg n-(V_\prim-m_{\prim,i})\lg (U-n)+O(\lg V_\prim).
\]
Thus, Claim~\ref{cl_eqn2_eqn1} implies that if $m_{\prim,i}\notin[\kappa^{2c-3}+\kappa^c/3,\kappa^{2c-3}+2\kappa^c/3]$, i.e., the primary block is bad, then the data structure size for the primary block is at most
\[
	V_{\prim}\lg U-m_{\prim,i}\lg n-(V_\prim-m_{\prim,i})\lg(U-n)-\Omega(\kappa^3)
\]
(since $O(m_{\prim,i})\ll m_{\prim,i}\lg\kappa$ and $O(1)\ll \kappa^3$), and otherwise, it is at most
\[
	V_{\prim}\lg U-m_{\prim,i}\lg n-(V_\prim-m_{\prim,i})\lg (U-n)+O(\lg V_\prim).
\]

By applying the same argument to the secondary blocks, we conclude that if $m_{\rand,i}\notin[\kappa^{c+1},3\kappa^{c+1}]$, i.e., the secondary block is bad, then the data structure size for the secondary block is at most
\[
	V_{\rand}\lg U-m_{\rand,i}\lg n+(V_\rand-m_{\rand,i})\lg (U-n)-\Omega(\kappa^3),
\]
and otherwise, it is at most
\[
	V_{\rand}\lg U-m_{\rand,i}\lg n+(V_\rand-m_{\rand,i})\lg(U-n)+O(\lg V_\prim).
\]

Summing up the two bounds, and by the fact that at least one of the primary and secondary block is bad, the data structure size for the $i$-th bad block pair is at most
\begin{align*}
	V_\bl\lg U-(m_{\prim,i}+m_{\rand,i})\lg n-( V_\bl -m_{\prim,i}-m_{\rand,i})\lg(U-n)-\Omega(\kappa^3),
\end{align*}
since $\kappa^3\gg \lg V_\prim$.

Now we sum up the size for all $N_{\bad}$ bad blocks, which in total contain $n_{\bad}$ keys, the total size is at most
\begin{align*}
	N_{\bad}\cdot V_\bl\lg U-n_\bad\lg n-(N_{\bad}\cdot  V_\bl -n_\bad)\lg(U-n)-\Omega(N_{\bad}\cdot\kappa^3).
\end{align*}

Therefore, the total size of the data structure is at most
\begin{align*}
	&\,\OPT_{(N_\bl-N_\bad) V_\bl ,n-n_\bad}+O(N_\bad\lg N_\bl)+N_{\bad}\cdot V_\bl\lg U\\
	&\quad -n_\bad\lg n-(N_{\bad}\cdot  V_\bl -n_\bad)\lg(U-n)-\Omega(N_{\bad}\cdot\kappa^3) \\
	=&\, (N_\bl-N_\bad) V_\bl\lg ((N_\bl-N_\bad) V_\bl)-(n-n_\bad)\lg (n-n_\bad) \\
	&\quad -((N_\bl-N_\bad) V_\bl -(n-n_\bad))\lg ((N_\bl-N_\bad) V_\bl -(n-n_\bad))\\
	&\quad +N_{\bad}\cdot V_\bl\lg U-n_\bad\lg n-(N_{\bad}\cdot  V_\bl -n_\bad)\lg(U-n)-\Omega(N_{\bad}\cdot\kappa^3), \\
	\intertext{which by Claim~\ref{cl_eqn2_eqn1} and the fact that $N_\bad\geq 1$, is at most}
	=&\, (N_\bl-N_\bad) V_\bl\lg U-(n-n_\bad)\lg n-((N_\bl-N_\bad) V_\bl -(n-n_\bad))\lg (U-n)\\
	&\quad +N_{\bad}\cdot V_\bl\lg U-n_\bad\lg n-(N_{\bad}\cdot  V_\bl -n_\bad)\lg(U-n)-\Omega(\kappa^3), \\
	=&\, U\lg U-n\lg n-(U-n)\lg (U-n)-\Omega(\kappa^3) \\
	\leq&\, \lg\binom{U}{n}-\Omega(\kappa^3) \\
	=&\, \OPT_{U,n}-\Omega(\kappa^3),
\end{align*}
as we claimed.

\paragraph{Hash functions.}
For all $x$ in the good blocks, we simply use their hash value according to Lemma~\ref{lem_no_bad_block}, for which, $h$ takes values in $[n-n_{\bad}]$ and $\oh$ takes values in $[V_\bl\cdot (N_\bl-N_\bad)-(n-n_{\bad})]$.
For $x$ in the $i$-th bad pair with hash value $v$, let $s_i$ be the total number of keys in first $i-1$ bad pairs (which is explicitly stored in the data structure), then if $x\in S$, we set $h(x):=n-n_{\bad}+s_i+v$; if $x\notin S$, we set $\oh(x):=V_\bl\cdot (N_\bl-N_\bad+(i-1))-(n-n_{\bad}+s_i)+v$.

That is, all elements in good blocks take the smallest values, and elements in bad blocks take the rest according to the order of the blocks.
By definition, they are both bijections.

\paragraph{Lookup tables.}
We include the lookup table from the data structure for no bad blocks, as well as all tables $\tblS_{V,m}$ from Lemma~\ref{lem_small_bad_block} for $V=V_\prim$ or $V=V_\rand$, and $1\leq m\leq \kappa^{3c}$.
The total lookup table size is $\tilde{O}(2^{\epsilon\kappa})$.
It is at most $n^\epsilon$ by readjusting the constant $\epsilon$.

\paragraph{Query algorithm.}
Given a query $x$, suppose $x$ is in the $i$-th block pair.
We first query the hash table to check if it is one of the bad pairs.
If the block pair is bad, we follow the pointer and query the data structure for the primary block or the secondary block depending on which block $x$ is in.
Its hash value can be computed according to the definition.
It takes constant time.

If the block pair is good, we spend constant time to find out where $i$-th block pair is mapped to, in the first $N_\bl-N_\bad$ pairs.
Then we run $\mathtt{qalgG}$ for good blocks, which takes constant time in expectation.
This proves Lemma~\ref{lem_with_bad_block}.

\subsection{Final data structure for medium size sets}
Consider the following preprocessing algorithm for general $U$ and $U^{1/{12}}\leq n\leq U/2$ ($U$ not necessarily a multiple of $V_\bl$).
We first construct a data structure for the block pairs, and fuse the two cases (with or without bad blocks) together.
\begin{code}{perfHash}[preprocessing algorithm][(U,n,S,\cR)]
	\item compute $V_\prim$, $V_\rand$ and $\kappa$
	\item compute $N_\bl:=U\Div  V_\bl $ and $V:=U\Mod  V_\bl $
	\item divide the universe $[U]$ into $N_\bl$ block pairs and a last block of size $V$
	\item if all $N_\bl$ block pairs are good
	\item\quad set $i:=0$, apply Lemma~\ref{lem_no_bad_block} on the $N_\bl$ block pairs, and obtain a data structure $\cD_0$
	\item else
	\item\quad set $i:=1$, apply Lemma~\ref{lem_with_bad_block} on the $N_\bl$ block pairs, and obtain a data structure $\cD_1$
	\item\label{step_dict_8} apply Proposition~\ref{prop_fuse} to fuse $i$ into $\cD_i$, and obtain a data structure $\cD_\bl$ \ctn
\end{code}
Suppose there are $n_\bl$ keys in the first $N_\bl$ block pairs.
By Lemma~\ref{lem_no_bad_block}, $\cD_0$ has length at most
\[
	\OPT_{N_\bl V_\bl ,n_\bl}+n_\bl\cdot 2^{-\kappa/2+2}.
\]
By Lemma~\ref{lem_with_bad_block}, $\cD_1$ has length at most
\[
	\OPT_{N_\bl V_\bl ,n_\bl}-\Omega(\kappa^3).
\]
Thus, by Proposition~\ref{prop_fuse}, $\cD_\bl$ has length at most
\begin{align*}
	&\, \OPT_{N_\bl V_\bl ,n_\bl}+\lg (2^{n_\bl\cdot 2^{-\kappa/2+2}}+2^{-\Omega(\kappa^3)})+2^{-\kappa+2} \\
	\leq&\, \OPT_{N_\bl V_\bl ,n_\bl}+\lg (1+n_\bl\cdot 2^{-\kappa/2+2}+2^{-\Omega(\kappa^3)})+2^{-\kappa+2} \\
	\leq&\, \OPT_{N_\bl V_\bl ,n_\bl}+n_\bl\cdot 2^{-\kappa/2+3} \\
	=&\, \OPT_{U-V,n_\bl}+n_\bl\cdot 2^{-\kappa/2+3}.
\end{align*}

Then we construct a separate data structure for the last block using Lemma~\ref{lem_small_bad_block} or Lemma~\ref{lem_large_bad_block} based on the number of keys in it.
\begin{code}{perfHash}
	\item if $n-n_\bl\leq \kappa^{3c}$
	\item \quad construct $\cD_{\la}$ for the last block using Lemma~\ref{lem_small_bad_block}
	\item\label{step_dict_13}\quad apply Proposition~\ref{prop_concat} to concatenate $\cD_{\bl}$ and $\cD_\la$, and obtain $\cD'$
	\item \quad let $n'_\bl=n_\bl$
	\item else
	\item \quad construct $\cD_{\la}$ for the last block using Lemma~\ref{lem_large_bad_block}
	\item \quad spend $\lceil\lg n\rceil$ bits to store $n_\bl$, 
	\item \quad round both $\cD_{\bl}$ and $\cD_\la$ to integral lengths and concatenate them
	\item \quad spend $\lceil\lg n\rceil$ bits to store a point to $\cD_\la$, let the resulting data structure be $\cD'$
	\item \quad let $n'_\bl=n-\kappa^{3c}-1$
	\item\label{step_dict_14} apply Proposition~\ref{prop_fuse} to fuse the value of $n'_\bl$ into $\cD'$ for $n'_\bl\in [n-\kappa^{3c}-1,n]$, and obtain $\cD$
	\item return $\cD$
\end{code}
We do not fuse the whole value of $n_\bl$ into $\cD'$, as its range is large and Proposition~\ref{prop_fuse} requires a large lookup table to do this.
However, we only fuse its value if $n_\bl\geq n-\kappa^{3c}$, and otherwise only indicate that it is smaller than $n-\kappa^{3c}$ (by setting $n'_\bl$ to $n-\kappa^{3c}-1$).
This is fine because we spend extra $\lg n$ bits to explicitly store its value in this case.
We will show in the following that the total space is close the optimum.

There are $n-n_\bl$ keys in the last block.
If $n-n_\bl\leq \kappa^{3c}$, $\cD_\la$ has size at most
\[
	\OPT_{V,n-n_\bl}+(n-n_\bl-1)\cdot 2^{-\kappa/2+1}.
\]
In this case, the length of $\cD'$ is at most
\begin{align}
	&\,\OPT_{U-V,n_\bl}+\OPT_{V,n-n_\bl}+n_\bl\cdot 2^{-\kappa/2+3}+(n-n_\bl-1)\cdot 2^{-\kappa/2+1}+2^{-\kappa+2}\nonumber\\
	\leq&\, \OPT_{U-V,n_\bl}+\OPT_{V,n-n_\bl}+n\cdot 2^{-\kappa/2+3} \nonumber\\
	=&\, \lg\binom{U-V}{n_\bl}\binom{V}{n-n_\bl}+n\cdot 2^{-\kappa/2+3}.\label{eqn_size_cD_good}
\end{align}

If $n-n_\bl>\kappa^{3c}$, $\cD_\la$ has size at most
\[
	\OPT_{V,n-n_\bl}+O(n-n_\bl+\lg n).
\]
The length of $\cD'$ is at most
\begin{equation}\label{eqn_size_cD_bad}
	\lg\binom{U-V}{n_\bl}\binom{V}{n-n_\bl}+O(n-n_\bl+\lg n).
\end{equation}
By Sterling's formula, the first term is at most
\begin{align*}
	\lg\binom{U-V}{n_\bl}\binom{V}{n-n_\bl}&=\lg\frac{(U-V)!V!}{n_{\bl}!(U-V-n_\bl)!(n-n_\bl)!(V-n+n_\bl)!} \\
	&\leq (U-V)\lg (U-V)-n_{\bl}\lg n_\bl-(U-V-n_\bl)\lg(U-V-n_\bl)\\
	&\quad +V\lg V-(n-n_\bl)\lg(n-n_\bl)-(V-n+n_\bl)\lg (V-n+n_\bl)+O(\lg U) \\
	\intertext{which again by the fact that $f(x)=m\lg x+(V-m)\lg (V-x)$ is maximized at $x=m$, is at most}
	&\leq (U-V)\lg (U-V)-n_{\bl}\lg \frac{(U-V)n}{U}-(U-V-n_\bl)\lg(U-V-\frac{(U-V)n}{U})\\
	&\quad +V\lg V-(n-n_\bl)\lg(n-n_\bl)-(V-n+n_\bl)\lg (V-n+n_\bl)+O(\lg U) \\
	&=(U-V)\lg U-n_{\bl}\lg n-(U-V-n_\bl)\lg(U-n)\\
	&\quad +V\lg V-(n-n_\bl)\lg \frac{Vn}{U}-(V-n+n_\bl)\left(V-\frac{Vn}U\right) \\
	&\quad -(n-n_\bl)\lg\frac{U(n-n_\bl)}{Vn}-(V-n+n_\bl)\lg \frac{V-n+n_\bl}{V-\frac{Vn}U}+O(\lg U) \\
	&=U\lg U-n\lg n-(U-n)\lg(U-n)\\
	&\quad -(n-n_\bl)\lg\frac{U(n-n_\bl)}{Vn}+(V-n+n_\bl)\lg \left(1+\frac{n-n_\bl-\frac{Vn}U}{V-n+n_\bl}\right)+O(\lg U) \\
	\intertext{which by the fact that $\ln(1+x)\leq x$, is at most}
	&\leq \lg\binom{U}{n}-(n-n_\bl)\lg\frac{U(n-n_\bl)}{Vn}+(n-n_\bl)\lg e+O(\lg U) \\
	\intertext{which by the fact that $n-n_\bl\geq \kappa^{3c}$, is at most}
	&\leq \lg\binom{U}{n}-(n-n_\bl)\lg\frac{U\kappa^{3c}}{eVn}+O(\lg U) \\
	\intertext{which by the fact that $U/(eVn)\geq \kappa^{-2c}$ and $c\geq 1$, is at most}
	&\leq \lg\binom{U}{n}-(n-n_\bl)\lg \kappa+O(\lg U).
\end{align*}
Therefore, when $n-n_\bl>\kappa^{3c}$, the size of $\cD'$ is at most
\begin{align*}
	\eqref{eqn_size_cD_bad}&\leq \lg\binom{U}{n}-\Omega((n-n_\bl)\lg\kappa)+O(\lg U+\lg n)\\
	&\leq \OPT_{U,n}-\Omega(\kappa^3).
\end{align*}

Finally together with Equation~\eqref{eqn_size_cD_good}, by Proposition~\ref{prop_fuse}, the size of $\cD$ is at most
\begin{align*}
	&\, \lg\left(2^{\OPT_{U,n}-\Omega(\kappa^3)}+\sum_{n_\bl=n-\kappa^{3c}}^n\binom{U-V}{n_\bl}\binom{V}{n-n_\bl}\cdot 2^{n\cdot 2^{-\kappa/2+3}} \right)+n\cdot 2^{-\kappa+2} \\
	\leq &\, \lg\left(2^{\OPT_{U,n}-\Omega(\kappa^3)}+\sum_{n_\bl=0}^n\binom{U-V}{n_\bl}\binom{V}{n-n_\bl}\cdot 2^{n\cdot 2^{-\kappa/2+3}} \right)+n\cdot 2^{-\kappa+2} \\
	= &\, \lg\left(2^{\OPT_{U,n}-\Omega(\kappa^3)}+\binom{U}{n}\cdot 2^{n\cdot 2^{-\kappa/2+3}} \right)+n\cdot 2^{-\kappa+2} \\
	= &\, \OPT_{U,n}+\lg\left(2^{-\Omega(\kappa^3)}+2^{n\cdot 2^{-\kappa/2+3}} \right)+n\cdot 2^{-\kappa+2} \\
	\leq &\, \OPT_{U,n}+\lg\left(2^{-\Omega(\kappa^3)}+1+n\cdot 2^{-\kappa/2+3} \right)+n\cdot 2^{-\kappa+2} \\
	\leq &\, \OPT_{U,n}+2^{-\Omega(\kappa^3)}+n\cdot 2^{-\kappa/2+4}+n\cdot 2^{-\kappa+2} \\
	\leq &\, \OPT_{U,n}+U^{-1}.
\end{align*}

\paragraph{Hash functions.}
For all $x$ in the first $N_{\bl}$ block pairs, we simply use their hash values defined by $\cD_\bl$ (from Lemma~\ref{lem_no_bad_block} or Lemma~\ref{lem_with_bad_block}), for which, $h$ takes values from $[n_{\bl}]$ and $\oh$ takes values from $[V_\bl \cdot N_\bl-n_\bl]$.
For all $x$ in the last block, let $v$ be its hash value defined by $\cD_{\la}$.
If $x\in S$, let $h(x):=n_\bl+v$; if $x\notin S$, let $\oh(x):=V_\bl\cdot N_\bl-n_\bl+v$.
By definition, $h$ and $\oh$ are both bijections.

\paragraph{Lookup table.}
We include the lookup tables in Section~\ref{sec_good_block} and in Section~\ref{sec_bad_block}, which both have size $n^{\epsilon}$.
Then we include the lookup tables needed by Proposition~\ref{prop_concat} and Proposition~\ref{prop_fuse} in line~\ref{step_dict_8},~\ref{step_dict_13} and~\ref{step_dict_14}.
The total size is $n^{\epsilon}$.

\paragraph{Query algorithm.}
Given a query $x$, we decode all the components, and query the part based on the value of $x$.
\begin{code}{qAlg}[query algorithm][(x)]
	\item compute $V_\prim,V_\rand,N_\bl$ and $V$
	\item apply Proposition~\ref{prop_fuse} to recover $n'_\bl$ and decode $\cD'$ from $\cD$
	\item if $n'_\bl>n-\kappa^{3c}-1$
	\item\quad let $n_\bl:=n'_\bl$	
	\item\quad apply Proposition~\ref{prop_concat} to decode $\cD_\bl$ and $\cD_\la$
	\item else
	\item\quad recover $n_\bl$ from $\cD'$
	\item\quad decode $\cD_\bl$ and $\cD_\la$
	\item if $x\geq U-V$
	\item\quad query $x$ in $\cD_\la$, and obtain $(b, v)$
	\item\quad if $b=1$, then return $(1, n_\bl+v)$
	\item\quad if $b=0$, then return $(0, V_\bl\cdot N_\bl-n_\bl+v)$
	\item else
	\item\quad apply Proposition~\ref{prop_fuse} to recover $i$ and decode $\cD_i$ from $\cD_\bl$
	\item\quad query $x$ in $\cD_b$ using the corresponding query algorithm, and return the outcome
\end{code}
All subroutines run in constant time, the overall query time is constant.
This proves Theorem~\ref{thm_medium_set}.

\section{Data Structure Pair for Block Pair}\label{sec_block}

In this section, we prove our main technical lemma (Lemma~\ref{lem_block}), which constructs a pair of data structures for a pair of blocks.

\begin{restate}[Lemma~\ref{lem_block}]
	\contlemblock
\end{restate}

As mentioned in the overview, we first improve the rank data structure of P\v{a}tra\c{s}cu~\cite{Pat08}.
We show that if a block has $\kappa^c$ keys, then there is a rank data structure with constant query time and negligible extra space.
The $\rank$ problem asks to preprocess a set $S$ of $m$ keys into a data structure, supporting
\begin{compactitem}
	\item $\rank_S(x)$: return the number of keys that are at most $x$.
\end{compactitem}
In particular, by computing both $\rank_S(x)$ and $\rank_S(x-1)$, one can decide if $x\in S$.

\newcommand{\contlemblockrank}{
	Let $c$ be any constant positive integer and $\epsilon$ be any positive constant.
	There is a preprocessing algorithm $\rankPre$, query algorithm $\qalgr$ and lookup tables $\tblr_{V,m}$ of sizes $\tilde{O}(2^{\epsilon \kappa})$, such that for any integers $V\leq 2^{\kappa/2}, m\leq \kappa^c$, given a set $S\subset [V]$ of size $m$, $\rankPre(V,m,S)$ outputs a data structure $\cD$ of length 
	\[\lg \binom{V}{m}+ (m-1)\cdot 2^{-\kappa/2}.\]
	Given $x\in[V]$, $\qalgr(V,m,x)$ computes $\rank_S(x)$ in constant time, by accessing $\cD$ and $\tblr_{V,m}$.
	In particular, by computing both $\rank_S(x)$ and $\rank_S(x-1)$, one can decide if $x\in S$ in constant time.
	The algorithms run on a random access machine with word-size $w=\Theta(\kappa)$.
}
\begin{lemma}\label{lem_block_rank}
	\contlemblockrank
\end{lemma}
We prove this lemma in Section~\ref{sec_small_set}.
Note that a rank data structure easily defines hash functions that map the keys to $[m]$, and the non-keys to $[V-m]$: 
For each key $x$, we set $h(x):=\rank(x)-1$;
for each non-key $x$, we set $\oh(x):=x-\rank(x)-1$.
Lemma~\ref{lem_block} designs a pair of two data structure, where the size of the main data structure \emph{does not} depend on $m$ (it only assumes that $m$ is in a small range).
Most queries can be answered by accessing only the main data structure, also \emph{without} knowing the value of $m$.

To construct the two data structures, we first apply Lemma~\ref{lem_block_rank} to construct a rank data structure for the secondary block $S_\sm\subset[V_\sm]$.
Denote this data structure by $\cD_\sm$.
Then we pick a set of $\kappa^{2c-3}$ keys from $S$, as well as $V-\kappa^{2c-3}-\kappa^c$ non-keys from $[V]\setminus S$, and construct the above rank data structure, which will be the main data structure.
The remaining $\kappa^c$ elements in $[V]$ will correspond to the ``unknowns.''
We pick the two sets based on the bits in $\cD_\sm$.
That is, we apply Proposition~\ref{prop_divide} and divide $\cD_\sm$ into a string of length $\approx\lg\binom{m}{\kappa^{2c-3}}$, a string of length $\approx\lg\binom{V-m}{V-\kappa^{2c-3}-\kappa^c}$ and the remaining bits.
Then we apply the following lemma to \emph{interpret} the first string as a subset of $S$ and the second string as a subset of $[V]\setminus S$.
The final auxiliary data structure consists of the remaining bits of $\cD_\sm$, as well as a data structure for the ``unknowns.''
\newcommand{\contlemsubsetinterpret}{
	Let $c\geq 2$ be a constant positive integer.
	There is a preprocessing algorithm $\intPre$, a query algorithm $\qalgi$ and lookup tables $\tblint_{V,m}$ of sizes $O(\kappa^{c+2})$, such that for any integers $V$ and $m$ where $V\leq 2^{\kappa/2}$ and $m\leq \kappa^c$, given a (double-ended) string $\cD=(K_h,M,K_t)$ of length 
	\[
		s\leq\lg\binom{V}{m}-m(V-1)2^{-\kappa+2},
	\]
	$\intPre(V,m,\cD)$ outputs a set $S\subseteq [V]$ of size $m$.
	For any $-1\leq a_1\leq a_2\leq |M|$ and $a_2<a_1+\kappa$, $\qalgi(0,V,m,\range{K_h},|M|,\range{K_t},a_1,a_2)$ computes $\cD[a_1, a_2]$ in $O(\kappa^4)$ time using $O(\kappa^2)$ $\rank$ queries to $S$, assuming it can make random accesses to the lookup table $\tblint_{V,m}$.}
\begin{lemma}\label{lem_subset_interpret}
	\contlemsubsetinterpret
\end{lemma}
The lemma guarantees that even if the original string is not stored explicitly as part of the final data structure, it can still be accessed implicitly assuming $\rank$ queries to the sets, which the above dictionary data structures support naturally.
We prove the above lemma in Section~\ref{sec_data_interpret}.

In the remainder of the section, we show how to construct the pair of data structures in detail.

\begin{proof}[Proof of Lemma~\ref{lem_block}]
	We begin by presenting the preprocessing algorithm.
	\paragraph{Preprocessing algorithm.}
	In the preprocessing algorithm, we first construct a rank data structure for the secondary block, and divide it into three substrings.
	\begin{code}{perfHashBlk}[Preprocessing algorithm][(V,m,V_{\sm},m_{\sm},S,S_{\sm},\cR)]
		\item run $\cD_{\sm}:=\rankPre(V_{\sm},m_{\sm},S_{\sm})$ (from Lemma~\ref{lem_block_rank}) to construct a rank data structure for $S_{\sm}$ using 
		\[
			s_{\sm}\leq \lg\binom{V_{\sm}}{m_{\sm}}+(m_{\sm}-1)\cdot 2^{-\kappa/2}
		\] bits
		\item apply Proposition~\ref{prop_divide} twice to divide $\cD_{\sm}$ into
		\begin{compactitem}
			\item $\cD_{\sm,1}$: length $\leq\lg\binom{m}{m-\kappa^{2c-3}}-\kappa^c(m-1) 2^{-\kappa+2}$
			\item $\cD_{\sm,2}$: length $\leq\lg\binom{V-m}{\kappa^{2c-3}+\kappa^c-m}-\kappa^c (V-m-1)2^{-\kappa+2}$
			\item $\cD_{\sm,3}$: length $\leq s_{\sm}-\lg\binom{m}{m-\kappa^{2c-3}}-\lg\binom{V-m}{\kappa^{2c-3}+\kappa^c-m}+\kappa^c V 2^{-\kappa+2}$ \ctn
		\end{compactitem}
	\end{code}
	Note that $m-\kappa^{2c-3}\geq \kappa^c/3$ and $\kappa^{2c-3}+\kappa^c-m\geq \kappa^c/3$, therefore, both $\cD_{\sm,1}$ and $\cD_{\sm,2}$ have length at least $4\kappa$.
	For $\cD_{\sm,3}$, we have $s_{\sm}\geq \lg\binom{V_{\sm}}{m_{\sm}}\geq \kappa^{c+1}$, and thus,
	\begin{align*}
		&\, s_{\sm}-\lg\binom{m}{m-\kappa^{2c-3}}-\lg\binom{V-m}{\kappa^{2c-3}+\kappa^c-m}+\kappa^c V 2^{-\kappa+2} \\
		\geq&\, \kappa^{c+1}-(m-\kappa^{2c-3})\lg V-(\kappa^{2c-3}+\kappa^c-m)\lg V \\
		\geq&\, \kappa^{c+1}-\kappa^c\lg V \\
		\geq&\, 4\kappa.
	\end{align*}
	The premises of Proposition~\ref{prop_divide} are satisfied.
	In order to store two \emph{random} subsets in the main data structure, we ``XOR'' $\cD_{\sm,1}$ and $\cD_{\sm,2}$ with the random string $\cR$.
	\begin{code}{perfHashBlk}
		\item compute $\cD_{\sm,1} \oplus \cR$ and $\cD_{\sm,2}\oplus \cR$\ctn
	\end{code}
	For double-ended string $\cD=(K_h,M,K_t)$, $\cD\oplus \cR$ is defined as follows: compute the bitwise XOR of $M$ and $\cR[1, |M|]$, treat $\cR[|M|+1, |M|+2\kappa]$ and $\cR[|M|+2\kappa+1, |M|+4\kappa]$ as two $2\kappa$-bit integers, and compute $(K_h+\cR[|M|+1, |M|+2\kappa])\Mod \range{K_h}$ and $(K_t+\cR[|M|+2\kappa+1, |M|+4\kappa])\Mod\range{K_t}$;
	$\cD\oplus\cR$ is the double-ended string (with the same length as $\cD$), formed by the outcomes.
	In particular, since $\range{K_h}$ and $\range{K_t}$ are both smaller than $2^{\kappa+1}\ll2^{2\kappa}$, when $\cR$ is uniformly random, $\cD\oplus\cR$ is very close to uniform.
	We have the following claim by standard information theory.
	\begin{claim}\label{cl_entropy_xor}
	For any fixed $\cD=(K_h,M,K_t)$ and uniformly random $\cR$, we have
	\begin{align*}
		H(\cD\oplus \cR)&\geq (\lg(\range{K_h})+|M|+\lg(\range{K_t}))(1-2^{-\kappa+2}) \\
		&\geq (|\cD|-2^{-\kappa+2})(1-2^{-\kappa+2}).
	\end{align*}
	\end{claim}
	Also, $K_h, K_t$ and any $O(\kappa)$ consecutive bits of $M$ can be computed in constant time, given random access to $\cD\oplus\cR$ and $\cR$.
	Next, we interpret the $\cD_{\sm,1}\oplus \cR$ and $\cD_{\sm,2}\oplus\cR$ as two subsets using Lemma~\ref{lem_subset_interpret}.
	\begin{code}{perfHashBlk}
		\item run $S_1:=\intPre(m,m-\kappa^{2c-3},\cD_{\sm,1}\oplus\cR)$ (from Lemma~\ref{lem_subset_interpret}) to interpret $\cD_{\sm,1}\oplus\cR$ as a set $S_1\subseteq [m]$ of size $m-\kappa^{2c-3}$\\
		run $S_2:=\intPre(V-m,\kappa^{2c-3}+\kappa^c-m,\cD_{\sm,2}\oplus\cR)$ to interpret $\cD_{\sm,2}\oplus\cR$ as a set $S_2\subseteq [V-m]$ of size $\kappa^{2c-3}+\kappa^c-m$
		\item compute $S_{\unk}\subseteq S$ according to $S_1$ \\
		compute $\overline{S}_{\unk}\subseteq [V]\setminus S$ according to $S_2$\ctn
	\end{code}
	More specifically, for each $i\in[m]$, $S_{\unk}$ contains the $(i+1)$-th smallest element in $S$ if and only if $i\in S_1$.
	Similarly, $\overline{S}_{\unk}$ contains the $(i+1)$-th smallest element in $[V]\setminus S$ if and only if $i\in S_2$.
	They are the keys and non-keys that are \emph{not} to be stored in the main data structure, i.e., the ``unknowns.''

	Then, we compute the main data structure $\cD_{\main}$.
	\begin{code}{perfHashBlk}
		\item\label{step_concat_main} apply Proposition~\ref{prop_concat} to concatenate the following two data structures: and obtain $\cD_\main$:
		\begin{compactitem}
			\item $\cD_{\main,1}:=\rankPre(V,\kappa^{2c-3},S\setminus S_{\unk})$ (from Lemma~\ref{lem_block_rank}), a rank data structure
			\item $\cD_{\main,2}:=\rankPre(V-\kappa^{2c-3},\kappa^c,$``$S_{\unk}\cup \overline{S}_{\unk}$''$)$ a rank data structure for $S_{\unk}\cup \overline{S}_{\unk}$ over $[V]\setminus (S\setminus S_{\unk})$ (see below) \ctn
		\end{compactitem}
	\end{code}

	For $\cD_{\main,2}$, before running $\rankPre$, we first remove all $\kappa^{2c-3}$ elements in $S\setminus S_{\unk}$ from both the universe $[V]$ and $S_{\unk}\cup\overline{S}_{\unk}$, and keep the order of the remaining elements.
	Thus, the new universe becomes $[V-\kappa^{2c-3}]$.
	In the other words, $\cD_{\main,2}$ supports queries of form ``return \# of elements in $S_{\unk}\cup\overline{S}_{\unk}$ that are no larger than $i$-th smallest element in $[V]\setminus (S\setminus S_{\unk})$''.


	Finally, we compute the auxiliary data structure $\cD_{\aux}$.
	\begin{code}{perfHashBlk}
		\item\label{step_concat_aux} apply Proposition~\ref{prop_concat} to concatenate the following two data structures and obtain $\cD_\aux$:
		\begin{compactitem}
			\item $\cD_{\aux,1}:=\rankPre(\kappa^c,m-\kappa^{2c-3},$``$S_{\unk}$''$)$, a rank data structure for $S_{\unk}$ over $S_{\unk}\cup \overline{S}_{\unk}$
			\item $\cD_{\aux,2}:=\cD_{\sm,3}$
		\end{compactitem}
	\end{code}

	Similarly, $\cD_{\aux,1}$ supports queries of form ``return \# of elements in $S_{\unk}$ that are no larger than the $i$-th smallest element in $S_{\unk}\cup \overline{S}_{\unk}$.''

	\paragraph{Space analysis.}
	Next, we analyze the length of $\cD_{\main}$ and $\cD_{\aux}$.
	$\cD_{\main}$ is the concatenation of $\cD_{\main,1}$ and $\cD_{\main,2}$.
	For $\cD_{\main,1}$, its length is at most
	\[
		\lg\binom{V}{\kappa^{2c-3}}+(\kappa^{2c-3}-1)2^{-\kappa/2}
	\]
	by Lemma~\ref{lem_block_rank}.
	For $\cD_{\main,2}$, its length is at most
	\[
		\lg\binom{V-\kappa^{2c-3}}{\kappa^c}+(\kappa^c-1)2^{-\kappa/2}.
	\]
	By Proposition~\ref{prop_concat}, the length of $\cD_{\main}$ is at most
	\begin{align*}
		&\, \lg\binom{V}{\kappa^{2c-3}}+(\kappa^{2c-3}-1)2^{-\kappa/2}+\lg\binom{V-\kappa^{2c-3}}{\kappa^c}+(\kappa^c-1)2^{-\kappa/2}+2^{-\kappa+4} \\
		\leq &\, \lg\binom{V}{\kappa^{2c-3},\kappa^c}+\kappa^{2c-3}2^{-\kappa/2+1}.
	\end{align*}

	$\cD_{\aux}$ is the concatenation of $\cD_{\aux,1}$ and $\cD_{\aux,2}$.
	For $\cD_{\aux,1}$, its length is at most
	\[
		\lg\binom{\kappa^c}{m-\kappa^{2c-3}}+(m-\kappa^{2c-3}-1)2^{-\kappa/2}.
	\]
	For $\cD_{\aux,2}$, which is $\cD_{\sm,3}$, its length is at most
	\begin{align*}
		&\, s_{\sm}-\lg\binom{m}{m-\kappa^{2c-3}}-\lg\binom{V-m}{\kappa^{2c-3}+\kappa^c-m}+\kappa^cV2^{-\kappa+2} \\
		\leq&\, \lg\binom{V_{\sm}}{m_{\sm}}-\lg\binom{m}{m-\kappa^{2c-3}}-\lg\binom{V-m}{\kappa^{2c-3}+\kappa^c-m}+3\kappa^{c+1}2^{-\kappa/2},
	\end{align*}
	since $V\leq 2^{\kappa/2}$ and $m_{\sm}\leq 3\kappa^{c+1}$.
	Summing up the lengths and by Proposition~\ref{prop_concat}, the length of $\cD_{\aux}$ is at most
	\begin{align*}
		&\, \lg\binom{\kappa^c}{m-\kappa^{2c-3}}+(m-\kappa^{2c-3}-1)2^{-\kappa/2}+\lg\binom{V_{\sm}}{m_{\sm}}\\
		&\quad -\lg\binom{m}{m-\kappa^{2c-3}}-\lg\binom{V-m}{\kappa^{2c-3}+\kappa^c-m}+3\kappa^{c+1}2^{-\kappa/2}+2^{-\kappa+4}\\
		\leq&\, \lg\frac{\kappa^c!\kappa^{2c-3}!(V-\kappa^{2c-3}-\kappa^c)!}{m!(V-m)!}+\lg\binom{V_{\sm}}{m_{\sm}}+(m-\kappa^{2c-3}+3\kappa^{c+1})2^{-\kappa/2} \\
		\leq&\, \lg\binom{V}{m}+\lg\binom{V_{\sm}}{m_{\sm}}-\lg\binom{V}{\kappa^{2c-3},\kappa^c}+\kappa^{c+1}2^{-\kappa/2+2},
	\end{align*}
	as we claimed.
	This proves item (i) and (ii) in the statement.

	\paragraph{Hash functions.}
	For $x\in S\cup S_{\sm}$, we define $h(x)$ as follows.
	\begin{compactitem}
		\item For $x\in S\setminus S_{\unk}$, let $h(x):=\rank_{S\setminus S_{\unk}}(x)-1$; they are mapped to $[\kappa^{2c-3}]$.
		\item For $x\in S_{\unk}$, let $h(x):=\kappa^{2c-3}+\rank_{S_{\unk}}(x)-1$; they are mapped to $\{\kappa^{2c-3},\ldots,m-1\}$.
		\item For $x\in S_{\sm}$, let $h(x):=m+\rank_{S_{\sm}}(x)-1$; they are mapped to $\{m,\ldots,m+m_{\sm}-1\}$.
	\end{compactitem}
	Similarly, for $x\notin S\cup S_{\sm}$, we define $\oh$ as follows.
	\begin{compactitem}
		\item For $x\in ([V]\setminus S)\setminus \overline{S}_{\unk}$, let $\oh(x):=\rank_{[V]\setminus S\setminus \overline{S}_{\unk}}(x)-1$; they are mapped to $[V-\kappa^{2c-3}-\kappa^c]$.
		\item For $x\in \overline{S}_{\unk}$, let $\oh(x):=V-\kappa^{2c-3}-\kappa^c+\rank_{\overline{S}_\unk}(x)-1$; they are mapped to $\{V-\kappa^{2c-3}-\kappa^c,\ldots,V-m-1\}$.
		\item For $x\in \{V,\ldots,V+V_{\sm}-1\}\setminus S_{\sm}$, let $\oh(x):=V-m+\rank_{\{V,\ldots,V+V_{\sm}-1\}\setminus S_{\sm}}(x)-1$; they are mapped to $\{V-m,\ldots,V+V_{\sm}-m-m_{\sm}-1\}$.
	\end{compactitem}
	Overall, $h$ is a bijection between $S\cup S_{\sm}$ and $[m+m_{\sm}]$, and $\oh$ is a bijection between $[V+V_{\sm}]\setminus (S\cup S_{\sm})$ and $[V+V_{\sm}-m-m_{\sm}]$.
	Moreover, $h(S)\supset [\kappa^{2c-3}]$ and $\oh([V]\setminus S)\supset [V-\kappa^{2c-3}-\kappa^c]$.
	This proves item (iii) in the statement.

	\paragraph{Lookup table.}
	We store the following information in the lookup table.
	\begin{code}{tableBlk}[lookup table][][tableBlk_{\mathnormal{V,V_{\sm}}}]
		\item lookup table for line~\ref{step_concat_main} from Proposition~\ref{prop_concat}
		\item $\tblr_{V,\kappa^{2c-3}}$, $\tblr_{V-\kappa^{2c-3},\kappa^c}$ from Lemma~\ref{lem_block_rank}
		\item for \emph{all} $m\in [\kappa^{2c-3}+\kappa^c/3,\kappa^{2c-3}+2\kappa^c/3]$ and $m_{\sm}\in[\kappa^{c+1},2\kappa^{c+1}]$
		\begin{compactitem}
			\item lookup tables for line~\ref{step_concat_aux} from Proposition~\ref{prop_concat}
			\item $\tblr_{\kappa^c,m-\kappa^{2c-3}}$ and $\tblr_{V_{\sm},m_{\sm}}$ from Lemma~\ref{lem_block_rank}
			\item $\tblint_{m,m-\kappa^{2c-3}}$ and $\tblint_{V-m,\kappa^{2c-3}+\kappa^c-m}$ from Lemma~\ref{lem_subset_interpret}
		\end{compactitem}
	\end{code}

	Each $\tblr$ has size $\tilde{O}(2^{\epsilon \kappa})$ and each $\tblint$ has size $O(\kappa^{c+2})$.
	The total size of $\tblB_{V,V_{\sm}}$ is $\tilde{O}(2^{\epsilon \kappa})$.

	\paragraph{The main query algorithm.}
	We show how to answer each query $x$ in constant time with high probability, by querying only the main data structure (and without knowing $m$).
	We begin by decoding the two data structures $\cD_{\main,1}$ and $\cD_{\main,2}$ from $\cD_{\main}$, and query $\cD_{\main,1}$.
	\begin{code}{qalgBM}[query algorithm][(V,x)][\qalgB_{\main}]
		\item decode $\cD_{\main,1}$ and $\cD_{\main,2}$ from $\cD_{\main}$ using Proposition~\ref{prop_concat}
		\item $x_r:=\cD_{\main,1}.\qalgr(V,\kappa^{2c-3},x)$ (from Lemma~\ref{lem_block_rank})
		\item\label{step_x_known} if $x_r>\cD_{\main.1}.\qalgr(V,\kappa^{2c-3},x-1)$
		\item\quad return $(1, x_r-1)$ \ctn
	\end{code}
	$x_r$ is the number of elements in $S\setminus S_{\unk}$ that are at most $x$.
	Line~\ref{step_x_known} checks if $x\in S\setminus S_{\unk}$.
	If it is, then $x$ is the $x_r$-th element in $S\setminus S_{\unk}$, and we return its hash value according by the definition of $h$.

	\begin{code}{qalgBM}
		\item $x_{\unk}:=\cD_{\main,2}.\qalgr(V-\kappa^{2c-3},\kappa^c,x-x_r)$
		\item\label{step_x_unk} if $x_{\unk}>\cD_{\main,2}.\qalgr(V-\kappa^{2c-3},\kappa^c,x-x_r-1)$
		\item\quad return ``unknown''
		\item return $(0, x-x_r-x_{\unk})$
	\end{code}
	If $x\notin S\setminus S_{\unk}$, we query $\cD_{\main,2}$ to check if $x\in S_{\unk}\cup\overline{S}_{\unk}$ in line~\ref{step_x_unk}.
	Note that $x$ is the $(x-x_r+1)$-th element in $[V]\setminus (S\setminus S_{\unk})$.
	If it is, we return ``unknown''.
	Otherwise, we know that $x\notin S$, and it is the $(x-x_r-x_\unk+1)$-th element in $[V]\setminus (S\cup \overline{S}_\unk)$, we return its $\oh$-value.
	Since $\qalgr$ has constant query time, $\qalgB$ also runs in constant time.
	Clearly, $\qalgB$ outputs $\hash(x)$ when $x\in S$ and $h(x)\in [\kappa^{2c-3}]$, or $x\notin S$ and $\oh(x)\in [V-\kappa^{2c-3}-\kappa^c]$, and otherwise it outputs ``unknown''.
	This proves item (iv) in the statement.

	\bigskip

	Next, we show that the probability that it outputs ``unknown'' is small.
	To this end, let us fix the input data $S,S_{\sm}$ and query $x$, and let $\cR$ be uniformly random.
	We will show that $S_{\unk}$ is close to a uniformly random subset of $S$ of size $m-\kappa^{2c-3}$, and $\overline{S}_{\unk}$ is close to a uniformly random subset of $[V]\setminus S$ of size $\kappa^{2c-3}+\kappa^c-m$.
	By Claim~\ref{cl_entropy_xor}, we have $H(\cD_{\sm,1}\oplus \cR)\geq (|\cD_{\sm,1}|-2^{-\kappa+2})(1-2^{-\kappa+2})$.
	Since the division operation in Proposition~\ref{prop_divide} is an injection, we have
	\begin{align*}
		|\cD_{\sm,1}|&\geq s_{\sm}-|\cD_{\sm,2}|-|\cD_{\sm,3}| \\
		&\geq \lg\binom{m}{m-\kappa^{2c-3}}-\kappa^cV2^{-\kappa+3}.
	\end{align*}
	Therefore, $H(\cD_{\sm,1}\oplus\cR)\geq \lg\binom{m}{m-\kappa^{2c-3}}-\kappa^c2^{-\kappa/2+4}$.
	Furthermore, since $\intPre(m,m-\kappa^{2c-3},\cdot)$ is an injection, we have $H(S_1)\geq \lg\binom{m}{m-\kappa^{2c-3}}-\kappa^c2^{-\kappa/2+4}$, which in turn implies that 
	\[
		H(S_{\unk})\geq \lg\binom{m}{m-\kappa^{2c-3}}-\kappa^c2^{-\kappa/2+4},
	\]
	for any fixed $S$ and $S_{\sm}$.
	By Pinsker's inequality, the $\ell_1$ distance between $S_{\unk}$ and a uniformly random subset of $S$ of size $m-\kappa^{2c-3}$ is at most $O(\kappa^{c/2}2^{-\kappa/4})$.
	In particular, it implies that for any fixed $x\in S$, the probability that $x\in S_{\unk}$ is at most
	\[
		\frac{m-\kappa^{2c-3}}{m}+O(\kappa^{c/2}2^{-\kappa/4})\leq O(\kappa^{-c+3}).
	\]

	By applying the same argument to $\cD_{\sm,2}$, $S_2$ and $\overline{S}_{\unk}$, we conclude that for any fixed $x\notin S$, the probability that $x\in \overline{S}_{\unk}$ is at most
	\[
		\frac{\kappa^{2c-3}+\kappa^c-m}{V-m}+O(\kappa^{c/2}2^{-\kappa/4})\leq O(\kappa^{-c+3}).
	\]
	This proves item (v) in the statement.

	\paragraph{The general query algorithm.}
	Finally, we describe the query algorithm for all $x\in{[V+V_{\sm}]}$.
	We use two different algorithms for $x\in [V]$ and $x\in \{V,\ldots,V+V_\sm-1\}$.
	We begin by the $x\in [V]$ case ($x$ is in the primary block).
	\begin{code}{qalgBlk}[query algorithm][(V,m,V_{\sm},m_{\sm},x)]
		\item (if $x<V$)
		\item $(b,v):=\qalgB_{\main}(V,x)$
		\item if $(b,v)$ is not ``unknown''
		\item\quad return $(b,v)$
		\item let $x_{\unk}:=\rank_{S_\unk\cup \overline{S}_{\unk}}(x)$ (already computed in $\qalgB_{\main}(V,x)$) \ctn
	\end{code}
	When $\qalgB_{\main}$ returns ``unknown'', $x$ is the $x_{\unk}$-th element in $S_{\unk}\cup \overline{S}_{\unk}$.
	Next, we query $\cD_{\aux,1}$ to find out whether $x\in S_{\unk}$ or $x\in \overline{S}_{\unk}$ and its rank in the corresponding set.
	Then we return its $h$ or $\oh$ value according to the definition.
	\begin{code}{qalgBlk}
		\item apply Proposition~\ref{prop_concat} on $\cD_{\aux}$ to decode $\cD_{\aux,1}$
		\item $x_{\unk,r}:=\cD_{\aux,1}.\qalgr(\kappa^c,m-\kappa^{2c-3},x_{\unk}-1)$
		\item if $x_{\unk,r}>\cD_{\aux,1}.\qalgr(\kappa^c,m-\kappa^{2c-3},x_{\unk}-2)$
		\item\quad return $\kappa^{2c-3}+x_{\unk,r}-1$
		\item else
		\item\quad return $V-\kappa^{2c-3}-\kappa^c+(x_\unk-x_{\unk,r})-1$
	\end{code}
	Similarly to $\qalgB_{\main}$, we check if the $x_{\unk}$-th element is in $S_{\unk}$.
	if it is, then it is the $x_{\unk,r}$-th element in $S_{\unk}$.
	Otherwise, it is the $(x_{\unk}-x_{\unk,r})$-th element in $\overline{S}_{\unk}$.
	In this case ($x\in [V]$), the query algorithm runs in constant time.

	Next, we show how to handle $x\in\{V,\ldots,V+V_\sm-1\}$.
	To this end, let us first assume that we can make random access to $\cD_{\sm}$.
	\begin{code}{qalgBlk}
		\item (if $x\geq V$)
		\item apply Proposition~\ref{prop_concat} on $\cD_{\aux}$ to decode $\cD_{\aux,2}$
		\item $x_r:=\cD_{\sm}.\qalgr(V_{\sm},m_{\sm},x-V)$ (from Lemma~\ref{lem_block_rank})
		\item if $x_r>\cD_{\sm}.\qalgr(V_{\sm},m_{\sm},x-V-1)$
		\item\quad return $m+x_r-1$
		\item else
		\item\quad return $V-m+(x-V-x_r)$
	\end{code}
	If we had access to $\cD_{\sm}$, then the query algorithm would be similar to the previous cases, and it runs in constant time.
	However, $\cD_{\sm}$ is not stored in the data structure explicitly.
	In the following, we show how $\qalgB$ accesses $\cD_{\sm}$ from its implicit representation.

	More specifically, $\qalgB$ only needs to access $\cD_{\sm}$ when it runs the query algorithm $\qalgr$ on $\cD_{\sm}$.
	By Lemma~\ref{lem_block_rank}, $\qalgr$ runs on a RAM with word-size $\Theta(\kappa)$, i.e., it may request $\Theta(\kappa)$ consecutive bits of the data structure $\cD_\sm$ during its runtime.
	To implement such access requests, we first apply Proposition~\ref{prop_divide} to reduce each access to $O(1)$ accesses to $\cD_{\sm,1}$, $\cD_{\sm,2}$ and $\cD_{\sm,3}$.
	$\cD_{\sm,3}$ is stored as $\cD_{\aux,2}$, which has been decoded.
	Each access to it can be implemented in constant time.
	For $\cD_{\sm,1}$, $\cD_{\sm,1}\oplus \cR$ is interpreted as a set $S_1\subseteq [m]$ of size $m-\kappa^{2c-3}$.
	Lemma~\ref{lem_subset_interpret} guarantees that each access to $\cD_{\sm,1}\oplus \cR$ can be implemented in $O(\kappa^4)$ time and $O(\kappa^2)$ $\rank$ queries to $S_1$, which by the previous argument, implies that each access to $\cD_{\sm,1}$ can also be implemented in the same time and number of $\rank$ queries.

	On the other hand, the way the preprocessing algorithm ``encodes'' $S_1$ guarantees that $\rank_{S_1}(k)$ queries can be implemented efficiently.
	To see this, recall that $S_{\unk}\subset S$ is determined according to $S_1$.
	$\rank_{S_1}(k)$ is exactly the number of elements in $S_{\unk}$ that are no larger than the $(k+1)$-th smallest element in $S$.
	We first do a binary search to find the $(k+1)$-th smallest element in $S$.
	\begin{code}{rankS1}[implementing $\rank$ queries on $S_1$][(k)][rank\_S_1]
		\item decode $\cD_{\main,1},\cD_{\main,2}$ and $\cD_{\aux,1}$
		\item binary search for $(k+1)$-th element $x^*$ in $S$:
		given $x\in [V]$,
		\begin{compactenum}[(i)]
			\item $x_r:=\cD_{\main,1}.\qalgr(V,\kappa^{2c-3},x)$
			\item $x_{\unk}:=\cD_{\main,2}.\qalgr(V-\kappa^{2c-3},\kappa^c,x-x_r)$
			\item $\rank_S(x):=x_r+\cD_{\aux,1}.\qalgr(\kappa^c,m-\kappa^{2c-3},x_{\unk}-1)$
		\end{compactenum}
	\end{code}
	$x_r$ is the number of elements in $S\setminus S_{\unk}$ that are at most $x$.
	$x_{\unk}$ is the number of elements in $S_{\unk}\cup\overline{S}_{\unk}$ that are at most $x$.
	$\cD_{\aux,1}.\qalgr(\kappa^c,m-\kappa^{2c-3},x_{\unk}-1)$ computes the number of elements in $S_{\unk}$ that are at most $x$.
	By summing up $x_r$ and $\cD_{\aux,1}.\qalgr(\kappa^c,m-\kappa^{2c-3},x_{\unk})$, we compute $\rank_S(x)$, the number of elements in $S$ that are at most $x$, in constant time.
	Being able to compute $\rank_S(x)$ for any given $x$ allows us to binary search for the $(k+1)$-th smallest element $x^*$ in $S$ in $O(\lg V)=O(\kappa)$ time, which then allows us to compute $\rank_{S_1}(k)$.
	\begin{code}{rankS1}
		\item $x^*_r:=\cD_{\main,1}.\qalgr(V,\kappa^{2c-3},x^*)$
		\item return $\rank_{S_1}(k):=k-x^*_r+1$
	\end{code}
	This shows that $\rank_{S_1}(k)$ can be computed in $O(\kappa)$ time, and thus, each access to $\cD_{\sm,1}$ can be implemented in $O(\kappa^4+\kappa\cdot\kappa^2)=O(\kappa^4)$ time.

	Similarly, each access to $\cD_{\sm,2}$ can be implemented in $O(\kappa^4)$ time:
	Lemma~\ref{lem_subset_interpret} reduces it to $O(\kappa^2)$ $\rank$ queries to $S_2$ and $O(\kappa^4)$ processing time;
	For $\rank_{S_2}(k)$, we do binary search to find the $(k+1)$-th element in $[V]\setminus S$;
	By querying $\cD_{\main,1}$, $\cD_{\main,2}$ and $\cD_{\aux,1}$, we compute $\rank_{S_2}(k)$.

	Overall, the above algorithms allow us to access $\cD_{\sm,1}$, $\cD_{\sm,2}$ and $\cD_{\sm,3}$ in $O(\kappa^4)$ time, which in turn, allows us to access $\cD_{\sm}$ in $O(\kappa^4)$.
	Thus, $\qalgB$ runs in $O(\kappa^4)$ time.
	This proves item (vi) in the statement.
\end{proof}

\subsection{Data interpretation}\label{sec_data_interpret}
In this subsection, we prove Lemma~\ref{lem_subset_interpret}, showing how to represent a string as a set which allows us to locally decode the string given access to a $\rank$ oracle of the resulting set.

\begin{restate}[Lemma~\ref{lem_subset_interpret}]
	\contlemsubsetinterpret
\end{restate}
To construct set $S$ from the input string $\cD$, the main idea is to apply Proposition~\ref{prop_sep} and~\ref{prop_divide}, and then recurse on the two halves of $[V]$.
We extract an integer $m_1\in[m+1]$ from $\cD$ using Proposition~\ref{prop_sep}, which encodes the number of elements in the first half of $[V]$.
Then we divide $\cD$ into two data structure $\cD_1$ and $\cD_2$ such that the length of $\cD_1$ is approximately $\lg\binom{V/2}{m_1}$ and the length of $\cD_2$ is approximately $\lg\binom{V/2}{m-m_1}$.
Then the set $S$ is recursively constructed in the two halves.
When the $m$ gets sufficiently small and the $\cD$ has length $O(\kappa^2)$, we continue the recursion without applying the two propositions about fraction-length strings.
Instead, we take the whole string as an integer smaller than $2^{O(\kappa^2)}$, and use the integer to encode a set (also decode the whole integer at the decoding time).
See below for the formal argument.

\paragraph{Encoding and decoding an integer using a set.}
We first show that given an integer $Z\leq \binom{V}{m}$, how to turn it into a set of size $m$ in $[V]$, such that $Z$ can be recovered using $\rank$ queries.
\begin{code}{encSet}[encoding algorithm][(V,m,Z)]
	\item if $m=0$, return $\emptyset$
	\item if $m=V$, return $[V]$
	\item $V_1:=\lfloor V/2\rfloor$ and $V_2:=\lceil V/2\rceil$
	\item compute the largest $0\leq j\leq m$ such that $Z\geq \sum_{i=0}^{j-1}\binom{V_1}{i}\binom{V_2}{m-i}$
	\item $Y:=Z-\sum_{i=0}^{j-1}\binom{V_1}{i}\binom{V_2}{m-i}$
	\item $Z_1:= Y\Div\binom{V_2}{m-j}$ and $Z_2:=Y\Mod \binom{V_2}{m-j}$
	\item return $\encset(V_1,j,Z_1)\cup (\encset(V_2,m-j,Z_2)+V_1)$
\end{code}
To construct the set, the algorithm is a standard recursion.
All possible sets are listed in the increasing order of the number of elements in the $[V_1]$.
We compute this number, and then recurse into the two halves.
$Z$ can be recovered by the following algorithm, assuming the set generated is in the universe $[X,X+V)$.
For technical reasons that we will encounter later, sometimes we may only have access to the \emph{complement} of the set.
The bit $b$ indicates whether we should take the complement.
\begin{code}{decSet}[decoding algorithm][(X,V,m,b)]
	\item if $m=0$ or $m=V$, return $0$
	\item $V_1:=\lfloor V/2\rfloor$ and $V_2:=\lceil V/2\rceil$
	\item $j:=\rank_S(X+V_1-1)-\rank_S(X-1)$
	\item if $b$, then $j:=V_1-j$
	\item $Z_1:=\decset(X,V_1,j,b)$ and $Z_2:=\decset(X+V_1,V_2,m-j,b)$
	\item return $Z:=\sum_{i=0}^{j-1}\binom{V_1}{i}\binom{V_2}{m-i}+Z_1\cdot \binom{V_2}{m-j}+Z_2$
\end{code}
We will store the sum $\sum_{i=0}^{j-1}\binom{V_1}{i}\binom{V_2}{m-i}$ and the binomial coefficient $\binom{V_2}{m-j}$ in the lookup table.
Since the recursion terminates when $m=0$ and the value of $V$ decreases by a factor of two each time, the size of the recursion tree is $O(m\lg V)$.
Thus, we have the following claim.
\begin{claim}\label{cl_decset}
$\decset$ uses $O(m\lg V)$ arithmetic operations on $O(\lg\binom{V}{m})$-bit integers, as well as $O(m\lg V)$ $\rank$ queries.
\end{claim}

\paragraph{Preprocessing into a set.}
Given a string $\cD=(K_h,M,K_t)$ of length at most $\lg\binom{V}{m}-m(V-1)2^{-\kappa+2}$, we preprocess it into a set $S\subseteq [V]$ of size $m$.

\begin{code}{prepIntoSet}[preprocessing algorithm][(V, m, \cD=(K_h,M,K_t))]
	\item if $2m>V$
	\item\quad return $[V]\setminus\intPre(V, V-m, \cD)$
	\item if $m\leq 24\kappa$
	\item\quad rewrite $\cD$ as an nonnegative integer $Z<\range{K_h}\cdot\range{K_t}\cdot 2^{|M|}$ 
	\item\quad return $\encset(V,m,Z)$
	\item if $|\cD|\leq 24\kappa$
	\item\quad return $\intPre(48\kappa,24\kappa,\cD)\cup \{V-(m-24\kappa),\ldots,V-1\}$ \ctn
\end{code}
If $m$ is larger than $V/2$, we simply work on the complement.
If $m$ is $O(\kappa)$, we view the entire string $\cD$ as an integer, and call $\encset$.
If the string is too short while $m$ (and $V$) are still large, we manually decrease $m$ and $V$, and reduce it to the $m=O(\kappa)$ case.
Note that $\binom{V}{m}$ may be at most $2^{O(\kappa^2)}$, $Z$ occupies $O(\kappa)$ words (as $\kappa=\Theta(w)$).

Otherwise, we extract an integer $j$ from $\cD$.
\begin{code}{prepIntoSet}
	\item $V_1:=\lfloor V/2\rfloor$ and $V_2:=\lceil V/2\rceil$
	\item compute $s_j:=\lg\binom{V_1}{j}+\lg\binom{V_2}{m-j}-m(V-2)2^{-\kappa+2}$
	\item\label{step_prepintoset_sep} apply Proposition~\ref{prop_sep} for $j\in\{\lfloor m/3\rfloor+1,\ldots,2\lfloor m/3\rfloor\}$ and $C=\lfloor m/3\rfloor$, encode $\cD$ using a pair $(j, \cD_j)$ such that $\cD_j$ has length at most $s_j$
	\item let $(S_1,S_2):=\intPreT(V_1,V_2,j,m-j,\cD_j)$
	\item return $S_1\cup(S_2+V_1)$
\end{code}
$\intPreT$ preprocesses $\cD_j$ into two sets of sizes $j$ and $m-j$ over the two halves of the universe (see below).
Then we return their union.
Proposition~\ref{prop_sep} requires that the length of $\cD$ is at least $3\kappa+2$ and at most $\lg(\sum_j 2^{s_j})-(C-1)2^{-\kappa+2}$.
This is true, because on one hand, the length of $\cD$ is at least $24\kappa$; on the other hand,
\begin{align*}
	2^{s_1}+\cdots+2^{s_C}&=\sum_{j=\lfloor m/3\rfloor+1}^{2\lfloor m/3\rfloor}\binom{V_1}{j}\cdot \binom{V_2}{m-j}\cdot 2^{-m(V-2)2^{-\kappa+2}}\\
	&\geq 2^{-m(V-2)2^{-\kappa+2}}\cdot \left(\binom{V}{m}-\frac{2m}{3}\binom{V_1}{\lfloor m/3\rfloor}\binom{V_2}{\lceil 2m/3\rceil}\right) \\
	&= 2^{-m(V-2)2^{-\kappa+2}}\cdot \left(\binom{V}{m}-\frac{2m}{3}\binom{V_1}{\lfloor m/2\rfloor}\binom{V_2}{\lceil m/2\rceil}\cdot\prod_{j=\lfloor m/3\rfloor+1}^{\lfloor m/2\rfloor}\frac{j(V_2-m+j)}{(V_1-j+1)(m-j+1)} \right) \\
	&\geq 2^{-m(V-2)2^{-\kappa+2}}\cdot \binom{V}{m}\cdot\left(1-\frac{2m}{3}\cdot\prod_{j=\lfloor m/3\rfloor+1}^{\lfloor m/2\rfloor}\frac{j}{m-j+1} \right) \\
	&\geq 2^{-m(V-2)2^{-\kappa+2}}\cdot \binom{V}{m}\cdot\left(1-\frac{2m}{3}\cdot e^{-\sum_{j=\lfloor m/3\rfloor+1}^{\lfloor m/2\rfloor}\frac{m-2j+1}{m-j+1}} \right) \\
	&\geq 2^{-m(V-2)2^{-\kappa+2}}\cdot \binom{V}{m}\cdot\left(1-\frac{2m}{3}\cdot e^{-m/24} \right).
\end{align*}
Therefore, by the fact that $m\geq 24\kappa$, we have
\begin{align*}
	\lg(2^{s_1}+\cdots+2^{s_C})&\geq \lg\binom{V}{m}-m(V-2)2^{-\kappa+2}-m 2^{-\kappa}\\
	&\geq s+m2^{-\kappa+2}-m2^{-\kappa} \\
	&\geq s+(C-1)\cdot 2^{-\kappa+2}.
\end{align*}
Thus, the premises of Proposition~\ref{prop_sep} are also satisfied.
Next, we describe $\intPreT$.
\begin{code}{prepIntoTwoSet}[preprocessing algorithm][(V_1,V_2, m_1, m_2, \cD)][\intPreT]
	\item let $s_1:=\lg\binom{V_1}{m_1}-m_1(V_1-1)2^{-\kappa+2}$ and $s_2:=\lg\binom{V_2}{m_2}-m_2(V_2-1)2^{-\kappa+2}$
	\item\label{step_preintosettwo_divide} apply Proposition~\ref{prop_divide}, divide $\cD$ into $\cD_1$ and $\cD_2$ of lengths at most $s_1$ and $s_2$ respectively
	\item let $S_1:=\intPre(V_1,m_1,\cD_1)$ and $S_2:=\intPre(V_2,m_2,\cD_2)$
	\item return $(S_1,S_2)$
\end{code}
Proposition~\ref{prop_divide} requires that the length of $\cD$ is at most $s_1+s_2-2^{-\kappa+2}$ (and at least $3\kappa$), and $s_1,s_2\geq 3\kappa$.
It is easy to verifier the former.
For the latter, because $m_1+m_2\leq V/2$, $m_2/2\leq m_1\leq 2m_2$ and $m_1+m_2>24\kappa$, and in particular, we have $V_1\geq 24\kappa$ and $m_1\in [V_1/3,2V_1/3]$.
Hence, we have
\[
	\binom{V_1}{m_1}\geq 3^{8\kappa},
\]
and it implies $s_1\geq 8\kappa$.
Similarly, we also have $s_2\geq 8\kappa$.

\paragraph{Lookup table.}
We store the following lookup table.
\begin{code}{tableInt}[lookup table][][tableInt_{\mathnormal{V,m}}]
	\item if $m\leq 24\kappa$
	\item\quad $\sum_{i=0}^{j-1}\binom{V_1}{i}\binom{V_2}{m-i}$ for $j=0,\ldots,m$
	\item\quad $\binom{V_2}{j}$ for all $j=0,\ldots,m$
	\item else
	\item\quad lookup table from Proposition~\ref{prop_sep} for line~\ref{step_prepintoset_sep} of $\intPre$
	\item include all tables $\tblint_{V',m'}$ for $V'=\lfloor V/2^i\rfloor$ or $V'=\lceil V/2^i\rceil$ for $i\geq 1$, and $0\leq m'\leq m$
\end{code}
The lookup table $\tblint_{V,m}$ itself has size at most $O(\kappa)$ words for $m>24\kappa$ and $O(\kappa^2)$ words for $m\leq 24\kappa$.
Including the smaller tables, its total size is at most $O(\kappa^2m+\kappa^4)\leq O(\kappa^{c+2})$ words for $m\leq \kappa^c$ and $c\geq 2$.

\paragraph{Access the string.}
Suppose $S$ is the set generated from a string $\cD$ using the above preprocessing algorithm.
In the following, we show how to access $\cD[a_1, a_2]$ for $a_2-a_1<\kappa$, assuming $\rank$ queries can be computed efficiently on $S$.
Assuming the set $S$ restricted to $[X, X+V)$ (with $m$ elements in this range) is generated from a string $\cD=(K_h,M,K_t)$, $\qalgi(X, V, m,\range{K_h}, |M|, \range{K_t}, a_1, a_2, b)$ recovers $\cD[a_1, a_2]$, where $b$ indicates if we take the complement of $S$.
\begin{code}{access}[accessing algorithm][(X,V,m,\range{K_h},|M|,\range{K_t},a_1,a_2, b)][\qalgi]
	\item if $2m>V$
	\item\quad $m:=V-m$ and $b:=\neg b$
	\item if $m\leq 24\kappa$
	\item\quad $Z:=\decset(x,V,m,b)$
	\item\quad rewrite $Z$ as a string $\cD=(K_h,M,K_t)$
	\item\quad return $\cD[a_1, a_2]$
	\item if $\lg(\range{K_h})+|M|+\lg(\range{K_t})\leq 24\kappa$
	\item\quad return $\qalgi(X,48\kappa,24\kappa,\range{K_h},|M|,\range{K_t},a_1,a_2, b)$ \ctn
\end{code}
If $S$ has small size, we recover the whole data structure using $\decset$.
If $\cD$ is too short, we reduce $m$ and $V$.
\begin{code}{access}
	\item $V_1:=\lfloor V/2\rfloor$, $V_2:=\lceil V/2\rceil$
	\item ask $\rank$ queries and compute $j:=\rank_S(X+V_1-1)-\rank_S(X-1)$
	\item if $b$, then $j:=V_1-j$
	\item find the size of $\cD_j=(K_{j,h},M_j,K_{j,t})$ in the lookup table
\end{code}
We compute $j$, the integer extracted from $\cD$, which encodes the number of elements in the first half.
We recover the size of $\cD_j$, and use the fact that $(M_j,K_{j,t})$ is a suffix of $\cD$ (by Proposition~\ref{prop_sep}) to recurse.
\begin{code}{access}
	\item if $a_1\geq |M|-|M_j|$
	\item\quad return $\qalgiT(X,V_1,V_2,j,m-j,\range{K_{j,h}},|M_j|,\range{K_{j,t}},a_1-(|M|-|M_j|),a_2-(|M|-|M_j|),b)$
	\item else
	\item\quad recover $\cD_j[-1, a_2-(|M|-|M_j|)]:=$\\
	\phantom{\quad}\hfill$\qalgiT(X,V_1,V_2,j,m-j,\range{K_{j,h}}, |M_j|,\range{K_{j,t}},-1,a_2-(|M|-|M_j|),b)$
	\item\quad compute $\cD[a_1, a_2]$ using Proposition~\ref{prop_sep}
\end{code}
$\qalgiT$ recovers the requested substring of $\cD$ assuming $\rank$ queries to the set generated from $\intPreT$.
Since $(M_j,K_{j,t})$ is a suffix of $\cD$, if $\cD[a_1, a_2]$ is entirely contained in this range, we simply recursive on $\cD_j$.
Otherwise, Proposition~\ref{prop_sep} guarantees that the remaining bits can be recovered from $j$ and $K_{j,h}$.
Note that in either case, the difference $a_2-a_1$ does not increase.

Next, we describe $\qalgiT$.
\begin{code}{accessT}[accessing algorithm][(X,V_1,V_2,m_1,m_2,\range{K_{j,h}}, |M_j|,\range{K_{j,t}},a_1,a_2,b)][\qalgiT]
	\item compute the sizes of $\cD_1=(K_{1,h},M_1,K_{1,t})$ and $\cD_2=(K_{2,h},M_2,K_{2,t})$, which $\cD$ is divided into \\ \strut\ctn
\end{code}
Suppose $S$ restricted to $[X,X+V_1)$ and $[X+V_1,X+V_1+V_2)$ is generated from $\cD$ using $\intPreT$.
Then by Proposition~\ref{prop_divide}, $(K_{1,h},M_1)$ is a prefix of $\cD$ and $(M_2,K_{2,t})$ is a suffix.
\begin{code}{accessT}
	\item\label{step_suffix} if $a_1\geq |M|-|M_2|$
	\item\quad return $\qalgi(X+V_1,V_2,m_2,\range{K_{2,h}},|M_2|,\range{K_{2,t}},a_1-(|M|-|M_2|),a_2-(|M|-|M_2|),b)$
	\item\label{step_prefix} if $a_2<|M_1|$
	\item\quad return $\qalgi(X,V_1,m_1,\range{K_{1,h}},|M_1|,\range{K_{1,t}},a_1,a_2,b)$ \ctn
\end{code}
If the requested bits $\cD[a_1, a_2]$ are entirely contained in $\cD_1$ or $\cD_2$, we simply recurse on the corresponding substring.
In this case, the difference $a_2-a_1$ does not change either.
\begin{code}{accessT}
	\item\label{step_two_call_a} recover $\cD_1[a_1, |M_1|]:=\qalgi(X,V_1,m_1,\range{K_{1,h}},|M_1|,\range{K_{1,t}},a_1,|M_1|,b)$
	\item\label{step_two_call_b} recover $\cD_2[-1, a_2-(|M|-|M_2|)]:=$\\
	\phantom{\quad}\hfill$\qalgi(X+V_1,V_2,m_2,\range{K_{2,h}},|M_2|,\range{K_{2,t}},-1,a_2-(|M|-|M_2|),b)$
	\item reconstruct $\cD[a_1, a_2]$ using Proposition~\ref{prop_divide}
\end{code}
Finally, if the requested bits $\cD[a_1, a_2]$ split across both substrings, then we make two recursive calls.

\paragraph{Query time.}
Next, we analyze the query time.
First observe that $\qalgi$ has at most $O(\lg m)$ levels of recursion before we call $\decset$.
This is because each time $m$ is reduced at least by a factor of $1/3$ by the preprocessing algorithm.
The only place that the whole recursion makes more than one recursive calls is line~\ref{step_two_call_a} and line~\ref{step_two_call_b} in $\qalgiT$.
In all other cases, the algorithm makes at most one recursive call with the same (or smaller) difference $a_2-a_1$.
Moreover, we claim that those two lines can only be executed at most once throughout the whole recursion.
\begin{claim}\label{cl_branch_once}
	Line~\ref{step_two_call_a} and line~\ref{step_two_call_b} in $\qalgiT$ can at most be executed once throughout the whole recursion.
\end{claim}
\begin{proof}
	When these two lines are executed, the two recursive calls will both request either a prefix or a suffix of the substring.
	Also, as we observed above, the difference $a_2-a_1$ never increases throughout the recursion.
	The recursive call that requests a prefix will have $a_1=-1$ and $a_2<\kappa-1$.
	Thereafter, any subsequence recursive calls in this branch will have $a_1=-1$ and $a_2<\kappa-1$.
	Since Proposition~\ref{prop_divide} always generates two strings of length at least $3\kappa$, line~\ref{step_prefix} in $\qalgiT$ is always true (as $|M_1|\geq\kappa-1$).
	Line~\ref{step_two_call_a} and line~\ref{step_two_call_b} will hence not be executed in this branch.
	The recursive branch that requests a suffix is similar, in which line~\ref{step_suffix} in $\qalgiT$ is alway true.
	This proves the claim.
\end{proof}
Claim~\ref{cl_branch_once} implies that the whole recursion tree has at most $O(\lg m)$ nodes, and at most two leaves.
In each node, the algorithm spends constant time, and makes two $\rank$ queries.
In each leaf, the algorithm makes one call to $\decset$.
As we argued earlier, the integer $Z\leq\lg\binom{V}{m}$ has at most $O(\kappa^2)$ bits (and $O(\kappa)$ words).
Since $m\leq O(\kappa)$ when $\decset$ is called, by Claim~\ref{cl_decset}, each $\decset$ takes $O(\kappa^4)$ time ($O(\kappa)$-word numbers take $O(\kappa^2)$ time to multiply or divide), and makes $O(\kappa^2)$ $\rank$ queries.
Combining the above facts, we conclude that $\qalgi$ runs in $O(\kappa^4)$ time, and it makes at most $O(\kappa^2)$ $\rank$ queries.
This proves Lemma~\ref{lem_subset_interpret}.

\subsection{Small sets}\label{sec_small_set}

In this section, we prove Lemma~\ref{lem_block_rank}, which constructs a succinct rank data structure for sets of size $\kappa^{O(1)}$, with constant query time.
We first show that the fusion trees~\cite{FW93} can be implemented succinctly.
This gives us a data structure for small sets with a sublinear, although large, redundancy.
\begin{lemma}\label{lem_large_redundancy}
	Let $c$ be any constant positive integer and $\epsilon$ be any positive constant.
	There is a preprocessing algorithm $\rankPreL$, a query algorithm $\qalgrL$ and lookup tables $\tblrL_{V,m}$ of sizes $2^{\epsilon \kappa}$ such that for any integers $V, m$ such that $V\leq 2^\kappa$ and $m\leq \kappa^c$, given a set $S\subset [V]$ of size $m$, $\rankPreL$ preprocesses it into a data structure using 
	\[\lg \binom{V}{m}+\frac{1}{8}m\lg \kappa\] bits of space.
	Given any $x\in[V]$, $\qalgrL$ compute $\rank_S(x)$ in constant time, by accessing the data structure and $\tblrL{}_{V,m}$.
	The algorithms run on a random access machine with word-size $w\geq \Omega(\kappa)$.
\end{lemma}
Since the main ideas are similar, we may omit the proof of a few claims in the construction, and refer the readers to the original fusion trees for details (\cite{FW93}).
\begin{proof}(sketch)
	Let $S=\{y_1,\ldots,y_m\}$ and $y_1<y_2<\cdots<y_m$.
	Let us first show how to construct such a data structure using 
	\[
		m\lceil \lg V\rceil+m\lceil\lg \kappa\rceil
	\]
	bits when $m\leq \kappa^{1/4}$.
	We view each $y_i$ as a $\lceil \lg V\rceil$-bit binary string, and consider the first bit where $y_i$ and $y_{i+1}$ differ, for every $i=1,\ldots,m-1$.
	Let $W$ be this set of bits, i.e., $j\in W$ if and only there exists some $i$ such that $j$-th bit is the first bit where $y_i$ and $y_{i+1}$ differ.
	Then $|W|\leq m-1$.
	Similar to fusion trees, let $\skt(y)$ be $y$ restricted to $W$.
	Observe that we must have $\skt(y_1)<\skt(y_2)<\cdots<\skt(y_m)$.

	The data structure first stores $W$ using $m\lceil\lg \kappa\rceil$ bits.
	Then it stores $\skt(y_1),\ldots,\skt(y_m)$.
	Finally, the data structure stores the remaining bits of each $y_i$, for $i=1,\ldots,m$ and from the top bits to the low bits.
	It is clear that the data structure occupies $m\lceil \lg V\rceil+m\lceil\lg \kappa\rceil$ bits of space.

	To answer a query $x\in[V]$, $\qalgrL_{V,m}$ first breaks $x$ into $\skt(x)$ and the remaining bits.
	That is, it generates two strings: $x$ restricted to $W$ (a $|W|$-bit string), and the remaining bits (a $(\lceil \lg V\rceil-|W|)$-bit string).
	It can be done in constant time using a lookup table of size $2^{O(\epsilon \kappa)}$, e.g., we divide the bits of $x$ into chunks of length $\epsilon \kappa$, and store in $\tblrL_{V,m}$ for each chunk, every possible set $W$ and every possible assignment to the bits of $x$ in the chunk, their contribution to $\skt(x)$ and the remaining bits (note that there are only $2^{o(\kappa)}$ different sets $W$).
	Summing over all chunks gives us $\skt(x)$ and the remaining bits.
	The query algorithm then finds the unique $i$ such that $\skt(y_i)\leq \skt(x)<\skt(y_{i+1})$.
	This can be done by storing a lookup table of size at most $2^{(m+1)|W|}\leq 2^{\kappa^{1/2}}$, since $(\skt(y_1),\ldots,\skt(y_m))$ has only $m|W|$ bits, and $\skt(x)$ has $|W|$ bits.
	However, we might not necessarily have $y_i\leq x<y_{i+1}$, but similar to the arguments in fusion trees, $x$ has the longest common prefix (LCP) with either $y_i$ or $y_{i+1}$ (among all $y\in S$).
	$\qalgrL_{V,m}$ next computes the LCP between $x$ and $y_i$ and the LCP between $x$ and $y_{i+1}$.
	Both can be done in constant time, since to compute the LCP between $x$ and $y_i$, it suffices to compute the LCP between $\skt(x)$ and $\skt(y_i)$ and the LCP between their remaining bits.
	Suppose $x$ and $y_{i^*}$ have a longer LCP ($i^*=i$ or $i+1$).
	If $x=y_{i^*}$, then $\rank_S(x)=i^*$.
	Otherwise, let their common prefix be $x'$.
	If $x>y_{i^*}$, then let $j$ be the unique index such that $\skt(y_j)\leq \skt(x'111\cdots11)<\skt(y_{j+1})$.
	The argument from fusion trees shows that we must have $y_j<x<y_{j+1}$, i.e., $\rank_S(x)=j$.
	Likewise, if $x<y_{i^*}$, then let $j$ be the unique index such that $\skt(y_j)<\skt(x'000\cdots00)\leq\skt(y_{j+1})$.
	We must have $y_j<x<y_{j+1}$.
	By computing the value of $j$ using the lookup table again, we find the number of elements in $S$ that is at most $x$.
	Note that this data structure also allows us to retrieve each $y_i$ in constant time.

	\bigskip

	Next, we show that the above data structure generalizes to any $m\leq \kappa^c$, and uses space
	\[
		m(\lg V+(c+3)\lg \kappa)\leq \lg\binom{V}{m}+(2c+3)m\lg\kappa.
	\]
	When $m>\kappa^{1/4}$, let $B=\lfloor \kappa^{1/4}\rfloor$, we take $B$ evenly spaced elements from $S$, i.e., $y_{\lceil im/B\rceil}$ for $i=1,\ldots,B$.
	Denote the set of these $B$ elements by $S'=\{y'_1,\ldots,y'_B\}$, where $y'_i=y_{\lceil im/B\rceil}$.
	We apply the above data structure to $S'$, using space
	\[
		B\lceil \lg V\rceil+B\lceil\lg\kappa\rceil<B(\lg V+\lg\kappa+2).
	\]
	Then, we recurse on all $B$ subsets between elements in $S'$, where the $i$-th subset has $\lceil im/B\rceil-\lceil (i-1)m/B\rceil-1$ elements.
	Then the final data structure stores
	\begin{itemize}
		\item the data structure for $S'$;
		\item $B$ data structures for all subsets between elements in $S'$;
		\item an array of $B$ pointers, pointing to the starting locations of the above $B$ data structures.
	\end{itemize}
	We assign $(c+3/2)\lg \kappa$ bits to each pointer.

	Suppose for each subset, we are able to (recursively) construct a data structure using
	\[
		(\lceil im/B\rceil-\lceil (i-1)m/B\rceil-1)(\lg V+(c+3)\lg \kappa)
	\]
	bits of space.
	The total space usage is
	\begin{align*}
		B(\lg V+\lg\kappa+2)+(m-B)(\lg V+(c+3)\lg\kappa)+B(c+3/2)\lg\kappa \leq m(\lg V+(c+3)\lg\kappa).
	\end{align*}
	On the other hand, assigning $(c+3/2)\lg \kappa$ bits to each pointer is sufficient, because
	\[
		\lg\left(m(\lg V+(c+3)\lg \kappa)\right)\leq \lg\left(m\kappa+(c+3)m\lg \kappa\right)\leq (c+1)\lg \kappa+1.
	\]

	To answer query $x$, we first query the data structure for $S'$, and find the $i$ such that $y'_i\leq x<y'_{i+1}$.
	Then we recurse into the $i$-th subset.
	The query time is constant, because the size of the set reduces by a factor of $B=\Theta(\kappa^{1/4})$ each time.
	Note that for any given $i$, this data structure can also return $y_i$ in constant time.

	\bigskip

	Finally, we show that the redundancy $(2c+3)m\lg \kappa$ can be reduced to $\frac{1}{8}m\lg \kappa$.
	To this end, let $S'$ be the subset of $S$ with gap $16(2c+3)$, i.e., $S'=\{y'_1,y'_2,\ldots\}$ such that $y'_i=y_{16(2c+3)\cdot i}$.
	Then $|S'|=\lfloor \frac{m}{16(2c+3)}\rfloor$.
	We construct a data structure for $S'$ using space
	\[
		|S'|(\lg V+(c+3)\lg \kappa).
	\]
	Naturally, $S'$ partitions $S$ into chunks of $16(2c+3)-1$ elements.
	We simply write them down using
	\[
		(16(2c+3)-1)\lceil\lg(y'_{i+1}-y'_i-1) \rceil
	\]
	bits for chunk $i$.
	The final data structure consists of
	\begin{enumerate}
		\item the data structure for $S'$,
		\item all other elements in $S$ encoded as above,
		\item $|S'|+1$ pointers to each chunk.
	\end{enumerate}
	We assign $\lceil (c+3/2)\lg \kappa\rceil$ bits to each pointer.
	By the concavity of $\lg x$, the total space usage is
	\begin{align*}
		&|S'|(\lg V+(c+3)\lg \kappa)+\sum_i(16(2c+3)-1)\lceil\lg(y'_{i+1}-y'_i-1) \rceil+(|S'|+1)\lceil (c+3/2)\lg \kappa\rceil \\
		\leq&\, |S'|\lg \frac{V}{m}+|S'|(3c+5)\lg\kappa+\sum_i(16(2c+3)-1)\lg\frac{V}{|S'|+1}+m \\
		\leq&\, |S'|\lg \frac{V}{m}+\frac{(3c+5)m}{16(2c+3)}\lg\kappa+\sum_i(16(2c+3)-1)\lg\frac{V}{m}+O(m) \\
		\leq&\, m\lg\frac{V}{m}+\frac{(3c+5)m}{16(2c+3)}\lg\kappa+O(m) \\
		\leq&\, \lg\binom{V}{m}+\frac{m}{8}\lg\kappa.
	\end{align*}

	To answer query $x$, we first query the data structure for $S'$, and find $i$ such that $y'_i\leq x<y'_{i+1}$.
	Then we go over the $16(2c+3)$ elements between $y'_i$ and $y'_{i+1}$, and compare each of them with $x$.
\end{proof}

Next, we show that if the sets are very small ($m\leq O(\kappa/\lg\kappa)$), then there is a data structure with constant query time and negligible extra bits.

\begin{lemma}\label{lem_small_set}
	Let $c\geq 2,\epsilon$ be two positive constants.
	There is a preprocessing algorithm $\rankPreS$, a query algorithm $\qalgrS$ and lookup tables $\tblrS_{V,m}$ of sizes $O(2^{\epsilon \kappa})$, such that for any integers $V\leq 2^\kappa$ and $m\leq c\cdot \kappa/\lg \kappa$,  such that given a set $S\subset [V]$ of size $m$, $\rankPreS$ preprocesses $S$ into a data structure using $\lg \binom{V}{m}+ 2^{-\kappa/2}$ bits of space.
	Given any $x\in[V]$, $\qalgrS$ computes $\rank_S(x)$ in constant time by accessing the data structure and $\tblrS_{V,m}$.
\end{lemma}
\begin{proof}
	Consider the binary trie over $\{0,\ldots,V\}$.\footnote{We write every integer in the set as a $\lceil \lg (V+1)\rceil$-bit string, then construct a trie over these $V+1$ binary strings.
	Note that $S$ is a subset of $\{0,\ldots,V-1\}$, while the trie has $V+1$ leaves.}
	Every element in $\{0,\ldots,V\}$ corresponds to a root-to-leaf path.
	Consider all paths corresponding to an element in $S\cup\{V\}$ ($V$ is included for technical reasons).
	Their union forms a subtree $T(S)$ of the binary trie with $m+1$ leaves.
	In the following, we construct a data structure assuming the \emph{topological structure} of $T(S)$ is known, then apply Proposition~\ref{prop_fuse} to fuse the topological structure into the data structure.

	Roughly speaking, the \emph{topological structure} of a subtree $T$ is the tree $T$ without specifying for each node with only one child, whether it is a left or a right child (see Figure~\ref{fig_1a}).
	Formally, it is defined by partitioning the set of such subtrees into equivalence classes, modulo the $\flip$ operation.
	Let $v$ be a node in $T$ with only a left [resp. right] child, let $\flip(v, T)$ be $T$ relocating $v$'s entire left [resp. right] subtree to its right [resp. left] child.
	We say two trees $T\sim T'$ if there is a (finite) sequence of $\flip$ operations that modifies $T$ to $T'$.
	It is easy to verify that $\sim$ is an equivalence relation, hence it partitions the set of all $T$ into equivalence classes.

	We call an edge in $T(S)$ \emph{a shared edge} if it has more than one leaf in its subtree.
	Equivalently, a shared edge is shared between at least two root-to-leaf paths.
	Note that if an edge is shared, then all edges on the path from root to it are shared.
	It turns out that the \emph{number of shared edges} in $T(S)$ is an important parameter, which is also invariant under $\flip$.
	Thus, for each equivalence class $\cT$, all $T\in\cT$ have the same number of shared edges (see Figure~\ref{fig_1b}).

	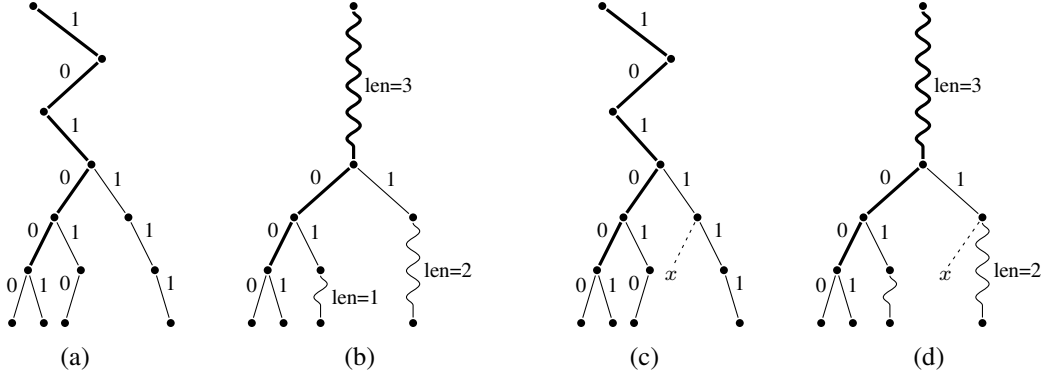
\begin{figure}[t]
		\begin{center}
			\begin{subfigure}{0.2\linewidth}
			\centering
			\begin{tikzpicture}[
				very thick,
				level/.style={level distance=20pt, sibling distance=60pt-#1 * 8pt},
				ver/.style={fill, inner sep=0, minimum size=3pt,circle}, 
				lc/.style={label={[yshift=-10pt]:\scriptsize 0}, xshift=-3pt, yshift=4pt}, 
				rc/.style={label={[yshift=-10pt]:\scriptsize 1}, xshift=3pt,yshift=4pt}]
				\node [ver] {}
					child {
						edge from parent [draw=none]
					}
					child {
						node [ver] {}
						child {
							node [ver] {}
							child {
								edge from parent [draw=none]
							}
							child {
								node [ver] {}
								child {
									node [ver] {}
									child {
										node [ver] {}
										child {
											node [ver] {}
											edge from parent [thin] node [lc] {}
										}
										child {
											node [ver] {}
											edge from parent [thin] node [rc] {}
										}
										edge from parent node [lc] {}
									}
									child {
										node [ver] {}
										child {
											node [ver] {}
											edge from parent node [lc]{}
										}
										child {
											edge from parent [draw=none]
										}
										edge from parent [thin] node [rc] {}
									}
									edge from parent node [lc] {}
								}
								child {
									node [ver] {}
									child {
										edge from parent [draw=none]
									}
									child {
										node [ver] {}
										child {
											edge from parent [draw=none]
										}
										child {
											node [ver] {}
											edge from parent node [rc] {}
										}
										edge from parent node [rc] {}
									}
									edge from parent [thin] node [rc] {}
								}
								edge from parent node [rc] {}
							}
							edge from parent node[lc] {}
						}
						child{
							edge from parent [draw=none]
						}
						edge from parent node[rc] {}
					};
			\end{tikzpicture}
			\caption{}\label{fig_1a}
			\end{subfigure}
			\quad
			\begin{subfigure}{0.2\linewidth}
			\centering
			\begin{tikzpicture}[
				very thick,
				level/.style={level distance=20pt, sibling distance=44pt-#1 * 8pt},
				curvyedge/.style={decorate, decoration={snake}},
				level 2/.style={sibling distance=45pt},
				ver/.style={fill, inner sep=0, minimum size=3pt,circle}, 
				lc/.style={label={[yshift=-10pt]:\scriptsize 0}, xshift=-3pt, yshift=4pt}, 
				rc/.style={label={[yshift=-10pt]:\scriptsize 1}, xshift=3pt,yshift=4pt}]
				\node [ver] {}
					child [level distance=60pt] {
						node [ver] {}
						child {
							node [ver] {}
							child {
								node [ver] {}
								child {
									node [ver] {}
									edge from parent [thin] node [lc] {}
								}
								child {
									node [ver] {}
									edge from parent [thin] node [rc] {}
								}
								edge from parent node [lc] {}
							}
							child {
								node [ver] {}
								child {
									node [ver] {}
									edge from parent [curvyedge] node [right] {\scriptsize len=1}
								}
								edge from parent [thin] node [rc] {}
							}
							edge from parent node [lc] {}
						}
						child {
							node [ver] {}
							child [level distance=40pt] {
								node [ver] {}
								edge from parent [curvyedge] node [right] {\scriptsize len=2}
							}
							edge from parent [thin] node [rc] {}
						}
						edge from parent [curvyedge] node [right] {\scriptsize len=3}
					};
			\end{tikzpicture}
			\caption{}\label{fig_1b}
			\end{subfigure}
			\quad
			\begin{subfigure}{0.2\linewidth}
			\centering
			\begin{tikzpicture}[
				very thick,
				level/.style={level distance=20pt, sibling distance=60pt-#1 * 8pt},
				ver/.style={fill, inner sep=0, minimum size=3pt,circle}, 
				lc/.style={label={[yshift=-10pt]:\scriptsize 0}, xshift=-3pt, yshift=4pt}, 
				rc/.style={label={[yshift=-10pt]:\scriptsize 1}, xshift=3pt,yshift=4pt}]
				\node [ver] {}
					child {
						edge from parent [draw=none]
					}
					child {
						node [ver] {}
						child {
							node [ver] {}
							child {
								edge from parent [draw=none]
							}
							child {
								node [ver] {}
								child {
									node [ver] {}
									child {
										node [ver] {}
										child {
											node [ver] {}
											edge from parent [thin] node [lc] {}
										}
										child {
											node [ver] {}
											edge from parent [thin] node [rc] {}
										}
										edge from parent node [lc] {}
									}
									child {
										node [ver] {}
										child {
											node [ver] {}
											edge from parent node [lc]{}
										}
										child {
											edge from parent [draw=none]
										}
										edge from parent [thin] node [rc] {}
									}
									edge from parent node [lc] {}
								}
								child {
									node [ver] {}
									child {
										node [inner sep=0, label={[xshift=0pt, yshift=5pt]below:\scriptsize $x$}] {}
										edge from parent [dotted]
									}
									child {
										node [ver] {}
										child {
											edge from parent [draw=none]
										}
										child {
											node [ver] {}
											edge from parent node [rc] {}
										}
										edge from parent node [rc] {}
									}
									edge from parent [thin] node [rc] {}
								}
								edge from parent node [rc] {}
							}
							edge from parent node[lc] {}
						}
						child{
							edge from parent [draw=none]
						}
						edge from parent node[rc] {}
					};
			\end{tikzpicture}
			\caption{}\label{fig_1c}
			\end{subfigure}
			\quad
			\begin{subfigure}{0.2\linewidth}
			\centering
			\begin{tikzpicture}[
				very thick,
				level/.style={level distance=20pt, sibling distance=44pt-#1 * 8pt},
				curvyedge/.style={decorate, decoration={snake}},
				level 2/.style={sibling distance=45pt},
				ver/.style={fill, inner sep=0, minimum size=3pt,circle}, 
				lc/.style={label={[yshift=-10pt]:\scriptsize 0}, xshift=-3pt, yshift=4pt}, 
				rc/.style={label={[yshift=-10pt]:\scriptsize 1}, xshift=3pt,yshift=4pt}]
				\node [ver] {}
					child [level distance=60pt] {
						node [ver] {}
						child {
							node [ver] {}
							child {
								node [ver] {}
								child {
									node [ver] {}
									edge from parent [thin] node [lc] {}
								}
								child {
									node [ver] {}
									edge from parent [thin] node [rc] {}
								}
								edge from parent node [lc] {}
							}
							child {
								node [ver] {}
								child {
									node [ver] {}
									edge from parent [curvyedge] node [right] {}
								}
								edge from parent [thin] node [rc] {}
							}
							edge from parent node [lc] {}
						}
						child {
							node [ver] {}
							child [sibling distance=14pt] {
								node [inner sep=0, label={[xshift=0pt, yshift=5pt]below:\scriptsize $x$}] {}
								edge from parent [dotted]
							}
							child [level distance=40pt] {
								node [ver] {}
								edge from parent [curvyedge] node [right] {\scriptsize len=2}
							}
							child {
								edge from parent [draw=none]
							}
							edge from parent [thin] node [rc] {}
						}
						edge from parent [curvyedge] node [right] {\scriptsize len=3}
					};
			\end{tikzpicture}
			\caption{}\label{fig_1d}
			\end{subfigure}
		\end{center}
		\caption{(b) is the \emph{topological structure} of (a), by getting rid of the information that for each single child, whether it is a left or a right child. The thick edges are \emph{shared}. Query $x$ branches off the tree from the dotted edge.}\label{fig_topo}
	\end{figure}
	
	Intuitively, for a typical set $S$, the corresponding $\cT\ni T(S)$ should have most of its degree-two nodes close to the root, i.e, it should have very \emph{few} shared edges. 
	Indeed, if we sample a uniformly random $S$, the number of shared edges is at most $O(\kappa)$ with probability at least $1-2^{-\Omega(\kappa)}$.
	As we will see below, on the inputs with few shared edges, it is relatively easy to construct data structures and answer queries.
	However, for the rare inputs with more than $\Omega(\kappa)$ shared edges, we can afford to use a different construction with a larger redundancy.
	Since they are sufficiently rare, the overall redundancy turns out to be small.

	\paragraph{Few shared edges.}
	Let us fix an equivalence class $\cT$, assume $\cT$ is known and consider all inputs $S$ such that $T(S)\in\cT$.
	Furthermore, assume the trees in $\cT$ have at most $(2c+1)\kappa$ shared edges.
	For each such $\cT$, we construct a lookup table $\tblrS_{V,m,\cT}$, and preprocess $S$ into a data structure using about $\lg|\cT|$ bits such that if the query algorithm is given access to this particular lookup table (specific for $\cT$), it answers $\rank$ queries in constant time.

	Since the tree $T(S)$ uniquely determines $S$, to construct the data structure for $S$, it suffices to encode for each edge in $T(S)$ that connects a single child and its parent, whether the child is left or right.
	The preprocessing algorithm constructs $T(S)$, then goes through all such edges in a \emph{fixed} order, and uses one bit to indicate whether the corresponding edge in $T(S)$ connects to a left child or a right child.
	To facilitate the queries (which we will describe in the next paragraph), all shared edges are encoded first in the \emph{depth-first-search} order, followed by all other edges encoded in the \emph{depth-first-search} order.
	This ensures that
	\begin{compactenum}
		\item if a shared edge $e_1$ is on the path from root to shared edge $e_2$, then $e_1$ is encoded before $e_2$;
		\item for each $y_i$, its non-shared edges (which is a suffix in the root-to-leaf path) are consecutive in the data structure.
	\end{compactenum}
	Note that this encoding is a \emph{one-to-one} mapping: Every $S$ such that $T(S)\in\cT$ is encoded to a different string; Every string has a corresponding $S$ with $T(S)\in\cT$ encoded to it.
	Thus, the algorithm constructs a data structure using exactly
	\[
		\lg |\{S:T(S)\in\cT\}|
	\]
	bits of space.

	Let $S=\{y_1,\ldots,y_m\}$ such that $y_1<y_2<\cdots<y_m$, and let $y_0=-1$ and $y_m=V$.
	Given a query $x\in\{0,\ldots,V-1\}$, the goal is to compute $i$ such that $y_i\leq x<y_{i+1}$.
	Let us consider the process of walking down the tree $T(S)$ following the bits in $x$.
	That is, we write $x$ also as a $\lceil \lg(V+1)\rceil$-bit string, and walk down the tree from the root: if the current bit in $x$ is $0$, we follow the left child, otherwise we follow the right child.
	The process stops when either the current node in $T(S)$ does not have a left (or right) child to follow, or we have reached a leaf.
	The location where it stops determines the answer to the query, in the same way for \emph{all} $T\in\cT$.
	See Figure~\ref{fig_1c} and~\ref{fig_1d} for a concrete example.
	Note that in the example, $x$ branches off the tree from left, which may not be possible at the same location for all $T\in\cT$, as some $T$ may have a left child there.
	But \emph{given} that $x$ branches off the tree at this location from left, all $T(S)\in\cT$ must have the same answer to $\rank_S(x)$.
	Thus, we store in $\tblrS_{V,m,\cT}$, for all nodes $v$ in the tree, $\textrm{ans}_{v,0}$ and $\textrm{ans}_{v,1}$, the answer to the query when the process branches off the tree from $v$ due to the lack of its left child (i.e., from left), and the answer when it branches off from $v$ due to the lack of its right child (i.e., from right) respectively.
	It takes $O(\kappa^2)$ words, since $m\leq \kappa$.

	Now the task is reduced to efficiently simulating this walk. 
	To this end, the query algorithm needs to compare the bits in $x$ with the corresponding bits of $T(S)$, which are stored in the data structure.
	It turns out that the difficult part is to compare $x$ with the shared edges, which are stored in the first (at most) $(2c+1)\kappa$ bits.
	The first step is to simulate the walk, and check if $x$ branches off $T(S)$ at a shared edge.
	We create lookup tables of size $2^{\epsilon \kappa}$ to compare $\epsilon \kappa$ bits at once.
	For now, let us focus on the first $\epsilon \kappa$ bits $x_{\leq \epsilon \kappa}$.
	These bits determine for all the degree-two nodes in the first $\epsilon \kappa$ levels, which child $x$ follows (note we have fixed $\cT$).
	Thus, it determines for all other bits, which bits in the data structure they should compare with.
	In the lookup table, we store for each of the $2^{\epsilon \kappa}$ possible values,
	\begin{compactitem}
		\item a $(2c+1)\kappa$-bit string, which permutes $x_{\leq \epsilon \kappa}$ to the same location as the bits they are comparing with;
		\item a $(2c+1)\kappa$-bit string, indicating for each shared edge in the data structure, whether they are being compared.
	\end{compactitem}

	With these two strings, the query algorithm is able to compare $x_{\leq \epsilon \kappa}$ with the first $\epsilon \kappa$ levels of $T(S)$.
	If they do not match, we could find the first edge where they differ (since edges are encoded in the DFS order), which is the location where $x$ branches off $T(S)$.
	If they all equal, we proceed and compare the next $\epsilon \kappa$ bits.
	Note that we may start the next chunk of the walk from different nodes depending on the value of $x_{\leq \epsilon \kappa}$, and we will need a different lookup for each starting location.
	However, $\cT$ can have at most $m$ nodes in each level, thus, only $m$ tables are needed for each chunk.
	We repeat the above process until we find a different bit, or we find out that $x$ matches all shared edges from the root.
	In the former case, as we argued above, the answer to the query can be found in the lookup table.
	In the latter case, by the definition of shared edges, we identified one $y_i$ which is the only element in $S$ that matches the prefix of $x$.
	Thus, it suffices to retrieve the remaining bits of $y_i$, which are stored consecutively in the data structure and take constant retrieval time, and compare $y_i$ with $x$.
	If $y_i\leq x$, then the query algorithm returns $i$, otherwise, it returns $i-1$.
	The query time is constant.

	\bigskip

	So far for every $\cT$ with at most $(2c+1)\kappa$ shared edges, we have designed a data structure that works for all inputs $S$ such that $S\in\cT$ using space $\lg|\{S:T(S)\in\cT\}|$ bits, constant query time and lookup table of size $2^{\epsilon \kappa}$.
	Next, we fuse $\cT$ into the data structure and merge all lookup tables, obtaining a single data structure
	that works for all $S$ such that $T(S)$ has at most $(2c+1)\kappa$ shared edge, which uses lookup table $\tblrS_{V,m,\textrm{few}}$.
	To this end, we fix an arbitrary ordering of all such equivalence classes $\cT$: $\cT_1,\ldots,\cT_C$, where $C$ is the number of equivalence classes.
	Let $s_i=\lg|\{S:T(S)\in\cT_i\}|$ be the size of the data structure for $\cT_i$.
	Then, $C\leq 2^{2m}\cdot \binom{(2c+1)\kappa+1}{m-1}\leq 2^{m\lg (\kappa/m)+O(m)}$.
	This is because there are at most $2^{2m}$ rooted binary trees with $m+1$ nodes (corresponding to the degree-two nodes).
	Each such tree can be extended to a class $\cT$ by specifying the distance from each child to its parent (adding the degree-one nodes).
	However, there are only $(2c+1)\kappa$ shared edges, thus, the sum of distances of all internal edges is at most $(2c+1)\kappa$, and there are $m-1$ internal edges.\footnote{An edge is internal if it does not connect to a leaf.}
	Hence, it is at most $\binom{\leq(2c+1)\kappa}{m-2}\leq\binom{(2c+1)\kappa+1}{m-1}$ choices.
	Once the distances on all internal edges are determined, the distance on each edge connecting to a leaf is also fixed, because all leaves are at depth $\lceil\lg (V+1)\rceil$.

	Given an input set $S$ such that $T(S)$ has at most $(2c+1)\kappa$ shared edges, the preprocessing algorithm computes $T(S)$ and finds the index $i$ such that $\cT_i\ni T(S)$.
	Then it runs the preprocessing algorithm for class $\cT_i$ on $S$, and computes a data structure $\cD_i$ of at most $s_i$ bits.
	Next, we use Proposition~\ref{prop_fuse} to store the pair $(i,\cD_i)$, using space at most
	\begin{align*}
		\lg \sum_{i=1}^C 2^{s_i}+C\cdot 2^{-\kappa+2} &\leq \lg\left(\sum_{i=1}^C |\{S:T(S)\in \cT_i\}|\right)+2^{m\lg (\kappa/m)+O(m)-\kappa+2} \\
		&<\lg\binom{V}{m}+2^{m\lg (\kappa/m)+O(m)-\kappa+2}\\
		&<\lg\binom{V}{m}+2^{-\frac{3}{4}\kappa}.
	\end{align*}

	The lookup table $\tblrS_{V,m,\textrm{few}}$ is simply the concatenation of all tables $\tblrS_{V,m,\cT_i}$ for $i=1,\ldots,C$, as well as the $O(C)$-sized table from Proposition~\ref{prop_fuse}.
	Thus, the total size is at most $2^{\epsilon \kappa}\cdot C+O(C)=2^{(\epsilon+o(1))\kappa}$.

	To answer a query $x$, Proposition~\ref{prop_fuse} allows us to decode $i$ and $\cD_i$ in constant time by storing a lookup table of size $O(C)$.
	Then, we find the corresponding lookup table $\tblrS_{V,m,\cT_i}$ and run the query algorithm for $\cT_i$ on query $x$ and data structure $\cD_i$.
	The query time is constant.

	\paragraph{Many shared edges.}
	Next, we construct a data structure that works for all $S$ such that $T(S)$ has more than $(2c+1)\kappa$ shared edges, using
	\[
		\lg\binom{V}{m}-\kappa
	\]
	bits of space.
	Note that this is possible, because there are very few such sets $S$ (a tiny fraction of all $\binom{V}{m}$ sets).
	We find the largest $k$ such that $T(S_{\leq k})$ has at most $(2c+1)\kappa$ shared edges, where $S_{\leq k}=\{y_1,\ldots,y_k\}$.
	Note that every element can introduce no more than $\kappa$ shared edges, thus, $T(S_{\leq k})$ has at least $2c\kappa$ shared edges.
	The data structure stores the (index of) equivalence class $\cT\ni T(S_{\leq k})$, then we run the preprocessing algorithm on $S_{\leq k}$.
	This encodes the first $k$ elements of $S$.
	For the next $m-k$ elements, we simply apply Lemma~\ref{lem_large_redundancy}.

	More specifically, for $k$ elements, there are at most $2^{k\lg (\kappa/k)+O(k)}$ equivalence classes, as we showed earlier.
	We construct the data structure as follows:
	\begin{enumerate}
		\item write down the index $k$ using $\lceil \lg m\rceil$ bits;
		\item write down the index $i$ such that $\cT_i\ni T(S_{\leq k})$ using $\lceil k\lg (\kappa/k)+O(k) \rceil$ bits;
		\item run the preprocessing algorithm on $S_{\leq k}$ and obtain a data structure of size
		\[
			\lg|\{S_{\leq k}:T(S_{\leq k})\in \cT_i\}|;
		\]
		\item run $\rankPreL$ on $\{y_{k+1},\ldots,y_m\}$ and obtain a data structure of size
		\[
			\lg\binom{V}{m-k}+\frac{1}{8}(m-k)\lg \kappa.
		\]
	\end{enumerate}
	Observe that Step 3 uses at most
	\[
		k\lceil\lg V\rceil-2c\kappa
	\]
	bits, because for any such $\cT_i$,
	\begin{compactitem}
		\item by construction, each bit of the data structure stores an input bit, i.e., one of the bits representing $\{y_1,\ldots,y_k\}$;
		\item each of the $\geq 2c\kappa$ shared edges corresponds to at least two input bits (since given $\cT$, these two input bits are always the same);
		\item each input bit is stored only once.
	\end{compactitem}
	Therefore, the preprocessing algorithm outputs a data structure using
	\begin{align*}
		&\ \lg m+k\lg (\kappa/k)+O(k)+(k\lg V-2c\kappa)+\left(\lg\binom{V}{m-k}+\frac{1}{8}(m-k)\lg \kappa\right)+k+2 \\
		\leq&\ \lg m+k\lg (\kappa/k)+(k\lg V-2c\kappa)+(m-k)\lg V+\frac{1}{8}m\lg \kappa+O(k) \\
		\leq&\ m\lg V-2c\kappa+\lg m+k\lg (\kappa/k)+\frac{1}{8}m\lg \kappa+O(k) \\
		\leq&\ \lg\binom{V}{m}+m\lg m-2c\kappa+\lg m+m\lg (\kappa/m)+\frac{1}{8}m\lg \kappa+O(m) \\
		\leq&\ \lg\binom{V}{m}-2c\kappa+\frac{9}{8}m\lg \kappa+O(m).
	\end{align*}
	By the fact that $m\leq c\kappa/\lg \kappa$ and $c\geq 2$, it is at most
	\[
		\lg\binom{V}{m}-\kappa.
	\]

	The lookup table includes $\tblrS_{V,k,\mathrm{few}}$ for all $k\leq m$, and has $2^{(\epsilon+o(1)) \kappa}$ size.

	To answer query $x$, the query algorithm reads $k$ and $i$.
	Then it runs the query algorithm for $\cT_i$ for query $x$ on the data structure for $S_{\leq k}$, as well as $\qalgrL$ for $x$ on the data structure for $\{y_{k+1},\ldots,y_m\}$.
	Both algorithms run in constant time.
	The answer to the query is simply the sum of the two answers.

	\paragraph{Combining the two cases.} Finally, we combine the two cases using Proposition~\ref{prop_fuse}, and construct a data structure that works for all sets $S$.
	Given set $S$, $\rankPreS$ computes $T(S)$ and the number of shared edges.
	If it has no more than $(2c+1)\kappa$ shared edges, it sets $b:=1$, runs the preprocessing algorithm for ``many shared edges'' and obtains a data structure $\cD_1$.
	Otherwise, it sets $b:=2$, runs the preprocessing algorithm for ``few shared edges'' and obtains a data structure $\cD_2$.
	At last, it applies Proposition~\ref{prop_fuse} to store the pair $(b,\cD_b)$.
	The space usage is
	\begin{align*}
		&\ \lg\left(\binom{V}{m}\cdot 2^{2^{-\frac{3}{4}\kappa}}+\binom{V}{m}\cdot 2^{-\kappa}\right)+2^{-\kappa+2} \\
		\leq&\ \lg\binom{V}{m}+2^{-\frac{3}{4}\kappa}+\lg(1+2^{-\kappa-2^{-\frac{3}{4}\kappa}})+2^{-\kappa+2} \\
		\leq&\ \lg\binom{V}{m}+2^{-\frac{1}{2}\kappa}.
	\end{align*}

	To answer query $x$, we simply decode $b$ and $\cD_b$ using Proposition~\ref{prop_fuse}, and use the corresponding query algorithm based on $b$.

	The lookup table $\tblrS_{V,m}$ also includes all $\tblrS_{V,k}$ for $k\leq m$, which has size $2^{O(\epsilon \kappa)}$.
	This proves the lemma.
\end{proof}

Finally, we prove Lemma~\ref{lem_block_rank}, which constructs a rank data structure for $m\leq \kappa^c$.

\begin{restate}[Lemma~\ref{lem_block_rank}]
	\contlemblockrank
\end{restate}
\begin{proof}
	The data structure construction is based on recursion.
	As the base case, if $m\leq 16\kappa/\lg \kappa$, we simply use the data structure from Lemma~\ref{lem_small_set}, and the statement holds.
	Otherwise for $m>16\kappa/\lg \kappa$, we divide $V$ into $B$ blocks of equal size, for $B=\lceil \kappa^{1/2}\rceil$.
	For a typical set $S$, we would expect each block to contain roughly $m/B$ elements.
	If it indeed happens, the size of $S$ would be reduced by a factor of $B$.
	Hence, we will reach the base case in constant rounds.
	On the other hand, input sets $S$ which have at least one block with significantly more than $m/B$ elements are very rare.
	If such blocks occur, we are going to apply Lemma~\ref{lem_large_redundancy} on them.
	Although Lemma~\ref{lem_large_redundancy} introduces a large redundancy, such cases occur sufficiently rarely, so that the overall redundancy is still small.
	
	We partition the input set $S$ into $B$ subsets $S_1,\ldots,S_B$ such that $S_i$ contains all elements of $S$ between $\left\lceil (i-1)V/B\right\rceil$ and $\lceil iV/B\rceil-1$.
	Let $V_i:=\lceil iV/B\rceil-\left\lceil (i-1)V/B\right\rceil$ be the size of the $i$-th block.
	By definition, $|S_1|+\cdots+|S_B|=m$ and $V_1+\cdots+V_B=V$.
	We construct a data structure for each $S_i$, over a universe of size $V_i$.
	Then we apply Proposition~\ref{prop_concat} to concatenate the $B$ data structures \emph{given} the sizes of $S_1,\ldots,S_B$.
	Finally, we apply Proposition~\ref{prop_fuse} to union all combinations of sizes.
	We present the details below.

	\paragraph{Preprocessing algorithm.}
	Given a set $S$ of size $m$, if $2m\geq V$, we take the complement.
	Note that the space bound stated in the lemma becomes smaller after taking the complement.
	It is also easy to derive the answer from the data structure for the complement.
	Then if $m=1$, we simply write down the element; if $m\leq 16\kappa/\lg\kappa$, we apply Lemma~\ref{lem_small_set}.
	\begin{code}{rankpre}[preprocessing algorithm][(V,m,S)][\rankPre]
		\item if $V\leq 2m$
		\item\quad $m:=V-m$ and $S:=[V]\setminus S$
		\item if $m=1$
		\item\quad return the only element in $S$
		\item if $m\leq 16\kappa/\lg \kappa$
		\item\quad return $\cD:=\rankPreS(V,m,S)$ using Lemma~\ref{lem_small_set} \ctn
	\end{code}

	If $m>16\kappa/\lg\kappa$, we divide $[V]$ into $\kappa^{1/4}$ chunks, and construct a data structure for each chunk.
	\begin{code}{rankpre}
		\item $B:=\lfloor \kappa^{1/4}\rfloor$
		\item compute $S_i:=S\cap [(i-1)V/B, iV/B)$ and $m_i:=|S_i|$
		\item let $V_i:=\left\lceil iV/B\right\rceil-\left\lceil (i-1)V/B\right\rceil$
		\item for $i=1,\ldots,B$
		\item\quad if $m_i>\max\{m\cdot \kappa^{-1/4},16\kappa/\lg\kappa\}$
		\item\quad\quad compute $\cD_i:=\rankPreL(V_i,m_i,S_i)$ using Lemma~\ref{lem_large_redundancy}
		\item\quad else
		\item\quad\quad compute $\cD_i:=\rankPre(V_i,m_i,S_i)$ recursively
	\end{code}
	If the chunk has too many elements, we apply Lemma~\ref{lem_large_redundancy} to construct a data structure with larger redundancy.
	Otherwise, the size of the set at least decreases by a factor of $\kappa^{1/4}$, and we recurse.

	Next, we concatenate the data structures for all chunks, and fuse the tuple $(m_1,\ldots,m_B)$ into the data structure.
	\begin{code}{rankpre}
		\item\label{step_concat} apply Proposition~\ref{prop_concat} to concatenate $\cD_1,\ldots,\cD_B$, and obtain $\cD_{\textrm{cat}}$
		\item let $C:=\binom{m+B-1}{B-1}$ be the number of different tuples $(m_1,\ldots,m_B)$ such that $m_i\geq 0$ and $m_1+\cdots+m_B=m$
		\item let $1\leq j\leq C$ be the index such that the current $(m_1,\ldots,m_B)$ is the $j$-th in the lexicographic order
		\item\label{step_union} apply Proposition~\ref{prop_fuse} to fuse $j$ into $\cD_{\textrm{cat}}$, and obtain $\cD$
		\item return $\cD$
	\end{code}

	\paragraph{Space analysis.} In the following, we analyze the size of the data structure.
	We will prove by induction that $\rankPre(V,m,S)$ outputs a data structure of size at most
	\[
		\lg\binom{V}{m}+(m-1)2^{-\kappa/2}.
	\]
	The base case when $m\leq 16\kappa/\lg \kappa$ is a direct implication of Lemma~\ref{lem_small_set} (or if $m=1$, the space usage if $\lg V=\lg\binom{V}{1}$).
	Now, let us consider larger $m$.

	To prove the inductive step, let us fix a $B$-tuple $(m_1,\ldots,m_B)$, and consider the size of $\cD_{\textrm{cat}}$ from line~\ref{step_concat}.
	By Proposition~\ref{prop_concat}, when all $m_i\leq \max\{m\cdot \kappa^{-1/4}, 16\kappa/\lg \kappa\}$, its size is at most
	\[
		s(m_1,\ldots,m_B):=\lg\prod_{i=1}^B\binom{V_i}{m_i}+(m-B)\cdot 2^{-\kappa/2}+(B-1)2^{-\kappa+4};
	\]
	otherwise, its size is at most
	\begin{equation}\label{eqn_space_gen}
		s(m_1,\ldots,m_B):=\lg\prod_{i=1}^B\binom{V_i}{m_i}+\sum_{i:m_i>\max\{m\cdot \kappa^{-1/4},16\kappa/\lg \kappa\}} \frac{1}{8}m_i\lg \kappa+B.
	\end{equation}
	It turns out that in the latter case, \eqref{eqn_space_gen} is \emph{significantly} smaller than $\lg\binom{V}{m}$.
	\begin{claim}\label{cl_lgvm}
		If there is at least one $m_i>\max\{m\cdot \kappa^{-1/4},16\kappa/\lg \kappa\}$, then \eqref{eqn_space_gen} is at most $\lg\binom{V}{m}-\kappa$.
	\end{claim}
	We defer its proof to the end.
	Assuming the claim, by Proposition~\ref{prop_fuse}, the size of $\cD$ from line~\ref{step_union} is at most
	\begin{equation}\label{eqn_space}
		\lg\left(\sum_{\stackrel{m_1,\ldots,m_B:}{\sum_i m_i=m}}2^{s(m_1,\ldots,m_B)}\right)+C\cdot 2^{-\kappa+2}.
	\end{equation}
	To bound the sum in the logarithm, we first take the sum only over all tuples such that $m_i\leq \max\{m\cdot \kappa^{-1/4},16\kappa/\lg \kappa\}$, the sum is at most
	\begin{align*}
		\sum 2^{s(m_1,\ldots,m_B)}&\leq \sum\prod_{i=1}^B\binom{V_i}{m_i}\cdot 2^{(m-B)\cdot 2^{-\kappa/2}+(B-1)2^{-\kappa+4}} \\
		&\leq \binom{V}{m}\cdot 2^{(m-B)\cdot 2^{-\kappa/2}+(B-1)2^{-\kappa+4}},
	\end{align*}
	where the second inequality uses the fact that $\sum_{m_1,\ldots,m_B:\sum m_i=m}\prod_{i=1}^B \binom{V_i}{m_i}\leq \binom{\sum_{i=1}^B V_i}{m}$, and we are taking this sum over a subset of all such $B$-tuples.
	By Claim~\ref{cl_lgvm}, $s(m_1,\ldots,m_B)\leq \lg\binom{V}{m}-\kappa$ for all other tuples.
	Thus, the sum in the logarithm is at most
	\[
		\binom{V}{m}\cdot 2^{(m-B)\cdot 2^{-\kappa/2}+(B-1)2^{-\kappa+2}}+\binom{V}{m}\cdot C\cdot 2^{-\kappa}.
	\]

	Finally, since $C\leq m^B$ and $m\leq \kappa^c$, \eqref{eqn_space} is at most
	\begin{align*}
		\eqref{eqn_space}&\leq \lg\left(\binom{V}{m}\cdot 2^{(m-B)\cdot 2^{-\kappa/2}+(B-1)2^{-\kappa+4}}+\binom{V}{m}\cdot m^B\cdot 2^{-\kappa}\right)+m^B\cdot 2^{-\kappa+2} \\
		&\leq \lg\binom{V}{m}+(m-B)2^{-\kappa/2}+(B-1)2^{-\kappa+4}+\lg(1+2^{-\kappa+B\lg m})+2^{-\kappa+B\lg m+2} \\
		&\leq \lg\binom{V}{m}+(m-B)2^{-\kappa/2}+(B-1)2^{-\kappa+4}+2^{-\kappa+c\kappa^{1/4}\lg \kappa+3} \\
		&\leq \lg\binom{V}{m}+(m-1)2^{-\kappa/2}.
	\end{align*}
	
	By induction, it proves the data structure uses space as claimed.

	\paragraph{Lookup table.} 
	We store the following information in the lookup table.
	\begin{code}{tblr}[lookup table][][\tblr_{\mathnormal{V,m}}]
		\item if $m\leq 16\kappa/\lg \kappa$, include $\tblrS_{V,m}$ from Lemma~\ref{lem_small_set}
		\item\label{tblr_line_2} the value of $C=\binom{m+B-1}{B-1}$
		\item for all $1\leq j\leq C$
		\item\quad the $j$-th $B$-tuple $(m_1,\ldots,m_B)$ in the lexicographic order
		\item\quad for $i=1,\ldots,B$
		\item\quad\quad $m_1+\cdots+m_i$
		\item lookup table for Proposition~\ref{prop_concat} in line~\ref{step_concat}, for all possible $B$-tuples $(m_1,\ldots,m_B)$
		\item\label{tblr_line_5} lookup table for Proposition~\ref{prop_fuse} in line~\ref{step_union}
		\item include all tables $\tblr_{V',m'}$ and $\tblrL_{V',m'}$ for $V'=\lfloor V/B^i\rfloor$ or $\lceil V/B^i\rceil$ for $i\geq 1$, and $m'\leq m$
	\end{code}

	Since $C=\binom{m+B-1}{B-1}\leq 2^{o(\kappa)}$, line~\ref{tblr_line_2} to \ref{tblr_line_5} all have size $2^{o(\kappa)}$.
	Finally, we are only including $\kappa^{O(1)}$ other tables in line 1 and 6, each taking at most $\tilde{O}(2^{\epsilon \kappa})$ bits by Lemma~\ref{lem_large_redundancy} and \ref{lem_small_set}.
	The total size of $\tblr_{V,m}$ is $\tilde{O}(2^{\epsilon \kappa})$.

	\paragraph{Query algorithm.}
	Given a query $x$, if $V\leq 2m$, we retreat the data structures as storing the complement of $S$, and use the fact that $\rank_{S}(x)=x+1-\rank_{[V]\setminus S}(x)$.
	Then if $m=1$, we simply compare it with $x$.
	If $m\leq 16\kappa/\lg\kappa$, we invoke the query algorithm from Lemma~\ref{lem_small_set}.
	\begin{code}{qalgr}[query algorithm][(V,m,x)][\qalgr]
		\item if $V\leq 2m$
		\item\quad $m:=V-m$
		\item\quad in the following, when about to return answer $r$, return $x+1-r$
		\item if $m=1$
		\item\quad retrieve the element, compare it with $x$, and return $0$ or $1$
		\item if $m\leq 16\kappa/\lg \kappa$, 
		\item\quad return $\qalgrS(V,m,x)$ (from Lemma~\ref{lem_small_set})\ctn
	\end{code}

	If $m>16\kappa/\lg\kappa$, we decode $j$, which encodes the tuple $(m_1,\ldots,m_B)$ and $\cD_{\mathrm{cat}}$.
	Then if $x$ is in the $i$-th chunk, we decode $m_i$ and the corresponding $\cD_i$.
	\begin{code}{qalgr}
		\item apply Proposition~\ref{prop_fuse} to decode $j$ and $\cD_{\textrm{cat}}$
		\item let $i$ be the chunk that contains $x$
		\item apply Proposition~\ref{prop_concat} to decode $\cD_i$
		\item retrieve $m_1+\cdots+m_{i-1}$ and $m_i$ for $j$-th tuple from the lookup table\ctn
	\end{code}

	Then depending on the value of $m_i$, we invoke the query algorithm from Lemma~\ref{lem_large_redundancy} or recurse.
	\begin{code}{qalgr}
		\item if $m_i>\max\{m\cdot \kappa^{-1/4},16\kappa/\lg\kappa\}$
		\item\quad return $(m_1+\cdots+m_{i-1})+\cD_i.\qalgrL(V_i,m_i,x-\lceil(i-1)V/B\rceil)$ (from Lemma~\ref{lem_large_redundancy})
		\item else
		\item\quad return $(m_1+\cdots+m_{i-1})+\cD_i.\qalgr(V_i,m_i,x-\lceil(i-1)V/B\rceil)$
	\end{code}
	
	The query algorithm recurses only when $m_i\leq m\cdot \kappa^{-1/4}$.
	In all other cases, the query is answered in constant time.
	On the other hand, $m\leq \kappa^c$.
	The level of recursion must be bounded by a constant.
	Thus, the data structure has constant query time, proving the lemma.
\end{proof}

Next, we prove the remaining claim.
\begin{proof}[Proof of Claim~\ref{cl_lgvm}]
	To prove the claim, let us first compare the first term with $\lg\binom{V}{m}$.
	We have
	\begin{align}
		&\, \lg\binom{V}{m}-\lg\prod_{i=1}^B\binom{V_i}{m_i} \nonumber\\
		=&\,\lg\frac{V!\cdot m_1!\cdots m_B!\cdot (V_1-m_1)!\cdots (V_B-m_B)!}{V_1!\cdots V_B!\cdot m!(V-m)!}, \nonumber\\
		\intertext{which, by Stirling's formula, is at least}
		\geq&\, \sum_{i=1}^B \left(V_i\lg\frac{V}{V_i}-m_i\lg\frac{m}{m_i}-(V_i-m_i)\lg\frac{V-m}{V_i-m_i}\right)-O(B)-\lg V, \nonumber\\
		\intertext{which by the fact that $f(\varepsilon)=\varepsilon\log 1/\varepsilon$ is concave and hence $V\cdot f(\frac{V_i}{V})\geq m\cdot f(\frac{m_i}{m})+(V-m)\cdot f(\frac{V_i-m_i}{V-m})$, is at least}
		\geq&\, \sum_{i:m_i>\max\{m\cdot \kappa^{-1/4},16\kappa/\lg \kappa\}} \left(V_i\lg\frac{V}{V_i}-m_i\lg\frac{m}{m_i}-(V_i-m_i)\lg\frac{V-m}{V_i-m_i}\right)-O(B)-\lg V.\label{eqn2}
	\end{align}
	For each term in this sum, we have
	\[
		V_i\lg\frac{V}{V_i}=V_i\lg B-V_i\lg\left(1+\frac{V_i-V/B}{V/B}\right)\geq V_i\lg B-O(1),
	\]
	since $|V_i-V/B|\leq 1$;
	and
	\begin{align*}
		(V_i-m_i)\lg\frac{V-m}{V_i-m_i}&= (V_i-m_i)\left(\lg B+\lg\left(1+\frac{m_i-m/B+(V/B-V_i)}{V_i-m_i}\right)\right) \\
		&\leq (V_i-m_i)\lg B+(V_i-m_i)\cdot\frac{m_i-m/B+1}{V_i-m_i}\cdot\lg e \\
		&\leq (V_i-m_i)\lg B+2m_i.
	\end{align*}
	Plugging into \eqref{eqn2}, we have
	\begin{align*}
		&\, \lg\binom{V}{m}-\lg\prod_{i=1}^B\binom{V_i}{m_i}\\
		\geq &\, \sum_{i:m_i>\max\{m\cdot \kappa^{-1/4},16\kappa/\lg \kappa\}} \left(V_i\lg B-m_i\lg\frac{m}{m_i}-(V_i-m_i)\lg B-2m_i\right)-O(B)-\lg V \\
		= &\, \sum_{i:m_i>\max\{m\cdot \kappa^{-1/4},16\kappa/\lg \kappa\}} m_i\left(\lg \frac{Bm_i}{m}-2\right)-O(B)-\lg V \\
		\geq &\, \sum_{i:m_i>\max\{m\cdot \kappa^{-1/4},16\kappa/\lg \kappa\}} m_i\left(\frac{1}{4}\lg \kappa-2\right)-O(B)-\lg V.
	\end{align*}

	Therefore, we have
	\begin{align*}
		\eqref{eqn_space_gen}&\leq \lg\binom{V}{m}-\sum_{i:m_i>\max\{m\cdot \kappa^{-1/4},16\kappa/\lg \kappa\}} m_i\left(\frac{1}{4}\lg \kappa-2\right)+O(B)+\lg V \\
		&\quad +\sum_{i:m_i>\max\{m\cdot \kappa^{-1/4},16\kappa/\lg \kappa\}} \frac{1}{8}m_i\lg \kappa \\
		&\leq \lg\binom{V}{m}-\sum_{i:m_i>\max\{m\cdot \kappa^{-1/4},16\kappa/\lg \kappa\}}m_i\left(\frac{1}{8}\lg \kappa-2\right)+O(B)+\lg V \\
		&\leq \lg\binom{V}{m}-\kappa.
	\end{align*}
	The last inequality is due to the fact that there is at least one $m_i$ that is larger than $\max\{m\cdot \kappa^{-1/4},16\kappa/\lg \kappa\}$ (in particular, $m_i>16\kappa/\lg \kappa$), $B=\Theta(\kappa^{1/2})$ and $\lg V\leq \kappa/2$.
\end{proof}

$\rank$ queries can be viewed as mapping that maps $S\rightarrow [m]$, and $[V]\setminus S\rightarrow [V-m]$.
Thus, Lemma~\ref{lem_small_bad_block} is an immediate corollary.
\begin{restate}[Lemma~\ref{lem_small_bad_block}]
	\contlemsmallbadblock
\end{restate}







\section{Perfect Hashing for Sets of Any Size}\label{sec_general}

In this section, we generalize the data structure from Section~\ref{sec_medium} to sets of all sizes, proving our main theorem.
\begin{theorem}[main theorem]\label{thm_main}
	For any constant $\epsilon>0$, there is a preprocessing algorithm $\ph$, a query algorithm $\qalg$ and lookup tables $\tbl_{U, n}$ of size $n^{\epsilon}$, such that given 
	\begin{compactitem}
		\item a set $S$ of $n$ keys over the key space $[U]$,
		\item a uniformly random string $\cR$ of length $O(\lg^{12} n)$,
	\end{compactitem}
	$\ph$ preprocesses $S$ into a data structure $\cD$ of (worst-case) length
	\[
		\OPT_{U,n}+O(\lg\lg U),
	\]
	such that $\cD$ defines a \emph{bijection} $h$ between $S$ and $[n]$ and a \emph{bijection} $\oh$ between $[U]\setminus S$ and $[U-n]$.
	Given access to $\cD$, $\cR$ and $\tbl_{U,n}$, for any key $x\in [U]$, $\qalg(U,n,x)$ outputs $\hash(x)$ on a RAM with word-size $w\geq \Omega(\lg U)$, in time
	\begin{compactitem}
		\item $O(1)$ with probability $1-O(\lg^{-7}U)$ and
		\item $O(\lg^7 U)$ in worst-case,
	\end{compactitem}
	where the probability is taken over the random $\cR$.
	In particular, the query time is constant in expectation and with high probability.
\end{theorem}

\begin{proof}
	Similar to the proof of Theorem~\ref{thm_medium_set}, we first assume $2n\leq U$ (otherwise, we take the complement of $S$).
	Then if $n\geq U^{1/12}$, Theorem~\ref{thm_medium_set} already gives the desired result.
	From now on, we assume $n<U^{1/12}$.

	We partition $[U]$ into $n^{12}$ blocks.
	A typical set $S$ has all the keys in different blocks.
	In this case, we may view the universe size being only $n^{12}$, and apply Theorem~\ref{thm_medium_set}.
	On the other hand, only roughly $1/n^{11}$-fraction of the inputs have at least one pair of keys in the same block, which suggests that we may use at least $10\lg n$ extra bits.

	\paragraph{No collision in blocks and last block empty.}
	More specifically, given $S$ such that $n=|S|<U^{1/{12}}$, let $V:=\lceil U\cdot n^{-{12}} \rceil$ be the block size.
	We partition $U$ into blocks: $(U\Div V)$ blocks of size $V$ and one last block of size $(U\Mod V$).
	Let us first only consider inputs that have at most one key in every block and no key in the last block.
	We apply Theorem~\ref{thm_medium_set} on the universe of all blocks.
	That is, let the universe size $U_\new=U\Div V$, number of keys $n_\new=n$.
	We construct the new set $S_\new$ such that $i\in S_\new$ if and only if block $i$ contains a key $x\in S$.
	By Theorem~\ref{thm_medium_set}, we construct a data structure of size
	\[
		\lg\binom{U_\new}{n_\new}+1/U_\new=\lg\binom{U\Div V}{n}+1/(U\Div V),
	\]
	which defines hash functions $h_\new$ and $\oh_\new$.
	Besides this data structure, we also apply Lemma~\ref{lem_change_base} to store for each $i\in S_\new$, the key $x$ within block $i$, according to $h_\new(i)$.
	That is, we store $x-(i-1)V$ in coordinate $h_\new(i)$.
	Hence, this part takes
	\[
		n\lg V+(n-1)2^{-\kappa+5}
	\]
	bits.
	Then we apply Proposition~\ref{prop_concat} to concatenate the two data structures.
	The total space is at most
	\begin{align*}
		&\, \lg\binom{U\Div V}{n}+1/(U\Div V)+n\lg V+n\cdot 2^{-\kappa+5}\\
		\leq&\, \lg \frac{V^n\prod_{i=0}^{n-1}(U\Div V-i)}{n!}+O(1/n) \\
		\leq&\, \lg \frac{\prod_{i=0}^{n-1}(U-i)}{n!}+O(1/n) \\
		=&\, \lg\binom{U}{n}+O(1/n).
	\end{align*}

	To define the hash functions in this case, for each $x\in S$, which is in block $i$, we simply let $h(x):=h_\new(i)$, the hash value of the block.
	For $x\notin S$ in block $i$, 
	\begin{compactitem}
		\item if $i\in S_\new$, let $x^*$ be the key in block $i$,
		\begin{compactitem}
			\item if $x<x^*$, we let $\oh(x):=(V-1)\cdot h_\new(i)+(x-(i-1)V)$,
			\item if $x>x^*$, we let $\oh(x):=(V-1)\cdot h_\new(i)+(x-(i-1)V-1)$,
		\end{compactitem}
		\item if $i\notin S_\new$, we let $\oh(x):=(V-1)n+V\cdot\oh_\new(i)+(x-(i-1)V)$.
		\item if $x$ is in the last block, we let $\oh(x):=(U\Div V)\cdot V-n+(x-(U\Div V)\cdot V)$
	\end{compactitem}
	That is, we order all non-keys in block $i$ for $i\in S_\new$ first, in the increase order of $(h_\new(i), x)$; then we order all non-keys not in the last block, in the increasing order of $(\oh_\new(i), x)$; finally we order all non-keys in the last block.

	To answer a query $x$ in block $i$, we first query if $i\in S_\new$.
	If $i\notin S_\new$, then we know $x$ is not a key, calculate $\oh(x)$ by its definition, and return.
	Otherwise, we query the $(h_\new(i)+1)$-th value in the second data structure, using Lemma~\ref{lem_change_base}, to retrieve the key in block $i$.
	If $x$ happens to be this key, we return $(1, h_\new(i))$.
	Otherwise, $x\notin S$, and $\oh(x)$ can be calculated by its definition.
	Finally, for queries $x$ in the last block, $x$ is not a key, and we calculate $\oh(x)$ according to its definition.

	\paragraph{Exist collision in blocks or keys in last block.}
	Next, we consider the case where at least one block contains more than one key, or the last block contains at least one key.
	We spend the first $3\lceil\lg n\rceil$ bits to store
	\begin{compactitem}
		\item $N$, the number of blocks with at least two keys (blocks with collisions, or simply \emph{collision blocks}),
		\item $n_{\cl}$, the total number of keys in all collision blocks,
		\item $n_{\la}$, the number of keys in the last block.
	\end{compactitem}
	Next, we apply Lemma~\ref{lem_large_bad_block}, and construct a membership data structure for $N$ collision blocks using
	\[
		\lg\binom{U\Div V}{N}+O(N)
	\]
	bits, which defines a bijection $h_\cl$ between all collision blocks and $[N]$, and a bijection $\oh_\cl$ between all other blocks (except for the last block) and $[(U\Div V)-N]$.

	The final data structure has three \emph{more} components:
	\begin{enumerate}
		\item store all keys in $N$ collision blocks using Lemma~\ref{lem_large_bad_block}, where each element $x$ in block $i$ is stored as $V\cdot h_\cl(i)+(x-(i-1)V)$, which uses at most 
		\[
			\lg\binom{NV}{n_{\cl}}+O(n_\cl+\lg\lg V)
		\]
		bits;
		\item store all other $(U\Div V)-N$ blocks using the data structure for no collisions, where each element $x$ in block $i$ is stored as $V\cdot \oh_\cl(i)+(x-(i-1)V)$, which uses at most
		\[
			\lg\binom{(U\Div V)V-NV}{n-n_\cl-n_\la}+1
		\]
		bits;
		\item store the last block using Lemma~\ref{lem_large_bad_block}, which uses
		\[
			\lg\binom{U\Mod V}{n_\la}+O(n_\la+\lg\lg V)
		\]
		bits.
	\end{enumerate}
	Summing up the sizes of these three data structures, we get
	\[
		\lg\binom{NV}{n_{\cl}}\binom{(U\Div V)V-NV}{n-n_\cl-n_\la}\binom{U\Mod V}{n_\la}+O(n_\cl+n_\la+\lg\lg V).
	\]
	By the fact that $\binom{n}{k}\leq (en/k)^k$ and $n_{\cl}\geq 2N$, the first term is at most
	\begin{align*}
		&\ n_\cl\lg\frac{NV}{n_{\cl}}+(n-n_\cl-n_\la)\lg\frac{(U\Div V)V-NV}{n-n_\cl-n_\la}+n_\la\lg\frac{U\Mod V}{n_\la}+n\lg e \\
		\leq&\ n_{\cl}\lg \frac{V}{2}+(n-n_\cl-n_\la)\lg\frac{U}{n-n_\cl-n_\la}+n_\la\lg V+n\lg e \\
		\leq&\ n\lg\frac{eU}{n}+n_{\cl}\lg\frac{nV}{2U}+(n-n_\cl-n_\la)\lg\frac{n}{n-n_\cl-n_\la}+n_\la\lg\frac{nV}{U} \\
		\intertext{which by the fact that $V\leq 2U\cdot n^{-{12}}$, is at most}
		\leq&\ n\lg\frac{eU}{n}+n_{\cl}\lg n^{-{11}}+(n-n_\cl-n_\la)\lg\left(1+\frac{n_\cl+n_\la}{n-n_\cl-n_\la}\right)+n_\la\lg (2n^{-{11}}) \\
		\leq&\ n\lg\frac{eU}{n}+n_{\cl}\lg n^{-{11}}+(n_\cl+n_\la)\lg e+n_\la\lg (2n^{-{11}}) \\
		\leq&\ n\lg\frac{eU}{n}-(n_{\cl}+n_{\la})({11}\lg n-O(1)).
	\end{align*}

	On the other hand, by Stirling's formula,
	\begin{align*}
		\lg\binom{U}{n}&=\lg\frac{U!}{n!(U-n)!} \\
		&\geq \lg\frac{\sqrt{U} U^U}{\sqrt{n} n^n\cdot \sqrt{U-n}(U-n)^{U-n}}-O(1) \\
		&\geq n\lg\frac{U}{n}+(U-n)\lg\frac{U}{U-n}-\frac{1}{2}\lg n-O(1) \\
		\intertext{which by the fact that $\ln(1+x)\geq x-x^2/2$ for $x\geq 0$, is at most}
		&\geq n\lg\frac{U}{n}+(U-n)\left(\frac{n}{U-n}-\frac{n^2}{2(U-n)^2}\right)\lg e-\frac{1}{2}\lg n-O(1) \\
		&=n\lg\frac{eU}{n}-\frac{n^2\lg e}{2(U-n)}-\frac{1}{2}\lg n-O(1) \\
		&\geq n\lg\frac{eU}{n}-\frac{1}{2}\lg n-O(1).
	\end{align*}

	Thus, the total size of the data structure when $N\geq 1$ is at most
	\begin{align*}
		&\ \lg\binom{U}{n}+\frac{1}{2}\lg n-(n_{\cl}+n_{\la})({11}\lg n-O(1))+3\lg n+\lg\binom{U\Div V}{N}+O(N+\lg\lg U) \\
		\leq &\ \lg\binom{U}{n}+\frac{1}{2}\lg n-(n_{\cl}+n_{\la})({11}\lg n-O(1))+3\lg n+12N\lg n+O(N+\lg\lg U) \\
		\intertext{which by the fact that $n_{\cl}\geq 2N$, is at most}
		\leq &\ \lg\binom{U}{n}+\frac{1}{2}\lg n-(n_{\cl}+n_{\la})({5}\lg n-O(1))+3\lg n+O(\lg\lg U) \\
		=&\ \OPT_{U,n}-\lg n+O(\lg\lg U).
	\end{align*}

	In this case, the hash functions are defined as follows.
	For both $h$ and $\oh$, we first order all elements in the $N$ collision blocks according to their hash values from component 1, which are mapped to $[n_\cl]$ and $[N\cdot V-n_\cl]$ respectively.
	Then we order all elements in the $(U\Div V)-N$ non-collision blocks according to their hash values from component 2, which are mapped to $\left\{n_\cl,\ldots,n-n_\la-1\right\}$ and $\left\{N\cdot V-n_\cl,\ldots,(U\Div V)\cdot V-(n-n_\la)-1\right\}$ respectively.
	Finally, we order all elements in the last block according to their hash values from component 3, which are mapped to $\left\{n-n_\la, n-1\right\}$ and $\left\{(U\Div V)\cdot V-(n-n_\la), U-n-1\right\}$ respectively.

	To answer a query $x$ in block $i$, we retrieve $N$, $n_\cl$ and $n_\la$, and query if $i$ is a collision block and $h(i)$ (or $\oh(i)$).
	If $i$ is a collision block, we query component 1; if $i$ is not a collision block, we query component 2; if $i$ is the last block, we query component 3.
	In any case, the hash value of $x$ can be computed according to its definition in constant time.

	\bigskip

	Finally, we apply Proposition~\ref{prop_fuse} to combine the two cases, by fusing a bit indicating whether there is any collision block.
	The final data structure has space bounded by
	\begin{align*}
		&\ \lg\left(2^{\OPT_{U,n}+O(1/n)}+2^{\OPT_{U,n}-\lg n+O(\lg\lg U)}\right)+2^{-\kappa+2} \\
		=&\ \OPT_{U,n}+\lg(2^{O(1/n)}+(\lg^{O(1)} U)/n)+2^{-\kappa+2} \\
		\leq&\ \OPT_{U,n}+O(\lg\lg U).
	\end{align*}

	The query algorithm is straightforward.
	To answer a query $x$, we apply Proposition~\ref{prop_fuse} to decode the data structure, and the bit indicating whether there is any collision block or any element in the last block.
	Then we apply the corresponding query algorithm as described above.
	This proves the theorem.
\end{proof}

\begin{remark}
	When the $O(\lg\lg U)$ term is at most $0.5\lg n$, the above data structure uses $\OPT+o(1)$ bits.
	To improve the $O(\lg\lg U)$ term when $U$ is large, we partition the universe into $\lg^{10} U$ blocks, and check if any block has at least two keys.
	In this case, the fraction of inputs with some block with at least two keys is only $1/\lg^{10} U$ fraction.
	Therefore, we can afford to ``waste'' about $10\lg\lg U$ bits, which dominates the $O(\lg\lg U)$ term.
	This strategy reduces the problem to storing $n$ non-empty blocks among a total of $\lg^{O(1)} U$ blocks, i.e., the universe size is reduced from $U$ to $\lg^{O(1)} U$.
	Thus, repeatedly applying it improves the $O(\lg\lg U)$ term to $O(\lg\lg\cdots\lg U)$ for logarithm iterated for any constant number of times.
\end{remark}




\section{Discussions and Open Problems}\label{sec_disc}
In this paper, we assumed that the word-size $w$ is at least $\Omega(\lg U+\lg \sigma)$, i.e., each (key, value) pair fits in $O(1)$ words.
When either the key or the value is larger than $\Theta(w)$ bits, it would take super-constant time to just read the query or write the output on a RAM.
The best query time one can hope for is $O((\lg U+\lg \sigma)/w)$.

When $\lg\sigma\gg w$, the only place being affected is Lemma~\ref{lem_change_base}, where we need to retrieve values longer than one word.
Our data structure naturally supports such long answers in optimal time.
When $\lg U\gg w$, a similar strategy to Section~\ref{sec_general} applies.
We view the first $O(w)$ bits of an element in $[U]$ as its ``hash value''.
If it turns out that all keys have different ``hash values'', it suffices to add the remaining bits of the key into its value.
Otherwise, if multiple keys share the same prefix, then we will be able to save $O(w)$ bits for every extra key with the same prefix.

\bigskip

Our dictionary data structure supports each query in constant expected time.
A major open question is to design \emph{deterministic} succinct dictionary with similar bounds, or to prove this is impossible.
Our approach crucially relies on sampling a small set of keys to be the ``hard queries''.
There is always a small portion of the data stored using the rank data structure of P\v{a}tra\c{s}cu, which takes $O(\lg n)$ time to decode.
``Derandomizing'' this data structure seems to require a completely different strategy.
On the other hand, proving lower bounds may also be challenging, as the common strategy of ``designing a hard distribution and proving average-case lower bound'' is doomed to fail.
For any fixed input distribution, we could always fix and hardwire the random bits in the data structure, thus, our data structure uses only $\OPT+1$ bits of space.

Our data structure only supports value-retrieval queries on a \emph{fixed} set of (key, value) pairs, i.e., it solves the \emph{static} dictionary problem.
The \emph{dynamic} dictionary problem further requires the data structure to support (key, value) insertions and deletions.
It seems non-trivial to extend our data structure to such updates to the data, even with good amortized expected time, although our data structure has $\tilde{O}(n)$ preprocessing time, thus using a hash table to store a ``buffer'' of size $n^{1-\epsilon}$ and using the technique of \emph{global-rebuilding}~\cite{Over83}, one can get $\tilde{O}(n^{1-\epsilon})$ redundancy, $n^{\epsilon}$ update time and expected constant query time.
On the other hand, it is also possible to update a (key, value) to a new value in our data structure, in $O(1)$ expected time.

Finally, the dependence on $U$ in the extra bits is intriguing.
In the RAM model, the dependence is very slow-growing, but still super constant.
We believe it is not necessary, but it is unclear how to remove this extra small term.
On the other hand, note that in the cell-probe model, it can actually be entirely removed (even for very large $U$).
This is because when $U$ is large enough so that $\lg\lg U$ becomes unignorable, we could simply apply Lemma~\ref{lem_small_bad_block}.
This strategy does not work on RAM, since it requires a large lookup table, which can only be hardwired in a cell-probe data structure.


\section*{Acknowledgment}
The author would like to thank anonymous reviewers for helpful comments.

\bibliography{ref}

\begin{thebibliography}{MHMP15}

\bibitem[BL13]{BL13}
Karl Bringmann and Kasper~Green Larsen.
\newblock Succinct sampling from discrete distributions.
\newblock In {\em Symposium on Theory of Computing Conference, STOC'13, Palo
  Alto, CA, USA, June 1-4, 2013}, pages 775--782, 2013.

\bibitem[BM99]{BM99}
Andrej Brodnik and J.~Ian Munro.
\newblock Membership in constant time and almost-minimum space.
\newblock {\em {SIAM} J. Comput.}, 28(5):1627--1640, 1999.

\bibitem[BMRV02]{BMRV02}
Harry Buhrman, Peter~Bro Miltersen, Jaikumar Radhakrishnan, and Srinivasan
  Venkatesh.
\newblock Are bitvectors optimal?
\newblock {\em {SIAM} J. Comput.}, 31(6):1723--1744, 2002.

\bibitem[CW79]{CW79}
Larry Carter and Mark~N. Wegman.
\newblock Universal classes of hash functions.
\newblock {\em J. Comput. Syst. Sci.}, 18(2):143--154, 1979.

\bibitem[DPT10]{DPT10}
Yevgeniy Dodis, Mihai P{\v a}tra{\c s}cu, and Mikkel Thorup.
\newblock Changing base without losing space.
\newblock In {\em Proc. 42nd ACM Symposium on Theory of Computing (STOC)},
  pages 593--602, 2010.

\bibitem[FKS84]{FKS84}
Michael~L. Fredman, J{\'{a}}nos Koml{\'{o}}s, and Endre Szemer{\'{e}}di.
\newblock Storing a sparse table with {O}(1) worst case access time.
\newblock {\em J. {ACM}}, 31(3):538--544, 1984.

\bibitem[FM95]{FM95}
Faith~E. Fich and Peter~Bro Miltersen.
\newblock Tables should be sorted (on random access machines).
\newblock In {\em Algorithms and Data Structures, 4th International Workshop,
  {WADS} '95, Kingston, Ontario, Canada, August 16-18, 1995, Proceedings},
  pages 482--493, 1995.

\bibitem[FN93]{FN93}
Amos Fiat and Moni Naor.
\newblock Implicit {O(1)} probe search.
\newblock {\em {SIAM} J. Comput.}, 22(1):1--10, 1993.

\bibitem[FNSS92]{FNSS92}
Amos Fiat, Moni Naor, Jeanette~P. Schmidt, and Alan Siegel.
\newblock Nonoblivious hashing.
\newblock {\em J. {ACM}}, 39(4):764--782, 1992.

\bibitem[FW93]{FW93}
Michael~L. Fredman and Dan~E. Willard.
\newblock Surpassing the information theoretic bound with fusion trees.
\newblock {\em J. Comput. Syst. Sci.}, 47(3):424--436, 1993.

\bibitem[GORR09]{GORR09}
Roberto Grossi, Alessio Orlandi, Rajeev Raman, and S.~Srinivasa Rao.
\newblock More haste, less waste: Lowering the redundancy in fully indexable
  dictionaries.
\newblock In {\em 26th International Symposium on Theoretical Aspects of
  Computer Science, {STACS} 2009, February 26-28, 2009, Freiburg, Germany,
  Proceedings}, pages 517--528, 2009.

\bibitem[Jac89]{Jacobson89}
Guy Jacobson.
\newblock Space-efficient static trees and graphs.
\newblock In {\em 30th Annual Symposium on Foundations of Computer Science,
  Research Triangle Park, North Carolina, USA, 30 October - 1 November 1989},
  pages 549--554, 1989.

\bibitem[MHMP15]{MHMP15}
A.~{Makhdoumi}, S.~{Huang}, M.~{M\'{e}dard}, and Y.~{Polyanskiy}.
\newblock On locally decodable source coding.
\newblock In {\em 2015 IEEE International Conference on Communications (ICC)},
  pages 4394--4399, 2015.

\bibitem[Mil96]{Mil96}
Peter~Bro Miltersen.
\newblock Lower bounds for static dictionaries on rams with bit operations but
  no multiplication.
\newblock In {\em Automata, Languages and Programming, 23rd International
  Colloquium, ICALP96, Paderborn, Germany, 8-12 July 1996, Proceedings}, pages
  442--453, 1996.

\bibitem[MNSW98]{MNSW98}
Peter~Bro Miltersen, Noam Nisan, Shmuel Safra, and Avi Wigderson.
\newblock On data structures and asymmetric communication complexity.
\newblock {\em J. Comput. Syst. Sci.}, 57(1):37--49, 1998.

\bibitem[Ove83]{Over83}
Mark~H. Overmars.
\newblock {\em The Design of Dynamic Data Structures}.
\newblock Lecture Notes in Economic and Mathematical Systems. Springer-Verlag,
  1983.

\bibitem[Pag01a]{Pagh01}
Rasmus Pagh.
\newblock Low redundancy in static dictionaries with constant query time.
\newblock {\em {SIAM} J. Comput.}, 31(2):353--363, 2001.

\bibitem[Pag01b]{Pagh01b}
Rasmus Pagh.
\newblock On the cell probe complexity of membership and perfect hashing.
\newblock In {\em Proceedings on 33rd Annual {ACM} Symposium on Theory of
  Computing, July 6-8, 2001, Heraklion, Crete, Greece}, pages 425--432, 2001.

\bibitem[P{\v a}t08]{Pat08}
Mihai P{\v a}tra{\c s}cu.
\newblock Succincter.
\newblock In {\em Proc. 49th IEEE Symposium on Foundations of Computer Science
  (FOCS)}, pages 305--313, 2008.

\bibitem[PT06]{PT06}
Mihai P{\v a}tra{\c s}cu and Mikkel Thorup.
\newblock Time-space trade-offs for predecessor search.
\newblock In {\em Proceedings of the 38th Annual {ACM} Symposium on Theory of
  Computing, Seattle, WA, USA, May 21-23, 2006}, pages 232--240, 2006.

\bibitem[PT07]{PT07}
Mihai P{\v a}tra{\c s}cu and Mikkel Thorup.
\newblock Randomization does not help searching predecessors.
\newblock In {\em Proceedings of the Eighteenth Annual {ACM-SIAM} Symposium on
  Discrete Algorithms, {SODA} 2007, New Orleans, Louisiana, USA, January 7-9,
  2007}, pages 555--564, 2007.

\bibitem[PV10]{PV10}
Mihai P{\v a}tra{\c s}cu and Emanuele Viola.
\newblock Cell-probe lower bounds for succinct partial sums.
\newblock In {\em Proceedings of the Twenty-First Annual {ACM-SIAM} Symposium
  on Discrete Algorithms, {SODA} 2010, Austin, Texas, USA, January 17-19,
  2010}, pages 117--122, 2010.

\bibitem[RRR07]{RRR07}
Rajeev Raman, Venkatesh Raman, and Srinivasa {Rao Satti}.
\newblock Succinct indexable dictionaries with applications to encoding
  \emph{k}-ary trees, prefix sums and multisets.
\newblock {\em {ACM} Trans. Algorithms}, 3(4):43, 2007.

\bibitem[SS90]{SS90}
Jeanette~P. Schmidt and Alan Siegel.
\newblock The spatial complexity of oblivious k-probe hash functions.
\newblock {\em {SIAM} J. Comput.}, 19(5):775--786, 1990.

\bibitem[Tho13]{Thorup13}
Mikkel Thorup.
\newblock Mihai {P}{\v a}tra{\c s}cu: Obituary and open problems.
\newblock {\em Bulletin of the {EATCS}}, 109:7--13, 2013.

\bibitem[TY79]{TY79}
Robert~Endre Tarjan and Andrew~Chi{-}Chih Yao.
\newblock Storing a sparse table.
\newblock {\em Commun. {ACM}}, 22(11):606--611, 1979.

\bibitem[Vio12]{Viola12a}
Emanuele Viola.
\newblock Bit-probe lower bounds for succinct data structures.
\newblock {\em {SIAM} J. Comput.}, 41(6):1593--1604, 2012.

\bibitem[VWY19]{VWY20}
Emanuele Viola, Omri Weinstein, and Huacheng Yu.
\newblock How to store a random walk.
\newblock {\em CoRR}, abs/1907.10874, 2019.

\bibitem[Yao81]{Yao81a}
Andrew~Chi{-}Chih Yao.
\newblock Should tables be sorted?
\newblock {\em J. {ACM}}, 28(3):615--628, 1981.

\bibitem[Yu19]{Yu19}
Huacheng Yu.
\newblock Optimal succinct rank data structure via approximate nonnegative
  tensor decomposition.
\newblock In {\em Proceedings of the 51st Annual {ACM} {SIGACT} Symposium on
  Theory of Computing, {STOC} 2019, Phoenix, AZ, USA, June 23-26, 2019.}, pages
  955--966, 2019.

\end{thebibliography}
\bibliographystyle{alpha}

\appendix

\section{Proofs for Fractional-length Strings}\label{sec_frac_length_app}

In the section, we prove the propositions from Section~\ref{sec_frac_length}.
We first show that two strings can be concatenated.

\newcommand{\contpropconcattwo}{
	Let $s_1,s_2\geq 0$.
	Given two binary strings $\cS_1=(M_1,K_1)$ and $\cS_2=(M_2,K_2)$ of $s_1$ and $s_2$ bits respectively, they can be concatenated into one string $\cS=(M,K)$ of length at most $s_1+s_2+2^{-\kappa+2}$, and both $M_1$ and $M_2$ are (consecutive) substrings of $M$.
	Moreover, given the values of $s_1$ and $s_2$, both $\cS_1$ and $\cS_2$ can be \emph{decoded} using constant time and one \emph{access} to $\cS$, i.e., a decoding algorithm recovers $K_1$ and $K_2$, and finds the starting locations of $M_1$ and $M_2$ within $M$ using constant time and one access to $\cS$.
}
\begin{proposition}\label{prop_concat_two}
	\contpropconcattwo
\end{proposition}
After decoding $\cS_1$ and $\cS_2$, any further access to the two strings can be performed as if they were stored explicitly.

\begin{proof}

	To concatenate two strings, let us first combine $K_1$ and $K_2$ into a single integer $K'\in\left[\range{K_1}\cdot\range{K_2}\right]$:
	\[
		K':=K_1\cdot \range{K_2}+K_2.
	\]
	If $s_1+s_2<\kappa+1$, we simply let $K=K'$, and let $M$ be the empty string.
	Then $\lg(\range{K})=s_1+s_2<\kappa+1$, $(M,K)$ is the concatenation.

	Next, we assume $s_1+s_2\geq \kappa+1$.
	In this case, in the final string $\cS=(M,K)$, $M$ will be the concatenation of $M_1$, $M_2$ and the lowest bits of $K'$.
	More specifically, let $|M|$ be $\lfloor s_1+s_2+2^{-\kappa+2}\rfloor-\kappa$, and let $\range{K}$ be $\lfloor 2^{\kappa+\fra(s_1+s_2+2^{-\kappa+2})}\rfloor$.\footnote{Recall that $\fra(x)=x-\lfloor x\rfloor$.}
	It is easy to verify that $|M|+\lg(\range{K})\leq s_1+s_2+2^{-\kappa+2}$ and $\range{K}\in[2^\kappa,2^{\kappa+1})$.

	We set 
	\[
		M:= M_1\circ M_2\circ (K'\Mod 2^{|M|-|M_1|-|M_2|})_2,
	\]
	where $(x)_2$ is the binary representation of $x$, and
	\[
		K:=K'\Div 2^{|M|-|M_1|-|M_2|}.
	\]

	To see why $K$ is at most $\range{K}-1$, we have
	\begin{align*}
		K&\leq (\range{K_1}\cdot\range{K_2}-1)\Div 2^{|M|-|M_1|-|M_2|} \\
		&=\left(2^{s_1-|M_1|+s_2-|M_2|}-1\right)\Div 2^{|M|-|M_1|-|M_2|} \\
		&\leq 2^{s_1+s_2-|M|} \\
		&\leq (\range{K}+1)\cdot 2^{-2^{-\kappa+2}} \\
		&\leq \range{K}-1,
	\end{align*}
	where the last inequality uses the fact that $2^{-2^{-\kappa+2}}\leq 1-2^{-\kappa+1}$ and $\range{K}\geq 2^\kappa$.

	\bigskip


	To decode $\cS_1$ and $\cS_2$, 
	observe that $\cS[|M_1|+|M_2|,|M|]$ encodes exactly $K'$.
	We access $\cS$ to retrieve its value, and compute $K_1$ and $K_2$ using $K_1:=K'\Div\range{K_2}$ and $K_2:=K'\Mod\range{K_2}$.
	By our construction, $M_1$ is $M\left[0,|M_1|-1\right]$ and $M_2$ is $M\left[|M_1|,|M_1|+|M_2|-1\right]$.
	Hence, we decode $\cS_1$ and $\cS_2$ in constant time and one access to $\cS$.
\end{proof}

Using ideas similar to~\cite{DPT10}, we show that multiple strings can be concatenated, allowing fast decoding of any given string.
\begin{proposition}\label{prop_concat_meta}
	Let $s_1,\ldots,s_B\geq \kappa$.
	Suppose there are numbers $\tT_0,\ldots,\tT_B$ such that
	\begin{compactitem}
		\item $\tT_0=0$ and $s_i+1\geq \tT_i-\tT_{i-1}\geq s_i+2^{-\kappa+2}$;
		\item each $\tT_i$ is of the form $\tT_i=\tm_i+\lg\tR_i$, where $\tR_i\in[2^{\kappa},2^{\kappa+1})$ and $\tm_i\geq 0$ are integers;
		\item for any given $i$, $\tT_i$ can be computed in $O(t)$ time.
	\end{compactitem}
	Then given $B$ strings $\cS_1,\ldots,\cS_B$, where $\cS_i=(M_i,K_i)$ has length $s_i$, they can be concatenated into one string $\cS=(M,K)$ of length $\tT_B$, and each $M_i$ is a (consecutive) substring of $M$.
	Moreover, for any given $i$, $\cS_i$ can be \emph{decoded} using $O(t)$ time and two accesses to $\cS$, i.e., a decoding algorithm recovers $K_i$, and finds the starting location of $M_i$ using $O(t)$ time and two accesses to $\cS$.	
\end{proposition}
\begin{proof}
	Without loss of generality, we assume $B$ is odd, since otherwise, we could first apply the following argument to the first $B-1$ strings, then apply Proposition~\ref{prop_concat_two} to concatenate the outcome with the last string $\cS_B$.

	To concatenate all strings, we break $\cS_2,\ldots,\cS_B$ into $(B-1)/2$ pairs, where the $j$-th pair consists of $\cS_{2j}$ and $\cS_{2j+1}$. 
	We start with the first string $\cS_1$, and repeatedly ``append'' the pairs to it.
	More specifically, let $\cS^{(0)}:=\cS_1$.
	Suppose we have concatenated $\cS_1$ and the first $j-1$ pairs into $\cS^{(j-1)}=(M^{(j-1)}, K^{(j-1)})$, such that $|M^{(j-1)}|=\tm_{2j-1}$ and $\range{K^{(j-1)}}=\tR_{2j-1}$.
	In particular, it has length $\tT_{2j-1}$.
	Now, we show how to ``append'' $\cS_{2j}$ and $\cS_{2j+1}$ to it.

	To this end, we combine $K_{2j}$ and $K_{2j+1}$ into a single integer $L_j$,
	\[
		L_j:= K_{2j}\cdot \range{K_{2j+1}}+K_{2j+1}.
	\]
	Thus, $\range{L_j}=\range{K_{2j}}\cdot \range{K_{2j+1}}$, and we have $\range{L_j}\in[2^{2\kappa}, 2^{2\kappa+2})$.
	Then, we re-break $L_j$ into a pair $(X_j, Y_j)$, such that the product of $\range{X_j}$ and $\range{K^{(j-1)}}$ is close to a power of two:
	we set
	\[
		\range{X_j}:=\left\lfloor \frac{2^{\lfloor \tT_{2j+1}\rfloor-\lfloor\tT_{2j-1}\rfloor-|M_{2j}|-|M_{2j+1}|}}{\range{K^{(j-1)}}}\right\rfloor,
	\]
	and
	\[
		\range{Y_j}:=\left\lceil \frac{\range{L_j}}{\range{X_j}}\right\rceil.
	\]
	Note that
	\[
		2\kappa\leq \lfloor \tT_{2j+1}\rfloor-\lfloor\tT_{2j-1}\rfloor-|M_{2j}|-|M_{2j+1}|\leq 2\kappa+4.
	\]
	To break $L_j$ into such a pair, we let $Y_j:=L_j\Div \range{X_j}$ and $X_j:=L_j\Mod \range{X_j}$.
	Next, we combine $K^{(j-1)}$ and $X_j$ into an integer $Z_j$ smaller than $2^{\lfloor \tT_{2j+1}\rfloor-\lfloor\tT_{2j-1}\rfloor-|M_{2j}|-|M_{2j+1}|}$: let $Z_j:=K^{(j-1)}\cdot \range{X_j}+X_j$.

	Finally, we let $\cS^{(j)}:=(M^{(j)},K^{(j)})$, where
	\[
		M^{(j)}:=M^{(j-1)}\circ (Z_j)_2\circ M_{2j}\circ M_{2j+1},
	\]
	and
	\[
		K^{(j)}:=Y_j.
	\]

	The length of $M^{(j)}$
	\begin{align*}
		|M^{(j)}|&=|M^{(j-1)}|+(\lfloor \tT_{2j+1}\rfloor-\lfloor\tT_{2j-1}\rfloor-|M_{2j}|-|M_{2j+1}|)+|M_{2j}|+|M_{2j+1}| \\
		&=\lfloor \tT_{2j+1}\rfloor-\kappa \\
		&=\tm_{2j+1}.
	\end{align*}
	The range of $K^{(j)}$ has size
	\begin{align*}
		\range{K^{(j)}}&<\frac{\range{L_j}}{\range{X_j}}+1 \\
		&\leq \frac{\range{K_{2j}}\cdot\range{K_{2j+1}}}{\frac{2^{\lfloor \tT_{2j+1}\rfloor-\lfloor\tT_{2j-1}\rfloor-|M_{2j}|-|M_{2j+1}|}}{\range{K^{(j-1)}}}-1}+1 \\
		&\leq \frac{\range{K_{2j}}\cdot\range{K_{2j+1}}\cdot \range{K^{(j-1)}}}{2^{\lfloor \tT_{2j+1}\rfloor-\lfloor\tT_{2j-1}\rfloor-|M_{2j}|-|M_{2j+1}|}}\cdot (1-2^{-\kappa+1})^{-1}+1 \\
		&=2^{s_{2j}+s_{2j+1}+\tT_{2j-1}-|M^{(j)}|}\cdot (1-2^{-\kappa+1})^{-1}+1 \\
		&\leq 2^{\kappa+\fra(\tT_{2j+1})-2^{\kappa+3}}\cdot (1-2^{-\kappa+1})^{-1}+1 \\
		&\leq 2^{\kappa+\fra(\tT_{2j+1})} \\
		&=\tR_{2j+1}.
	\end{align*}
	Thus, $\cS^{(j)}$ has length $\tT_{2j+1}$, and hence, the final string $\cS:=\cS^{((B-1)/2)}$ has length $\tT_{B}$.

	\bigskip

	Next, we show that each $\cS_i$ can be decoded in $O(t)$ time and two accesses to $\cS$.
	If $i=1$, we compute $\tT_3$ in $O(t)$ time, and then compute $\range{Z_1}$, $\range{X_1}$ and $|M_1|$.
	Thus, $M_1=M[0,|M_1|-1]$ and $Z_1$ is stored in $M$ immediately after $M_1$.
	By making one access to $\cS$, we recover the value of $Z_1$, and hence, $K_1$ can be computed using $K_1=Z_1\Div\range{X_1}$.
	This decodes $\cS_1$ in $O(t)$ time and one access to $\cS$.

	If $i>1$, let $j=\lfloor i/2\rfloor$, i.e., $\cS_i$ is in the $j$-th pair.
	We first compute $\tT_{2j-1}$ and $\tT_{2j+1}$ in $O(t)$ time.
	They determine $\range{Z_j}$, $\range{X_j}$ and $|M^{(j-1)}|$, as well as the starting location of $M_i$.
	Thus, $Z_j$ can be recovered with one access to $\cS$.
	$X_j$ can be computed using $X_j=Z_j\Mod \range{X_j}$.
	Similarly, we then recover $Z_{j+1}$, and $Y_j$ can be computed using $Y_j=K^{(j)}= Z_{j+1}\Div\range{X_{j+1}}$ (if $\cS_i$ is in the last pair, $Y_j$ is simply $K$ in the final string).
	This recovers both $X_j$ and $Y_j$.
	Next, we recover $L_j$ using $L_j=Y_j\cdot\range{X_j}+X_j$, and compute $K_{2j}$ and $K_{2j+1}$ using $K_{2j}=L_j\Div\range{K_{2j+1}}$ and $K_{2j+1}=L_j\Mod\range{K_{2j+1}}$.
	In particular, it recovers the value of $K_i$, and hence, it decodes $\cS_i$.
\end{proof}

\begin{restate}[Proposition~\ref{prop_concat_approx}]
	\contpropconcatapprox
\end{restate}

\begin{proof}
	Suppose we can compute $\tS_i=s_1+\cdots+s_i\pm2^{-\kappa}$.
	We set $\tm_i=\lfloor \tS_i+(i-1)\cdot 2^{-\kappa+3}\rfloor -\kappa$, $\tR_i=\lfloor 2^{\tS_i+(i-1)\cdot 2^{-\kappa+3}-\tm_i}\rfloor$ and $\tT_i=\tm_i+\lg\tR_i$.
	Then $\tT_i\leq\tS_i+(i-1)\cdot 2^{-\kappa+3}$ and $\tT_i>\tS_i+(i-1)\cdot 2^{-\kappa+3}-2^{-\kappa+1}$.
	Therefore,
	\begin{align*}
		\tT_i-\tT_{i-1}&\geq\tS_i-\tS_{i-1}+2^{-\kappa+3}-2^{-\kappa+1} \\
		&\geq s_i+2^{-\kappa+2}.
	\end{align*}
	Also, $\tT_i-\tT_{i-1}\leq s_i+2^{-\kappa+4}$.
	Finally, by Proposition~\ref{prop_concat_meta}, the size of the data structure is at most $\tT_B\leq \cS_B+(B-1)\cdot 2^{-\kappa+3}\leq s_1+\cdots+s_B+(B-1)\cdot 2^{-\kappa+4}$.
\end{proof}
In particular, by storing approximations of all $B$ prefix sums in a lookup table of size $O(B)$, the length of $\cS$ is at most $s_1+\cdots+s_B+(B-1)2^{-\kappa+4}$ and each $\cS_i$ can be decoded in $O(1)$ time.
\begin{restate}[Proposition~\ref{prop_concat}]
	\contpropconcat
\end{restate}
\begin{proof}
	If all $s_i\geq \kappa$, the proposition is an immediate corollary of Proposition~\ref{prop_concat_approx}, as we could simply store the approximations of all $B$ prefix sums.
	For general $s_i\geq 0$, we group the strings so that each group has length at least $\kappa$.

	We greedily divide all strings into groups:
	Pick the first $i_1$ such that $s_1+\cdots+s_{i_1}\geq \kappa$, then pick the first $i_2$ such that $s_{i_1+1}+\cdots+s_{i_2}\geq \kappa$, etc.
	Then each group has total length at least $\kappa$, possibly except for the last group.
	We store in the lookup table, which group each string belongs to, and the values of $i_1,i_2,\ldots$
	Then consider a group consisting of $\cS_a,\ldots,\cS_b$, we must have $s_{a}+\cdots+s_{b-1}<\kappa$, which means that they can be combined into one single integer smaller than $\prod_{i=a}^{b-1}\range{K_i}<2^{\kappa}$, e.g., 
	\[
		K:=\sum_{i=a}^{b-1}K_i\cdot \prod_{j=a}^{i-1}\range{K_j}.
	\]
	If we store $\prod_{j=a}^{i-1}\range{K_j}$ and $\range{K_i}$ in the lookup table for each $i$ in the group, then $K_i$ can be recovered from $K$ using
	\[
		K_i=(K \Div \prod_{j=a}^{i-1}\range{K_j})\Mod\range{K_i}.
	\]

	This concatenates all strings in the group except the last one.
	We then apply Proposition~\ref{prop_concat_two} to concatenate the last string in the group to it.
	Then we apply Proposition~\ref{prop_concat_approx} to concatenate the strings obtained from each group (except for the last group), using the lookup table.
	Finally, we concatenate the string obtained from the last group to it.

	Concatenating strings in each group loses at most $2^{-\kappa+2}$ due to Proposition~\ref{prop_concat_two}.
	The length of the final string is at most $s_1+\cdots+s_B+(B-1)2^{-\kappa+4}$.
	The lookup table has size $O(B)$.
\end{proof}


Next, we show that an integer $i\in[C]$ can be \emph{fused} into a string.

\begin{proposition}\label{prop_fuse_meta}
	Let $s_1,\ldots,s_C\geq 0$.
	Suppose there are numbers $\tT_1,\ldots,\tT_C$ such that
	\begin{compactitem}
		\item $2^{\tT_i}-2^{\tT_{i-1}}\geq 2^{s_i}$;
		\item each $\tT_i$ is of the form $\tT_i=\tm+\lg\tR_i$, where $\tm,\tR_i$ are integers;
		\item $\tT_C$ is a valid length, i.e., $\tm=0$ and $\tR_C\in[1, 2^{\kappa})$, or $\tm\geq 0$ and $\tR_C\in[2^{\kappa},2^{\kappa+1})$;
		\item for any given $K$, the largest $i\leq C$ such that $\tR_i\leq K$ can be computed in $O(t)$ time.
	\end{compactitem}
	Then given $i\in \{1,\ldots,C\}$ and string $\cS_i=(M_i, K_i)$ of length $s_i$, the pair $(i, \cS_i)$ can be stored in $\cS=(M,K)$ of length $\tT_C$, and $M_i$ is a (consecutive) substring of $M$.
	Moreover, we can recover the value of $i$ and \emph{decode} $\cS_i$ using $O(t)$ time and two accesses to $\cS$, i.e., a decoding algorithm recovers $i$, $K_i$, and finds the starting location of $M_i$ using $O(t)$ time and two accesses to $\cS$.
\end{proposition}

\begin{proof}
	Clearly, we have $s_i\leq \tT_C$ for all $i$, and hence, $|M_i|\leq \tm$.
	We first increase the length of $|M_i|$ to $\tm$ by appending the least significant bits of $K_i$ to it.
	That is, let
	\[
		M:=M_i\circ (K_i \Mod 2^{\tm-|M_i|})_2.
	\]
	Next, we encode the remaining information of $(i,\cS_i)$ in $K$, i.e., encode $i$ and the top bits of $K_i$:
	\[
		K:=\tR_{i-1}+(K_i\Div 2^{\tm-|M_i|}),
	\]
	where $\tR_0$ is assumed to be $0$.
	Note that we have
	\begin{align*}
		\tR_{i-1}+(\range{K_i}-1)\Div 2^{\tm-|M_i|}&<\tR_{i-1}+\range{K_i}\cdot 2^{|M_i|-\tm} \\
		&=\tR_{i-1}+2^{s_i-\tm} \\
		&=2^{-\tm}(2^{\tT_{i-1}}+2^{s_i}) \\
		&\leq 2^{\tT_i-\tm} \\
		&=\tR_i.
	\end{align*}
	That is, the value of $K$ determines both $i$ and $K_i\Div 2^{\tm-|M_i|}$, and $\range{K}$ is at most $\tR_C$.
	Thus, $\cS$ is a string of length $\tT_C$.

	\bigskip

	To decode $i$ and $\cS_i$, we first access $\cS$ to retrieve $K$.
	Then we compute the largest $i\leq C$ such that $\tR_i\leq K$ in $O(t)$ time.
	By the argument above, it recovers the value of $i$ and determines
	\[
		(K_i\Div 2^{\tm-|M_i|})=K-\tR_i.
	\]

	To decode $\cS_i$, observe that $M_i=M[0,|M_i|-1]$, and $M[|M_i|,\tm-1]$ stores the value of $K_i\Mod 2^{\tm-|M_i|}$.
	If $\tm-|M_i|\leq \kappa+1$, we retrieve its value using one access, and together with $K_i\Div 2^{\tm-|M_i|}$, it determines $K_i$.
	Otherwise, since $K_i<2^{\kappa+1}$, its value is entirely stored in $M$ (in its binary representation).
	We simply make one access to retrieve it.
	In both cases, we recover the value of $i$ and decode $\cS_i$ in $O(t)$ time and two accesses to $\cS$.
\end{proof}
\begin{restate}[Proposition~\ref{prop_fuse_approx}]
	\contpropfuseapprox
\end{restate}
\begin{proof}
	We compute $\tS_i=(2^{s_1}+\cdots+2^{s_i})\pm(2^{s_1}+\cdots+2^{s_C})\cdot 2^{-\kappa-3}$.
	If $2^{s_1}+\cdots+2^{s_C}<\kappa$, then the error term $(2^{s_1}+\cdots+2^{s_C})\cdot 2^{-\kappa-3}<1/8$.
	However, each $2^{s_1}+\cdots+2^{s_i}$ must be an integer by definition.
	$\tS_i$ rounded to the nearest integer is the \emph{accurate} value of $2^{s_1}+\cdots+2^{s_i}$.
	To apply Proposition~\ref{prop_fuse_meta}, we simply set $\tm:=0$, $\tR_i:=\lfloor\tS_i+1/2\rfloor$ for $i=1,\ldots,C$ and $\tT_i=\tm+\lg\tR_i$.
	It is easy to verify that $\tR_i-\tR_{i-1}\geq 2^{s_i}$; $\tT_C$ is a valid length.
	For any given $K$, by doing a binary search, the largest $i$ such that $\tR_i\leq K$ can be found in $O(t\lg C)$ time.
	Thus, by Proposition~\ref{prop_fuse_meta}, the pair $(i,\cS_i)$ can be stored using space 
	\[
		\tT_C=\lg(2^{s_1}+\cdots+2^{s_C}),
	\]
	and allowing $O(t\lg C)$ time for decoding.

	Next, we consider the case where $2^{s_1}+\cdots+2^{s_C}\geq\kappa$.
	To apply Proposition~\ref{prop_fuse_meta}, we let $\tT_C$ be the largest valid length smaller than $\lg\tS_C+C\cdot 2^{-\kappa+3}$.
	That is, we set
	\[
		\tm:=\lfloor \lg\tS_C+C\cdot 2^{-\kappa+3}\rfloor -\kappa.
	\]
	Then
	\[
		\tR_C:=\lfloor \tS_C\cdot2^{C\cdot 2^{-\kappa+3}}\cdot 2^{-\tm}\rfloor,
	\]
	and $\tT_C=\tm+\lg\tR_C$.
	Then for $i<C$, we let
	\[
		\tR_i:=\lfloor\tS_i\cdot 2^{-\tm}\rfloor+2(i-1),
	\]
	and $\tT_i=\tm+\lg\tR_i$.

	To apply Proposition~\ref{prop_fuse_meta}, we verify that $2^{\tT_i}-2^{\tT_{i-1}}\geq 2^{s_i}$.
	To see this, for $i<C$, we have
	\begin{align*}
		2^{\tT_i}-2^{\tT_{i-1}}&= 2^{\tm}\cdot \left(\tR_i-\tR_{i-1}\right) \\
		&\geq 2^{\tm}\cdot (\tS_i\cdot 2^{-\tm}-\tS_{i-1}\cdot 2^{-\tm}+1) \\
		&\geq 2^{s_i}+2^{\tm}-(2^{s_1}+\cdots+2^{s_C})\cdot 2^{-\kappa-2}.
	\end{align*}
	On the other hand, $\tS_C=(2^{s_1}+\cdots+2^{s_C})\cdot (1\pm2^{-\kappa-3})$, i.e., $2^{s_1}+\cdots+2^{s_C}=\tS_C\cdot (1\pm2^{-\kappa-3})^{-1}$.
	\begin{align*}
		2^{\tm}-(2^{s_1}+\cdots+2^{s_C})\cdot 2^{-\kappa-2}&\geq 2^{\tm}-\tS_C\cdot 2^{-\kappa-1}\geq 0.
	\end{align*}
	Thus, $2^{\tT_i}-2^{\tT_{i-1}}\geq 2^{s_i}$ for $i<C$.
	For $i=C$, it suffices to show $\lfloor \tS_C\cdot 2^{-\tm}\rfloor+2(C-1)\leq \tR_C$.
	Indeed, we have
	\begin{align*}
		\tR_C-(\lfloor \tS_C\cdot 2^{-\tm}\rfloor+2(C-1))&\geq \tS_C\cdot 2^{C\cdot 2^{-\kappa+3}}\cdot 2^{-\tm}-1-\tS_C\cdot 2^{-\tm}-2(C-1) \\
		&\geq \tS_C\cdot 2^{-\tm}\cdot (2^{C\cdot 2^{-\kappa+3}}-1)-2C \\
		&\geq \tS_C\cdot 2^{-\tm}\cdot C\cdot 2^{-\kappa+2}-2C.
	\end{align*}
	Since $\tm+\kappa\leq \lg\tS_C+1$, it is at least $0$.

	Since each $\tR_i$ can be computed in $O(t)$ time, by doing a binary search, for any given $K$, we can find the largest $i$ such that $\tR_i\leq K$ in $O(t\lg C)$ time.
	By Proposition~\ref{prop_fuse_meta}, we obtain a data structure of size
	\[
		\tT_C\leq \lg\tS_C+C\cdot 2^{-\kappa+3}\leq \lg(2^{s_1}+\cdots+2^{s_C})+C\cdot 2^{-\kappa+4}.
	\]
	This proves the proposition.
\end{proof}

Similar to the concatenation, the decoding algorithm takes constant time if we use a lookup table of size $O(C)$.
\begin{restate}[Proposition~\ref{prop_fuse}]
	\contpropfuse
\end{restate}
\begin{proof}
	Without loss of generality, assume $s_1\leq \cdots\leq s_C$, since otherwise, we simply sort $s_1,\ldots,s_C$ and store the permutation in the lookup table.

	To apply Proposition~\ref{prop_fuse_meta}, if $2^{s_1}+\cdots+2^{s_C}\leq 2^{\kappa}$, we set
	\[
		\tm:=0,
	\]
	\[
		\tR_i=2^{s_1}+\cdots+2^{s_i}
	\]
	and $\tT_i=\tm+\lg\tR_i$.
	Otherwise, if $2^{s_1}+\cdots+2^{s_C}>2^{\kappa}$, we set
	\[
		\tm:=\left\lfloor \lg(2^{s_1}+\cdots+2^{s_C})+C\cdot 2^{-\kappa+2} \right\rfloor-\kappa,
	\]
	for $i<C$, let
	\[
		\tR_i:=\lceil2^{s_1-\tm}\rceil+\cdots+\lceil2^{s_i-\tm}\rceil,
	\]
	and 
	\[
		\tR_C:=\max\left\{\lceil2^{s_1-\tm}\rceil+\cdots+\lceil2^{s_C-\tm}\rceil, 2^{\kappa}\right\}.
	\]
	Finally, let $\tT_i=\tm+\lg\tR_i$.
	Clearly, in both cases, we have $2^{\tT_i}-2^{\tT_{i-1}}\geq2^{\tm}\cdot 2^{s_i-\tm}\geq 2^{s_i}$.
	Also, we have $\tT_C\leq \lg(2^{s_1}+\cdots+2^{s_C})+C\cdot 2^{-\kappa+2}$.
	This is because
	\begin{align*}
		\tR_C&< \max\{(2^{s_1}+\cdots+2^{s_C})\cdot 2^{-\tm}+C,2^{\kappa}\} \\
		&=\max\{2^{\kappa+\fra(\lg(2^{s_1}+\cdots+2^{s_C})+C2^{-\kappa+2})-C2^{-\kappa+2}}+C,2^{\kappa}\} \\
		&\leq \max\{2^{\kappa+\fra(\lg(2^{s_1}+\cdots+2^{s_C})+C2^{-\kappa+2})}\cdot (1-C\cdot 2^{-\kappa+1})+C,2^{\kappa}\} \\
		&\leq 2^{\kappa+\fra(\lg(2^{s_1}+\cdots+2^{s_C})+C2^{-\kappa+2})}.
	\end{align*}
	Thus, $\tT_C=\tm+\lg\tR_C\leq \lg(2^{s_1}+\cdots+2^{s_C})+C\cdot 2^{-\kappa+2}$.
	\bigskip

	To apply Proposition~\ref{prop_fuse_approx}, we need to show that for any given $K$, the largest $i$ such that $\tR_i\leq K$ can be found in constant time.
	To this end, we store a \emph{predecessor search} data structure for the set $\{\tR_1,\ldots,\tR_C\}$.
	Note that the set of integers $\{\tR_1,\ldots,\tR_C\}$ has \emph{monotone gaps}.
	That is, the difference between adjacent numbers is non-decreasing.
	P\v{a}tra\c{s}cu~\cite{Pat08} showed that for such sets, there is a predecessor search data structure using linear space and constant query time, i.e., there is an $O(C)$-sized data structure such that given an integer $K$, the query algorithm can answer in constant time the largest value in the set that is at most $K$.
	This data structure is stored in the lookup table (it only depends on $s_1,\ldots,s_C$, but not the input string).
	To compute the index $i$ rather than $\tR_i$, we simply store another hash table using perfect hashing in the lookup table.
	Hence, the lookup table has size $O(C)$.

	The premises of Proposition~\ref{prop_fuse_approx} are all satisfied.
	The size of $\cS$ is $\tT_C\leq \lg(2^{s_1}+\cdots+2^{s_C})+C2^{-\kappa+2}$, and $i$ and $\cS_i$ can be decoded in constant time.
	This proves the proposition.
\end{proof}

Next, we show that it is possible to divide a binary string into two substrings. 
\begin{restate}[Proposition~\ref{prop_divide}]
	\contpropdivide
\end{restate}

\begin{proof}
	We first calculate the length of $M_1$ and $M_2$, let $|M_1|:=\lfloor s_1-\lg(\range{K_h})\rfloor-\kappa$ and $M_2:=\lfloor s_2-\lg(\range{K_t})\rfloor-\kappa$.
	Then let 
	\[
		(K_{h,1}, M_1):=\cS[-1,|M_1|-1]
	\]
	be a prefix, and 
	\[
		(M_2, K_{t,2}):=\cS[|M|-|M_2|,|M|]
	\] be a suffix.
	The remaining task is to divide the middle $|M|-|M_1|-|M_2|$ bits of $M$ into $K_{t,1}$ and $K_{h,2}$.
	
	To this end, we represent the middle bits as an integer $L$ in the range $[2^{|M|-|M_1|-|M_2|}]$.
	The sizes of ranges of $K_{t,1}$ and $K_{h,2}$ can be calculated using 
	\[
		\range{K_{t,1}}=\lfloor2^{s_1-\lg(\range{K_h})-|M_1|}\rfloor
	\]
	and
	\[
		\range{K_{h,2}}=\lfloor 2^{s_2-\lg(\range{K_t})-|M_2|}\rfloor.
	\]
	
	Then let $K_{t,1}:=L\Mod \range{K_{t,1}}$ and $K_{h,2}:=L\Div \range{K_{t,1}}$.
	Clearly, $K_{t,1}\in[\range{K_{t,1}}]$.
	It suffices to show that $K_{h,2}$ is in its range:
	\begin{align*}
		K_{h,2}&<\frac{2^{|M|-|M_1|-|M_2|}}{2^{s_1-\lg(\range{K_h})-|M_1|}-1} \\
		&=\frac{2^{|M|-|M_2|-s_1+\lg(\range{K_h})}}{1-2^{-s_1+\lg(\range{K_h})+|M_1|}} \\
		&=\frac{2^{s-|M_2|-s_1-\lg(\range{K_t})}}{1-2^{-\kappa}} \\
		&\leq \frac{2^{s_2-|M_2|-\lg(\range{K_t})-2^{-\kappa+2}}}{1-2^{-\kappa}} \\
		&<(\range{K_{h,2}}+1)\cdot\frac{2^{-2^{-\kappa+2}}}{1-2^{-\kappa}} \\
		&\leq\range{K_{h,2}}\cdot \frac{(1+2^{-\kappa}) (1-2^{-\kappa+1})}{1-2^{-\kappa}}\\
		&<\range{K_{h,2}}.
	\end{align*}
	Thus, $\cS_1$ has at most $s_1$ bits and $\cS_2$ has at most $s_2$ bits.
	This proves the proposition.
\end{proof}

Finally, we show that the inverse of fusion can be done efficiently.
\begin{restate}[Proposition~\ref{prop_sep}]
	\contpropsep
\end{restate}
\begin{proof}
	By setting $K_{i,t}:=K_t$, the task becomes to encode $(K_h, M)$ using $(i, (K_{i,h}, M_i))$.
	Next, we show how to determine $i$. 
	To this end, we divide the range of $K_{h}$ into $C$ disjoint intervals $\{[l_i,r_i)\}_{i=1,\ldots,C}$, such that the $i$-th interval has size at most
	\[
		\lfloor 2^{s_i-|M|-\lg(\range{K_t})}\rfloor.
	\]
	Such division is possible, because
	\begin{align*}
		\sum_{i=1}^C\lfloor 2^{s_i-|M|-\lg(\range{K_t})}\rfloor&>\sum_{i=1}^C 2^{s_i-|M|-\lg(\range{K_t})} -C \\
		&\geq 2^{-|M|-\lg(\range{K_t})}\cdot 2^{s+(C-1)\cdot 2^{-\kappa+2}}-C \\
		&\geq 2^{s-|M|-\lg(\range{K_t})}\cdot (2^{(C-1)\cdot 2^{-\kappa+2}}-C\cdot 2^{-\kappa})\\
		&\geq \range{K_h}\cdot (1+(C-1)2^{-\kappa+1}-C\cdot 2^{-\kappa}) \\
		&\geq \range{K_h}.
	\end{align*}
	Fix one such division, e.g., the $i$-th interval is from
	\[
		l_i:=\min\{\lfloor 2^{s_1-|M|-\lg(\range{K_t})}\rfloor+\cdots+\lfloor 2^{s_{i-1}-|M|-\lg(\range{K_t})}\rfloor, \range{K_h}-1\}
	\]
	to
	\[
		r_i:=\min\{\lfloor 2^{s_1-|M|-\lg(\range{K_t})}\rfloor+\cdots+\lfloor 2^{s_{i}-|M|-\lg(\range{K_t})}\rfloor, \range{K_h}-1\}
	\]
	excluding the right endpoint.
	We store all endpoints $l_i,r_i$ in the lookup table, taking $O(C)$ space.

	Now, find $i$ such that $K_h\in[l_i,r_i)$.
	Then compute $|M_i|=\lfloor s_i-\lg(\range{K_t})\rfloor-\kappa$, and let 
	\[
		M_i:=M[|M|-|M_i|,|M|-1].
	\]
	Finally, we view the first $|M|-|M_i|$ bits of $M$ as a nonnegative integer $Z\in[2^{|M|-|M_i|}]$
	and let
	\[
		K_{i,h}:=2^{|M|-|M_i|}\cdot \left(K_h-l_i\right)+Z.
	\]
	Observe that $K_{i,h}<\lfloor 2^{\kappa+\fra(s_i-\lg(\range{K_t}))}\rfloor$, because
	\begin{align*}
		K_{i,h}&<2^{|M|-|M_i|}\cdot (r_i-l_i) \\
		&\leq 2^{|M|-(\lfloor s_i-\lg(\range{K_t})\rfloor-\kappa)}\cdot 2^{s_{i}-|M|-\lg(\range{K_t})} \\
		&\leq 2^{\kappa+\fra(s_i-\lg(\range{K_t}))}.
	\end{align*}
	Thus, the length of $\cS_i=(K_{i,h},M_i,K_{i,t})$ is at most
	\[
		\lg(\range{K_{i,h}})+|M_i|+\lg(\range{K_{i,t}})\leq s_i.
	\]
	We also store the sizes of $\cS_i$ for every $i$ in the lookup table.

	\bigskip

	It is clear that $(M_i,K_{i,t})$ is a suffix of $\cS$.
	Given $i$ and $K_{i,h}$, we retrieve $l_i$ and $M_i$ from the lookup table.
	Then $\cS[-1]=K_h$ can be recovered using
	\[
		K_h=l_i+K_{i,h}\Div 2^{|M|-|M_i|}.
	\]
	Also, $Z$ can be recovered using
	\[
		Z=K_{i,h}\Mod 2^{|M|-|M_i|},
	\]
	which determines $\cS[0,|M|-|M_i|-1]$.
	This proves the proposition.
\end{proof}

\section{Approximating Binomial Coefficients}\label{sec_binom_approx}
In this section, we prove Claim~\ref{cl_dictrec_concat} and Claim~\ref{cl_dictrec_fuse} from Section~\ref{sec_good_block}.

\begin{restate}[Claim~\ref{cl_dictrec_concat}]
	\contcldictrecconcat
\end{restate}

\begin{proof}(sketch)
For Claim~\ref{cl_dictrec_concat}, the goal is essentially to efficiently approximate
\[
	s_1=\OPT_{(k-i+1) V_\bl ,m_1}-(k-i+1)\smain+(m_1-1)2^{-\kappa/2+2}
\]
and
\[
	s_2=\OPT_{(j-k) V_\bl ,m_2}-(j-k)\smain+(m_2-1)2^{-\kappa/2+2}.
\]


To approximate $s_1$ and $s_2$, we can store an approximation of $\smain$ up to $O(\kappa)$ bits of precision in the lookup table.
The task reduces to approximate the two $\OPT$s.
Recall that 
\[
	\OPT_{V,m}=\lg\binom{V}{m}.
\]
The problem further reduces to approximate $\lg\binom{(k-i+1)V_\bl}{m_1}$ and $\lg\binom{(j-k)V_\bl}{m_2}$.
In the following, we show that for any given $V,m\leq 2^{\kappa}$, it is possible to approximate $\lg\binom{V}{m}$ in $O(1)$ time.

$\lg\binom{V}{m}$ can be expanded to $\lg V!-\lg m!-\lg (V-m)!$.
We approximate each term separately.
By Stirling's formula, 
\[
	\ln k!=k\ln\left(\frac{k}{e}\right)+\frac{1}{2}\ln 2\pi n+\sum_{i=2}^{d}\frac{(-1)^iB_i}{i(i-1)k^{i-1}}+O(k^{-d}),
\]
where $B_i$ is the $i$-th Bernoulli number, and $d\geq 2$.
For any constant $\epsilon>0$, by setting $d\geq \Omega(1/\epsilon)$, the above approximation gives an error of $2^{-\Omega(\kappa)}$ for any $k\geq 2^{\epsilon\kappa}$.
We store the Bernoulli numbers in the lookup table, and the formula can be evaluated in constant time.
On the other hand, for all $k<2^{\epsilon\kappa}$, we simply store an approximation of $\lg k!$ in a global lookup table, taking $2^{\epsilon\kappa}$ size.
Finally, by approximating $\lg V!$, $\lg m!$ and $\lg(V-m)!$ independently with additive error $2^{-2\kappa-2}$, we obtain an estimation of $\lg\binom{V}{m}$ with additive error smaller than $2^{-2\kappa}$.
Note that each of the three values may be $2^{\omega(\kappa)}$, which takes super-constant words to store.
However, since the final value is guaranteed to be at most $2^\kappa$, we could safely apply mod $2^{\kappa}$ over the computation.
\end{proof}

\begin{restate}[Claim~\ref{cl_dictrec_fuse}]
	\contcldictrecfuse
\end{restate}
\begin{proof}(sketch)
	The goal is to approximate
	\[
		\sum_{i=0}^l \binom{V_1}{i}\binom{V_2}{m-i}
	\]
	up to additive error of $2^{-\kappa-3}\cdot \binom{V_1+V_2}{m}$, because
	\[
		2^{\OPT_{V_1,i}+\OPT_{V_2,m-i}}=\binom{V_1}{i}\binom{V_2}{m-i}.
	\]


	To this end, we shall use the following lemma from~\cite{Yu19} to approximate binomial coefficients.
	\begin{lemma}[\cite{Yu19}]\label{lem_approx_binom}
		For any large integers $V$, $d$ and $0<a\leq V/2$, such that $d\leq c\cdot a$, there is a polynomial $P$ of degree $d$, such that
		\[
			\binom{V}{a+x}\leq \binom{V}{a}\cdot \left(\frac{V-a}{a}\right)^x\cdot P_{V,d}(x)\leq \binom{V}{a+x}\cdot (1+2^{-\sqrt{d}+8}),
		\]
		for all integers $x\in[0, c\cdot \sqrt{a}]$, a (small) universal constant $c>0$.
		Moreover, given $V$ and $d$, the coefficients of $P_{V,d}$ can be computed in $O(d^{1.5})$ time.
	\end{lemma}
	This lemma allows us to approximate $\sum_{l=a}^b \binom{V_1}{l}\binom{V_2}{m-l}$ where $b-a\leq c\cdot \sqrt{a}$, up to a \emph{multiplicative} error of $1\pm 2^{-2\kappa}$ in $O(\kappa^4)$ time: it reduces approximating the sum to computing $\sum_{l} \alpha^l\cdot P_1(l)P_2(l)$ for two degree-$O(\kappa^2)$ polynomials $P_1,P_2$.

	Let $\overline{m}=\frac{V_1}{V_1+V_2}\cdot m$.
	For $l<\overline{m}-2\sqrt{\overline{m}\cdot \kappa}$, we return $0$ as the approximation;
	For $\overline{m}-2\sqrt{\overline{m}\cdot \kappa}\leq l\leq \overline{m}+2\sqrt{\overline{m}\cdot \kappa}$, we divide the range into chunks of size $O(\sqrt{\overline{m}})$, apply Lemma~\ref{lem_approx_binom} to approximate $\sum_l \binom{V_1}{l}\binom{V_2}{m-l}$ for each chunk in $O(\kappa^4)$ time, and return the sum;
	For $l>\overline{m}+2\sqrt{\overline{m}\cdot \kappa}$, we return (an approximation of) $\binom{V_1+V_2}{m}$ as the estimation.
	It is not hard to verify that in all cases we return an approximation with desired error.
	The details are omitted.
\end{proof}

\section{Dictionary with Linear Redundancy}\label{sec_proof_large_bad_block}
In this section, we show a proof sketch of Lemma~\ref{lem_large_bad_block}, and present a dictionary data structure that uses a linear number of extra bits.
Recall that $\OPT_{V,m}:=\lg\binom{V}{m}$.
For membership queries only, Pagh~\cite{Pagh01} already obtained a better data structure.
The data structure in this section is a generalization of Pagh's static dictionary.

\begin{restate}[Lemma~\ref{lem_large_bad_block}]
	\contlemlargebadblock
\end{restate}
We are going to use Pagh's static dictionary as a subroutine.
For this reason, let us first give an overview of this data structure.
The data structure uses a minimal perfect hashing of Schmidt and Siegel~\cite{SS90}.
The hashing has three levels.
In the first level, each key $x$ is mapped to $h_{k,p}(x)=(kx\Mod p)\Mod m^2$ \emph{with no collisions}, for a prime $p=\Theta(m^2\lg V)$ and $k\in[p]$.
A random pair $(k, p)$ works with constant probability, and it takes $O(\lg m+\lg\lg V)$ bits to encode the function.
This level effectively reduces the universe size from $V$ to $m^2$.
Each key $x\in S$ is then represented by a pair $(x^{(1)},x^{(2)})$ where $x^{(1)}\in[m^2]$ is the hash value, and $x^{(2)}=(x\Div p)\cdot \lceil p/m^2\rceil+(kx\Mod p)\Div m^2$ (called the quotient function in~\cite{Pagh01}).
Then $x^{(2)}\leq O(V/m^2)$ and $(x^{(1)},x^{(2)})$ uniquely determines $x$.

In the second level, we apply another hash function from the same family on $x^{(1)}$, $h_{k',p'}(x^{(1)})=(k'x^{(1)}\Mod p')\Mod m$ to map $x^{(1)}$ to $m$ buckets.
This time, we have $p'=\Theta(m^2)$ and $k'\in[p']$.
Let $A_i$ be the number of keys mapped to bucket $i$.
The hashing guarantees that for a random pair $(k', p')$, the expectation of each $A_i^2$ is bounded by $O(1)$.
Similarly, we can represent $x^{(1)}$ further as a pair such that the first component is the hash value in $[m]$, and the second component is the quotient function value, which is at most $O(m)$.

The third level hashing then hashes all keys in the same bucket to different integers.
It is applied on $x^{(1)}$: $g_{k_i,p_i}(x^{(1)})=(k_ix^{(1)}\Mod p_i)\Mod A_i^2$, for $p_i=\Theta(m^2)$ and $k_i\in[p_i]$ such that all keys in the bucket are mapped to different integers.
It turns out that a random pair $(k_i,p_i)$ works with constant probability.

The data structure stores the following for the hash functions: 
\begin{compactenum}
	\item the top-level hash functions $(k, p)$ and $(k', p')$,
	\item a list of $O(\lg m)$ (random) choices for the third-level hash functions $(k_1,p_1),(k_2,p_2),\ldots$,
	\item for each bucket $i$, the index $\pi_i$ of the first hash function in the list that works.
\end{compactenum}
It turns out that it is possible to use only $O(m)$ \emph{bits} to store the indices $\pi_i$.
This is because each second-level hash function works with constant probability, the entropy of each $\pi_i$ is a constant.
We can use the Huffman coding for each $\pi_i$ to achieve constant bits per index (which turns out to be the unary representation of $\pi_i$).

These hash functions map all $m$ input keys to $O(m)$ buckets with \emph{no} collisions.
By storing a rank data structure (e.g.,~\cite{Pat08}) among the $O(m)$ buckets using $O(m)$ bits of space, we further map all the non-empty buckets to $[m]$.
Finally, we store for each bucket, the \emph{quotient functions} of the input key mapped to it.
Hence, it take $\lg (V/m^2)+\lg m+O(1)=\lg (V/m)+O(1)$ bits to encode each key.
Thus, the total space is $m\lg (V/m)+O(m+\lg\lg V)=\lg\binom{V}{m}+O(m+\lg\lg V)$ bits.

\bigskip

This data structure supports membership queries, and naturally defines a bijection $h$ between $S$ and $[m]$, namely $h(x)$ simply being the bucket $x$ is mapped to.
To generalize the data structure and define an efficiently computable bijection $\oh$ between $[V]\setminus S$ and $[V-m]$, we apply an approach similar to Section~\ref{sec_bad_block}.
To this end, we first store the number of keys $m'$ in $[V-m]$.
This is also the number of non-keys in $\{V-m,\ldots,V-1\}$.
We are going to store a mapping that maps all $m'$ non-keys in $\{V-m,\ldots,V-1\}$ to all $m'$ keys in $[V-m]$.

We then store the above data structure for all keys in $[V-m]$, using 
\[
	m'\lg ((V-m)/m')+O(m'+\lg\lg V)\leq \lg\binom{V}{m}+O(m+\lg\lg V)
\]
bits, which defines a bijection $h'$ between $S\cap [V-m]$ and $[m']$.
Note that this data structure also allows us to ``randomly access'' all keys.
That is, given an index $i\in[m']$, it returns a key $x_i$, such that $\{x_1,\ldots,x_{m'}\}$ is the set of all $m'$ keys in $[V-m]$.
Then, we store a rank data structure for $\{V-m,\ldots,V-1\}$, such that given an $x\in\{V-m,\ldots,V-1\}$, the query algorithm returns if $x$ is a key, as well as its rank over the set of keys (or non-keys).
Hence, it maps all keys in $\{V-m,\ldots,V-1\}$ to $[m-m']$ and all non-keys to $[m']$.
The total space is $\OPT_{V,m}+O(m+\lg\lg V)$.

For each $x\in S$, we define $h(x)$ as follows.
\begin{compactitem}
	\item if $x<V-m$, let $h(x):=h'(x)$;
	\item if $x\geq V-m$, let $h(x)$ be $m'-1$ plus the rank of $x$ in $S\cap \{V-m,\ldots,V-1\}$.
\end{compactitem}
For $x\notin S$, we define $\oh(x)$ as follows.
\begin{compactitem}
	\item if $x<V-m$, let $h(x):=x$;
	\item if $x\geq V-m$, suppose the rank of $x$ in $\{V-m,\ldots,V-1\}\setminus S$ is $i$, then let $h(x):=x_i$.
\end{compactitem}
Having stored the above data structures, $h(x)$ or $\oh(x)$ can be computed in constant time.

\end{document}